\documentstyle[graphicx]{elsart}
\newcommand\ba{\begin{eqnarray}}
\newcommand\ea{\end{eqnarray}}
\newcommand\nn{\nonumber}
\newcommand{\be}{\begin{equation}}
\newcommand{\ee}{\end{equation}}

\begin{document}

\begin{frontmatter}

\title{Relativistically invariant analysis of $\Delta $--isobar
production in deuteron electrodisintegration:
$e^-+d\to e^-+\Delta +N:$
general analysis of polarization effects}
\author{G. I. Gakh }
\thanks[ ]{e-mail: \it gakh@kipt.kharkov.ua}
\address{\it National Science Centre "Kharkov Institute of Physics and Technology", Akademicheskaya 1,  61108 Kharkov, Ukraine }
\author{E. Tomasi-Gustafsson}
\address{IRFU,SPhN, Saclay, 91191 Gif-sur-Yvette France and \\
CNRS/IN2P3, Institut de Physique Nucl\'eaire, UMR 8608, 91405 Orsay, France}
\thanks[ ]{e-mail: {\it etomasi@cea.fr.}, corresponding author.}
\author{A. G. Gakh}
\address{Kharkov National University, 61077 4 Svobody Sq., Kharkov, Ukraine}

\begin{abstract}
The differential cross section
and the polarization observables for $\Delta $--isobar production in the
deuteron electrodisintegration process, $e^-+d\to e^-+\Delta +N$, are calculated in a general formalism based on structure functions. The obtained expressions have a general nature, hold  for one--photon--exchange, assuming P--invariance of the electromagnetic interaction and the conservation of the hadron electromagnetic current. The dependence of the differential cross section of the $e^-+d\rightarrow e^- +\Delta +N$ 
reaction on the vector and tensor polarizations of the deuteron target with 
unpolarized and longitudinally polarized electrons is considered. The general dependence of the asymmetries on two of five kinematic variables, the azimuthal angle $\varphi$ and $\epsilon$ 
(linear polarization of the virtual photon) is calculated. A similar analysis 
is performed for the polarization of the nucleon produced in $\gamma ^*d
\rightarrow \Delta N$ reaction provided the electron beam is unpolarized or 
longitudinally polarized. Polarization effects, which are due to the strong $\Delta N-$ 
interaction in the final state are calculated. The photoproduction of the $\Delta $--isobar on the deuteron target has been 
considered in detail, as a particular case. The differential cross section and various polarization 
observables have been derived in terms of the reaction amplitudes. The 
polarization observables due to the linear and circular polarizations of the 
photon, when the deuteron target is arbitrarily polarized have been derived 
in terms of the reaction amplitudes. The polarization of the final nucleon is 
also considered.
\end{abstract}
%\PACS 13.60.-r, 13.40.Gp, 13.88.+e, 25.30.Dh, 24.30.Gd
\end{frontmatter}
\maketitle
%%%%%%%%%%%%%%%%%%%%%%%%%%%%
\section{Introduction}
%%%%%%%%%%%%%%%%%%%%%%%%%%%%%%%%
It is well known, that the $\Delta $ resonance dominates in the pion production
processes and plays an important role in the physics driven by the strong interaction. Moreover, the mechanism of $\Delta $ resonance excitation has a dominant role in various
nuclear phenomena at energies higher than the pion--production threshold (for
the details see the reviews \cite{BW75}).

High--precision measurements of the $N\to \Delta $ transition induced by 
photon (real or virtual) became possible with the availability 
of  high--intensity GeV--energy electron beam facilities (such as Jefferson
Laboratory, Bates, ELSA, MAMI) and of the high performance spectrometers,
detectors and polarimeters (for recent review see \cite{PVY}). It was predicted
theoretically and proved experimentally that the electromagnetic $N\to \Delta $ 
transition is dominated by the magnetic ($M1$) dipole transition and that the two other quadrupole transitions (electric ($E2$) and Coulomb ($C2$)) are small. At moderate $Q^2$ (transfer momentum squared) the ratios $E2/M1$ and $C2/M1$ are at the level of a few percent.

The data on the $\pi^-p$ channel (both for the case of the photoproduction and
electroproduction) from a neutron (deuterium) target are rather scarce, but of current interest in various accelerators. For example, recent results have been obtained to measure cross sections and various polarization observables for the $\pi^+n$
\cite{V02} and $\pi^-p$ \cite{K02} channels and a program is ongoing at Jlab (CLAS collaboration). 

The processes of the $\Delta $--isobar excitation in the scattering of the
electrons by nuclei, $e^-+A\to e^-+\Delta +A'$, where $A$ and $A'$ are nuclear  states, as well as the processes $e^-+A\to e^-+N+A"$, involve both hadron electrodynamics and nuclear dynamics. Thus,
the simplest process of the $\Delta $--isobar production in $eA$--collisions,
$e^-+d\to e^-+\Delta +N$, brings information on the electromagnetic
form factors of the $\Delta^0\to n+\gamma^*$ transition ($\gamma^*$ is  the
virtual photon). The investigation of the pion photo-- and electroproduction 
off the neutron (deuteron) allows to determine the isotopic properties 
of the hadron electromagnetic current (namely, search for the isotensor 
contributions in the $\Delta $--isobar excitation), provided the background 
contributions can be sufficiently controlled \cite{KS03}. The amplitude for 
the excitation of the $\Delta $--isobar on the free proton and on the free 
neutron is identical as long as isotensor components can be neglected. 
The comparison of the electromagnetic form factors for two transitions  
$\Delta^+\to p+\gamma^*$ and $\Delta^0\to n+\gamma^*$ was used earlier to 
estimate the possible admixture of the isotensor component in this current 
\cite{SS71}.

For the purpose of nuclear physics itself, 
the process $e^-+A\to e^-+ \Delta +A'$ is important, first of all, for clarifying the 
old standing problem of non--nucleonic degrees of freedom in the nuclei and 
their role in various processes involving  nuclei \cite{WA78}. The problem of the $\Delta $--isobar behavior in the 
nuclear medium \cite{W77} is of great interest too. In particular, for the 
deuteron, despite the large precision and the new region of internal momentum 
explored, these questions (the relative role of different possible components 
in the deuteron wave function as the $\Delta\Delta $--configurations, 
$6q$--components, etc. are still under discussion \cite{TGR}).

When addressing the problem of non--nucleonic degrees of 
freedom (in particular, the signals of the $\Delta $--isobar excitation) in nuclear phenomena, it is natural to start these investigations from the most
simple nuclear system, the two--nucleon system (deuteron). Using this approach,
one may try to see whether the explicit inclusion of the resonance degrees of 
freedom (the most important $\Delta $--isobar) in the nuclear wave functions
improves our knowledge of the nuclear physics.

Due to the isospin conservation in the strong interactions, the $\Delta $--isobar can contribute to the deuteron ground wave function in form of 
$\Delta\Delta $--admixture. One should note that problems connected with high spin
of the $\Delta $--isobar, as the non--physical components of the
$\Delta $--isobar wave function are discussed in the literature \cite{LL87}. For on--mass--shell 
$\Delta $--isobar these components can be removed by imposing additional 
constraints on the Rarita--Schwinger field describing the $\Delta $--isobar 
\cite{JM92}. But in nuclei the $\Delta $--isobar must be off--mass--shell and, 
therefore, unphysical degrees of freedom are still possible. Therefore, to 
reduce the ambiguities inherent to any model--dependent analysis, it is natural 
to look for the $\Delta $--isobar in few--nucleon systems, and/or in processes 
where this degree of freedom is excited through the simpler and well--known 
electromagnetic and weak interactions. The $\Delta $--isobar can contribute 
both to the one--body current through direct coupling to the photon (real or 
virtual), and to the two--body currents, much in the same way as it comes into 
play in the intermediate nucleon--nucleon interaction. Exclusive reactions are 
expected to convey more information than simpler inclusive experiments, even if 
they are obviously more demanding for the theoretical analysis.

Pioneering calculations of the $\Delta\Delta $--admixture in the deuteron wave 
function have been done using static transition potentials with pion--exchange 
only \cite{ADW} and it was found that the  $\Delta $--percentage $P_{\Delta}$ is of the order of $1\%$. Later on some improvements of this simple approach were made: inclusion of the 
$\rho $--meson exchange \cite{GH74} and using a coupled--channel approach 
\cite{HGDD}. It was found that $P_{\Delta}\approx (0.4-0.8) \%$. The possible  
$\Delta $--isobar configurations in nuclear wave functions have been searched 
for the three--nucleon systems, $^3He$ and $^3H$.

As it was noted \cite{WA78}, resonances are a more efficient source of 
high momentum components in nuclear wave functions than the short--range 
repulsion in the nuclear force. As a result, a typical resonance Fermi momentum 
is about 0.3 to 0.6 GeV/c as compared to 0.1 GeV/c for a nucleon in the
deuteron.

The experimental investigation of the polarization effects in various processes
of the electron--deuteron scattering has been started some time ago. The 
experiments were done on the elastic scattering of unpolarized electrons by the
tensor polarized deuteron target and also on the measurement of the tensor 
polarization of the recoil deuteron (for the purpose to separate the charge 
and quadrupole deuteron electromagnetic form factors) (for the details see the 
reviews \cite{SGG}). To determine the neutron charge electromagnetic form
factor $G_{En}$, several polarization measurements were done for the 
deuteron electrodisintegration process (see, for example, \cite{GRTG}). 

The combination of $4\pi $ detectors with linearly and circularly polarized 
photon beams as well as polarized targets will provide access to new
 observables, very powerful for the extraction of specific resonance properties. The use of polarized proton and deuteron targets will allow measurement of 
double and triple polarization observables with polarized neutrons \cite{B07}.

In Ref. \cite{IOTF} the non--nucleonic degrees of freedom
in terms of the $\Delta $--isobar were investigated and the contribution of
the $\Delta\Delta $--component of the deuteron wave function was calculated in 
the framework of the Nambu--Jona--Lasinio model of light nuclei. It was found 
that $P_{\Delta}= 0.3 \%$. This prediction agrees well with the experimental 
estimate $P_{\Delta}\leq 0.4 \%$ at $90 \%$ of confidence level \cite{D86}.

The experimental measurement of the $\Delta\Delta $--admixture in the deuteron 
wave function was done in a number of experiments which investigate pure 
hadronic reactions and processes induced by leptons as well. The analysis of 
the final state $\bar NNN\pi $ in the interaction of the 
antiproton beam with deuteron target lead to the result $P_{\Delta}=16 \%$
\cite{B74}. The $dp\to NNN\pi $ reactions were studied in \cite{AL75} where an
upper limit $P_{\Delta}\leq (1.1\pm 0.3) \%$ was obtained. The processes of
interaction of the positive (negative) pion with deuteron \cite{E74}
(\cite{B76}) gave an upper limit of $0.8 \% (0.4 \%)$ for the 
$\Delta\Delta $--component of the deuteron wave function. In Ref. 
\cite{BS74} the process of the inclusive $\Delta $--isobar 
photoexcitation on the deuteron target, $\gamma d\to \Delta X$, was studied 
 obtaining $P_{\Delta}\approx 3 \%$ . The latest estimate $P_{\Delta}\leq 0.3\%$ was made
in the analysis of the neutrino--deuteron interaction \cite{D86}.

Note also that many neutrino oscillation experiments were done in the
kinematical region corresponding to a neutrino beam energy $\approx 1 GeV$. 
But in this kinematical region the inelastic processes (mainly the quasi free 
$\Delta $--isobar production) plays a significant role \cite{BFNSS}. The
theoretical predictions significantly underestimate the data in this region.  
On the other hand, the results of Ref. \cite{CS98} lead to the suggestion that
two--body currents may give sizable contribution in the dip region.

For high virtuality of the exchanged photon, the $^2H(e,e'p)n$ reaction is one 
of the simplest and best way to investigate the high--momentum components of
the deuteron  wave function, possible modifications to the internal structure 
of bound nucleons, and the nature of short--range nucleon correlations. It was 
 found that the $\Delta $--isobar production dominate 
over a large part of the phase space \cite{Eg07}.

It is highly desirable to perform measurements on the deuteron in kinematics 
where the short--distance structure is emphasized. Such information can be only accessed in the context of reaction models 
which include a quantitative description of final--state interaction, meson--exchange currents, isobar configurations and so on. Experiments have already been done in such kinematics \cite{Eg07}.

Theoretical studies of the inclusive electron--nucleus cross section at beam 
energies up to a few GeV show that, while the region of the quasielastic peak 
is quantitatively understood, the data in the $\Delta $--isobar region are 
largely underestimated. In view of the rapid development 
of neutrino physics, the treatment of nuclear effects in data analysis is 
now regarded as one of the main sources of systematic uncertainty. Much of the 
information needed for this analysis can be extracted from the results of 
experimental and theoretical studies of electron--nucleus scattering. In this 
kinematical regime  both quasielastic and inelastic processes, leading to the 
production of hadrons other than protons and neutrons, must be taken into 
account \cite{BM06}. 

Historically, the quasi--elastic cross section has been exploited in order 
to measure the neutron electric and magnetic form factors using mainly light 
$(A\leq 4)$ nuclear targets. Today the emphasis has shifted to the search of 
possible in medium modifications of the nucleon form factors. At large momentum 
transfer it appears that only the low--$\omega $ ($\omega $ is the electron energy 
loss) side of the quasi--elastic peak can be exploited, the large--$\omega $ 
side is obscured by the overlap with $\Delta $--isobar excitation \cite{BDS06}. 

In spite of a large complexity of the spin structure of the $e^-+d\to e^-+
\Delta +N$ reaction amplitude in comparison with the $e^-+d\to e^-+n+p$
reaction amplitude, the mechanism of $\Delta $--isobar production on the
deuteron is more simpler. The reason is that in the impulse approximation the
main mechanism for the $e^-+d\to e^-+\Delta +N$ reaction is described
only by one Feynman diagram whereas for the $e^-+d\to e^-+n+p$ reaction one
has to deal, at least, with four Feynman diagrams (due to the necessity to insure 
gauge invariance for the $\gamma^*+d\to n+p$ process \cite{GRTG}). Moreover, the 
single diagram in the impulse approximation for the $e^-+d\to e^-+\Delta +N$ reaction 
is determined, in good approximation, only by one form factor of the
$\Delta \to N+\gamma^*$ transition - by the transition of the magnetic dipole
type. All this makes the $e^-+d\to e^-+\Delta +N$ reaction more preferable
for the determination of the spin structure of the deuteron wave function in
comparison with the $e^-+d\to e^-+n+p$ reaction, from a theoretical point of view. 

No wonder that the role of the polarization experiments in the $e^-+d\to
e^-+\Delta +N$ and $e^-+d\to e^-+n+p$ reactions is also different. Although
the form factors of the $\Delta \to N+\gamma^*$ transition are not presently better  
known than the nucleon form factors, such uncertainty has no effect on the 
calculations of different asymmetries for the $e^-+d\to e^-+\Delta +N$ reaction due 
to the polarizations of the colliding  particles. One can expect that the various 
asymmetries in the $e^-+d\to e^-+\Delta +N$ reaction will be essentially constant, in 
the impulse approximation, over all the spectrum of the scattered electrons. This
expectation derives from a factorization of polarization effects for the
$e^-+N\to e^-+\Delta $ and $e^-+d\to e^-+\Delta +N$ processes, when the last one
is considered in the impulse approximation. So, the search of deviations
from this factorization deserves a special attention. Such deviations
may be originated by dibaryon resonances \cite{S80}, meson exchange currents
and contribution of the $\Delta\Delta $--configuration in the deuteron ground
state \cite{WA78}. The other (small) form factors of the $\Delta \to N+\gamma^*$ 
transition may also lead to such deviations.

The $e^-+d\to e^-+\Delta +N$ reaction has been investigated earlier in inclusive 
set--up: the spectra of the scattered electrons show two peaks, one from 
quasi--elastic electron--nucleon scattering and another 
corresponding to the $\Delta $--isobar excitation \cite{BBKA}. Experiments
on the $\Delta $--isobar excitation in the electron--nuclear scattering have
been also carried out \cite{BM}. The detailed investigation of the $\Delta $
--isobar (and other nucleon resonances) production which is planned in a number
of laboratories shows the importance of these processes in nuclear physics.

The theoretical analysis of the processes $\gamma +d\to N^* +N$ and $e^-+d\to
e^-+N^*+N$ has been done in a few papers \cite{IY67,SST,IO77,RGK}. The cross
sections of the reactions $e^-+d\to e^-+N^*+N$ ($N^*=\Delta (1232)$ and
$N^*(1480)$) have been calculated in the framework of the non--relativistic
impulse approximation \cite{IY67}. The $\gamma +d\to \Delta +N$ reaction was
considered in the relativistic impulse approximation with the help of the
$dnp$--vertex formalism \cite{SST}. A similar approach was used in Ref.
\cite{IO77} for the analysis of the $e^-+d\to e^-+\Delta +N$ reaction. Estimates 
of polarization effects were done in Ref. \cite{RGK} using the
formalism of the spin--density matrix of the virtual nucleon \cite{RGR79}.

For the study of polarization effects in the scattering of
electrons by hadrons and nuclei it is necessary to distinguish the general
analysis of the polarization phenomena on one side, and specific, model dependent, 
estimations of various asymmetries and polarizations, on the other side. The general 
analysis is based only on the
most general properties of the hadron electrodynamics, such as the
conservation of the hadron electromagnetic current, the invariance of the
hadron electromagnetic interactions with respect to the space reflections
and time reversal as well. The particular structure of the hadrons and
nuclei participating in the reaction, is not essential in this case. The
properties of the polarization phenomena, obtained in this way, are universal
for all reactions of the same type.

In this paper we perform a general analysis of the structure of the
differential cross section and various polarization observables for the
$e^-+d\to e^-+\Delta +N$ reaction. The observables related to the cases of an arbitrary polarized 
deuteron target, longitudinally polarized electron beam, polarization of the 
outgoing nucleon, as well as the polarization transfer from electron to final 
nucleon, and the correlation of the electron and deuteron polarizations are 
considered in detail. The particular case of the process of the photoproduction 
of the $\Delta $--isobar on the deuteron target has been considered in detail, 
separately. The differential cross section and various polarization observables have 
been derived in terms of the reaction amplitudes. The polarization observables due 
to the linear and circular polarizations of the photon, when the deuteron 
target is arbitrarily polarized, have been derived in terms of the reaction 
amplitudes. The polarization of the final nucleon is also considered. This 
analysis was done in frame of the structure function formalism.

The paper is organized as follows. In Section 2 the most general spin structure
of the matrix element of the reaction $\gamma^* +d\to \Delta +N$ is given. The 
general structure of the differential cross section when the scattered electron 
and one of the hadrons are detected in coincidence, when the electron beam is 
longitudinally polarized (the polarization states of the deuteron target and 
of the final nucleon can be any) is also given here. In Section 3 the 
polarization observables due to the longitudinally polarized electron beam 
and unpolarized deuteron target (Section 3.1), or vector (tensor) polarized 
deuteron target (Section 3.2 (Section 3.3)) are derived. Section 4 gives the  
expressions for the nucleon polarization for the unpolarized and longitudinally 
polarized electron beam. Section 5 contains the helicity amplitudes in terms of 
the reaction scalar amplitudes. In Section 6 we consider $\gamma +d\to \Delta +N$
reaction and the polarization observables with unpolarized and linear 
or circular polarized photon beam and unpolarized deuteron target (Section 6.1), 
or vector (tensor) polarized deuteron target (Section 6.2 (Section 6.3)) are 
derived. Section 6.4 gives the expressions of the nucleon polarization for the 
unpolarized and linear or circular polarized photon beam. The main results are 
summarized in the Conclusion. Technical details are given in the
Appendices.

%%%%%%%%%%%%%%%%%%%%%%%%%%%%%%%%%%%%%%%%%%%%%%%%%%%%%%%
\section{The matrix element and the differential cross section}
%%%%%%%%%%%%%%%%%%%%%%%%%%%%%%%%%%%%%%%%%%%%%%%%%%%%%%%%
The general structure of the differential cross section for the
$e^-+d\to e^-+\Delta +N$ reaction can be determined in the framework of the
one--photon--exchange mechanism. The formalism in this section is based on
the most general symmetry properties of the hadron electromagnetic
interaction, such as gauge invariance (the conservation of the hadronic and
leptonic electromagnetic currents) and P--invariance (invariance with respect
to space reflections) and does not depend on the deuteron structure and on
details of the reaction mechanism for $e^-+d\to e^-+\Delta +N$. In the
one--photon--exchange approximation, the matrix element of the $\Delta $
--isobar production in the deuteron electrodisintegration process
\be
e^-(k_1)+d(P)\to e^-(k_2)+\Delta (p_1)+N(p_2)
\label{eq:eq1}
\ee
(the four--momenta of the corresponding particles are indicated in
brackets) can be written as
\be
M_{fi}=\frac{e^2}{k^2}j_{\mu}J_{\mu},~
j_{\mu}=\bar u(k_2)\gamma _{\mu}u(k_1),
\label{eq:eq2}
\ee
where $k_1(k_2)$ is the four--momentum of the initial (final) electron,
$k=k_1-k_2,$ and $J_{\mu}$ is the electromagnetic current describing the
transition $\gamma ^*+d\rightarrow \Delta +N$ ($\gamma ^*$ is the virtual
photon).

The electromagnetic structure of nuclei, as probed by elastic and inelastic
electron scattering by nuclei, can be described by a set of response
functions or structure functions \cite{BJP}. Each of these structure functions
is determined by different combinations of the longitudinal and transverse
components of the electromagnetic current $J_{\mu}$, thus providing different
pieces of information about the nuclear structure or possible mechanisms of
the reaction under consideration. The ones which are determined by the
real parts of the bilinear combinations of the reaction amplitudes are nonzero
in impulse approximation, those which originate from the imaginary
part of the structure functions vanish in the absence of final state interaction.

The formalism of the structure functions is especially convenient for the
investigation of polarization phenomena in the reaction (\ref{eq:eq1}). As a starting
point, let us write the general structure of the differential cross section
of the reaction (\ref{eq:eq1}), when the scattered electron and one of the hadrons are
detected in coincidence, and the electron beam is longitudinally polarized
(the polarization states of the deuteron target and of the final nucleon can
be any):
\ba
\frac{d^3\sigma}{dE'd\Omega_ed\Omega_{\Delta}}&=&
N\biggl [H_{xx}+H_{yy}+\varepsilon\cos(2\varphi)(H_{xx}-H_{yy})+
\nn\\
&&
\varepsilon \sin(2\varphi)(H_{xy}+H_{yx})-2\varepsilon\frac{k^2}{k_0^2}H_{zz}-
\nn\\
&&
\frac{\sqrt{-k^2}}{k_0}\sqrt{2\varepsilon (1+\varepsilon)}
\cos\varphi(H_{xz}+H_{zx})-
\nn\\
&&
\frac{\sqrt{-k^2}}{k_0}\sqrt{2\varepsilon (1+\varepsilon)}
\sin\varphi(H_{yz}+H_{zy})\mp
\nn\\
&&
i\lambda\sqrt{(1-\varepsilon ^2)}(H_{xy}-H_{yx})\mp
\nn\\
&&
i\lambda \frac{\sqrt{-k^2}}{k_0}\sqrt{2\varepsilon (1-\varepsilon)}
\cos\varphi(H_{yz}-H_{zy})\pm 
\nn\\
&&
 i\lambda \frac{\sqrt{-k^2}}{k_0}\sqrt{2\varepsilon (1-\varepsilon)}
\sin\varphi(H_{xz}-H_{zx})\biggr ], \label{eq:eq3}
\ea
with
\ba
N&=&\frac{\alpha^2}{64\pi^3}\frac{E'}{E}\frac{p}{MW}\frac{1}{1-\varepsilon}
\frac{1}{(-k^2)}, 
\nn\\
|{\vec k}|&=&\frac{1}{2W}\sqrt{(W^2+M^2-k^2)^2-4M^2W^2},
\varepsilon^{-1}=1-2\frac{{\vec k}^2_{Lab}}{k^2}\tan^2\left (\frac{\theta_e}{2}\right ),~
\nn\\
p&=&\frac{1}{2W}\sqrt{(W^2+M_{\Delta}^2-m^2)^2-4M_{\Delta}^2W^2},~
H_{\mu\nu}=J_{\mu}J_{\nu}^*. 
\label{eq:eq4}
\ea
The $z$ axis is directed along the virtual photon momentum ${\vec k}$, the
momentum of the detected $\Delta $--isobar ${\vec p}$ lies in the $xz$ plane
(reaction plane); $E (E')$ is the energy of the initial (scattered) electron
in the deuteron rest frame (Lab system); $d\Omega_e$ is the solid angle
of the scattered electron in the Lab system, $d\Omega_{\Delta}(p)$
is the solid angle (value of the three-momentum) of the detected $\Delta $--isobar
in $\Delta N$--pair center--of--mass system (CMS), $M_{\Delta}, M$ and $m$
are the masses of the $\Delta $--isobar, deuteron and nucleon, respectively;
$\varphi$ is the azimuthal angle between the electron scattering plane and
the plane where the detected $\Delta $--isobar lies $(xz)$,
$k_0=(W^2+k^2-M^2)/2W$ is the virtual photon energy in the $\Delta N$--pair
CMS, $W$ is the invariant mass of the final hadrons, $W^2=M^2+k^2+2M(E-E')$;
$\lambda$ is the degree of the electron longitudinal polarization, $\varepsilon $
is the degree of the linear polarization of the virtual photon. The upper
(bottom) sign in this formula corresponds to the electron (positron)
scattering. This expression is valid for zero electron mass. Below we will
neglect it wherever possible.

As it is seen from Eq. (\ref{eq:eq3}), the differential cross section and various
polarization characteristics of the process under consideration are determined
only by the space components of the hadronic tensor $H_{\mu\nu}$.

Assuming the conservation of the leptonic $j_{\mu}$ and hadronic $J_{\mu}$
electromagnetic currents the matrix element can be written as
\be
M_{fi}=ee_{\mu}J_{\mu}=e{\vec l}\cdot {\vec J},~
e_{\mu}=\frac{e}{k^2}j_{\mu},~ {\vec l}=\frac{{\vec e}{\vec k}}
{k_0^2}{\vec k}-{\vec e}.
\label{eq:eq4a}
\ee
In CMS of the $\Delta $--isobar and final nucleon we get
$$M_{fi}=e{\vec \chi}_2^+{\vec F}\chi _1^c=e{\it F}, $$
where ${\vec \chi}_2^+ $ and $\chi _1^c$ are the $\Delta $--isobar vector
spinor and nucleon spinor, correspondingly. 

Let us introduce, the orthonormal system of basic unit
${\vec m}, {\vec n}$, and $\hat {\vec k}$ vectors which are built from the
momenta of the particles participating in the reaction under consideration
$$\hat {\vec k}=\frac{{\vec k}}{|{\vec k}|},~
{\vec n}=\frac{{\vec k}\times{\vec p}}{|{\vec k}\times{\vec p}|},~
{\vec m}={\vec n}\times\hat {\vec k}. $$
The unit vectors $\hat {\vec k}$ and ${\vec m}$ define the $\gamma^* +
d\rightarrow \Delta+N$ reaction $xz$--plane (the $z$ axis is directed along the
three--momentum of the virtual photon ${\vec k}$, the $x$ axis is directed along
the unit vector ${\vec m}$), and the unit vector ${\vec n}$ is perpendicular to
the reaction plane. 

In the analysis of polarization
phenomena, it is convenient to use the amplitude $F$ represented in the
above orthonormal basis. The amplitude $F$ can be chosen as
$${\vec F}={\vec m}F^{(m)}+\hat {\vec k}F^{(k)}, $$
\ba
F^{(m)}&=&{\vec l}\cdot {\vec m}(if_1{\vec U}\cdot {\vec m}
+if_2{\vec U}\cdot \hat {\vec k}+
f_3{\vec \sigma }\cdot {\vec n}{\vec U}\cdot {\vec m}+
f_4{\vec \sigma}\cdot {\vec n}{\vec U}\cdot \hat {\vec k}+
\nn\\
&&
f_5{\vec \sigma }\cdot {\vec m}{\vec U}\cdot {\vec n}+ 
f_6{\vec \sigma}\cdot \hat {\vec k}{\vec U}\cdot {\vec n})+
{\vec l}\cdot {\vec n}(if_7{\vec U}\cdot {\vec n}+
f_{8}{\vec \sigma }\cdot {\vec n}{\vec U}\cdot {\vec n}+
\nn\\
&&
f_9{\vec \sigma}\cdot {\vec m}{\vec U}\cdot {\vec m}+
f_{10}{\vec \sigma }\cdot \hat {\vec k}{\vec U}\cdot {\vec m}+
f_{11}{\vec \sigma}\cdot {\vec m}{\vec U}\cdot \hat {\vec k}+
f_{12}{\vec \sigma }\cdot \hat {\vec k}{\vec U}\cdot \hat {\vec k})+
\nn\\
&&
{\vec l}\cdot \hat {\vec k}(if_{13}{\vec U}\cdot {\vec m}+
if_{14}{\vec U}\cdot \hat {\vec k}+
f_{15}{\vec \sigma }\cdot {\vec n}{\vec U}\cdot {\vec m}+
f_{16}{\vec \sigma}\cdot {\vec n}{\vec U}\cdot \hat {\vec k}+
\nn\\
&&
f_{17}{\vec \sigma }\cdot {\vec m}{\vec U}\cdot {\vec n}+
f_{18}{\vec \sigma }\cdot \hat {\vec k}{\vec U}\cdot {\vec n}), 
\label{eq:eq5}
\\
F^{(k)}&=&{\vec l}\cdot {\vec m}(if_{19}{\vec U}\cdot {\vec m}
+if_{20}{\vec U}\cdot \hat {\vec k}+
f_{21}{\vec \sigma }\cdot {\vec n}{\vec U}\cdot {\vec m}+
f_{22}{\vec \sigma}\cdot {\vec n}{\vec U}\cdot \hat {\vec k}+
\nn\\
&&
f_{23}{\vec \sigma }\cdot {\vec m}{\vec U}\cdot {\vec n}+
 f_{24}{\vec \sigma}\cdot \hat {\vec k}{\vec U}\cdot {\vec n})+
{\vec l}\cdot {\vec n}(if_{25}{\vec U}\cdot {\vec n}+
f_{26}{\vec \sigma }\cdot {\vec n}{\vec U}\cdot {\vec n}+
\nn\\
&&
f_{27}{\vec \sigma}\cdot {\vec m}{\vec U}\cdot {\vec m}+
f_{28}{\vec \sigma }\cdot \hat {\vec k}{\vec U}\cdot {\vec m}+
f_{29}{\vec \sigma}\cdot {\vec m}{\vec U}\cdot \hat {\vec k}+
f_{30}{\vec \sigma }\cdot \hat {\vec k}{\vec U}\cdot \hat {\vec k})+
\nn\\
&&
{\vec l}\cdot \hat {\vec k}(if_{31}{\vec U}\cdot {\vec m}+
if_{32}{\vec U}\cdot \hat {\vec k}+
f_{33}{\vec \sigma }\cdot {\vec n}{\vec U}\cdot {\vec m}+ 
f_{34}{\vec \sigma}\cdot {\vec n}{\vec U}\cdot \hat {\vec k}+
\nn\\
&&
f_{35}{\vec \sigma }\cdot {\vec m}{\vec U}\cdot {\vec n}+
f_{36}{\vec \sigma }\cdot \hat {\vec k}{\vec U}\cdot {\vec n}), 
\label{eq:eq6}
\ea
where $f_i (i=1-36)$ are the scalar amplitudes, depending on three variables,
which completely determine the reaction dynamics. If we single out the
photon polarization vector ${\vec l}$, one can write the amplitude ${\it F}$
as ${\it F}={\it F_i}l_i $ and the hadronic tensor as 
$H_{ij}={\it F_i}{\it F_j^*}. $

%%%%%%%%%%%%%%%%%%%%%%%%%%%%%%%%%%%%%%%%%%%%%%%%%%
\section{Polarization of the deuteron target}
%%%%%%%%%%%%%%%%%%%%%%%%%%%%%%%%%%%%%%%%%%%%%%%%%%

In the general case the deuteron polarization is described by the spin--density matrix. 
Let us start from the following general expression
for the deuteron spin--density matrix in the coordinate representation
\cite{SC}
\be
\rho_{\mu\nu}=-\frac{1}{3}\left (g_{\mu\nu}-\frac{P_{\mu}P_{\nu}}{M^2}\right )-
\frac{i}{2M}\varepsilon _{\mu\nu\alpha\beta}s_{\alpha}P_{\beta}+S_{\mu\nu},
\label{eq:eq7}
\ee
where $s_{\alpha}$ is the four--vector describing the vector polarization of
the target, with $s^2=-1,$ $s\cdot P=0$. $S_{\mu\nu}$ is the tensor describing
the tensor (quadrupole) polarization of the target, with $S_{\mu\nu}=S_{\nu\mu},$
$P_{\mu}S_{\mu\nu}=0, $ $S_{\mu\mu}=0 $. Due to these properties the tensor
$S_{\mu\nu}$ has only five independent components. In Lab system all time
components of the tensor $S_{\mu\nu}$ are zero and the tensor polarization
of the target is described by five independent space components $(S_{ij}=
S_{ji}$, $S_{ii}=0$, $i,j=x,y,z)$. The four--vector $s_{\alpha}$ is related to
the unit vector ${\vec \xi}$ of the deuteron vector polarization in its rest
system: 
$$
s_0=-{\vec k}{\vec \xi}/M, ~{\vec s}={\vec \xi}+
{\vec k}({\vec k}{\vec \xi})/M(M+\omega),
$$ 
where $\omega $ is the deuteron energy
in the $\gamma ^*+d\rightarrow \Delta +N$ reaction CMS.

The hadronic tensor $H_{ij} (i,j=x,y,z)$ depends linearly on the target
polarization and it can be represented as follows
\be
H_{ij}=H_{ij}(0)+H_{ij}(\xi)+H_{ij}(S),
\label{eq:eq8}
\ee
where the term $H_{ij}(0)$ corresponds to the case of unpolarized deuteron
target, and the term $H_{ij}(\xi)$$(H_{ij}(S))$ corresponds to the case of the
vector(tensor-)-polarized target.

%%%%%%%%%%%%%%%%%%%%%%%%%%%%%%%%%%%%%%%%%%
\subsection{Unpolarized deuteron target}
%%%%%%%%%%%%%%%%%%%%%%%%%%%%%%%%%%%%%%%%%%
The general structure of the part of the hadronic tensor corresponding to unpolarized deuteron has the following form
\ba
H_{ij}(0)&=&\alpha_1\hat k_i\hat k_j+\alpha_2n_in_j+ \alpha_3m_im_j+
\alpha_4(\hat k_im_j+\hat k_jm_i)+\nn\\
&&
i\alpha_5(\hat k_im_j-\hat k_jm_i).
\label{eq:eq9}
\ea
The structure functions $\alpha_i$ are real and depend on three invariant variables
$s=W^2=(k+P)^2$, $k^2$ and $t=(k-P)^2$. Let us emphasize that the structure
function $\alpha_5$ is determined by the strong interaction effects of the
$\Delta $--isobar and the nucleon in the final state and it vanishes for the
pole diagram contribution in all kinematic range (independently on the
particular parametrization of the $\gamma^* \Delta N -$ and $dnp-$vertexes).
This is true for the non relativistic approach and for the relativistic one
as well, when describing the $\gamma^* +d\rightarrow \Delta +N$ reaction. The
scattering of longitudinally polarized electrons by unpolarized deuteron 
allows to determine the $\alpha_5$ contribution. Then, the corresponding
asymmetry is determined only by the strong interaction effects. More exactly,
it is determined by the effects arising from non pole mechanisms of various
nature (meson exchange currents can also induce nonzero asymmetry). Dibaryon resonances, 
if present, may also lead to nonzero asymmetry.

In the chosen coordinate system, the different hadron tensor components,
entering in the expression of the cross section (\ref{eq:eq9}), are related to the
structure functions $\alpha_i (i=1-5)$ by:
\ba
&&H_{xx}\pm H_{yy}=\alpha_3\pm \alpha_2,~  H_{zz}=\alpha_1,~
H_{xz}+ H_{zx}=2\alpha_4,
\nn \\
&&
H_{xz}- H_{zx}=-2i\alpha_5,~ 
H_{xy}\pm H_{yx}=0, ~
H_{yz}\pm H_{zy}=0. 
\label{eq:eq9a}
\ea
In the one--photon--exchange approximation, the general structure of the
differential cross section for the reaction $d({\vec e}, e'\Delta )N$ (in the
case of longitudinally polarized electron beam and unpolarized deuteron
target) can be written in terms of five independent contributions
\ba
\frac{d^3\sigma}{dE'd\Omega_ed\Omega_{\Delta}} &=&
N\biggl [\sigma_{T}+\varepsilon \sigma_{L}+
\varepsilon \cos(2\varphi)\sigma_{P}+
\sqrt{2\varepsilon (1+\varepsilon )}\cos\varphi \sigma_{I}+
\nn\\
&&
\lambda \sqrt{2\varepsilon (1-\varepsilon )}\sin\varphi \sigma'_{I}
\biggr ],
\label{eq:eq10}
\ea
where the individual contributions are related to the components of the
spin--independent hadronic tensor, Eq. (\ref{eq:eq9}), by:
\ba
\sigma_{T}&=&H_{xx}+H_{yy},~ \sigma_{P}=H_{xx}-H_{yy},~
\sigma_{L}=-2\frac{k^2}{k_0^2}H_{zz}, \nn \\
\sigma_{I}&=&-\frac{\sqrt{-k^2}}{k_0}(H_{xz}+H_{zx}),~
\sigma'_{I}=i\frac{\sqrt{-k^2}}{k_0}(H_{xz}-H_{zx}). 
\nn
\ea
From the above equations, one can define a single--spin asymmetry
which is due to the electron beam polarization:
\ba
\Sigma_e(\varphi )&=&\frac{d\sigma (\lambda =+1)-d\sigma (\lambda =-1)}
{d\sigma (\lambda =+1)+d\sigma (\lambda =-1)}= \nn\\
&&\frac{\sin\varphi \sqrt{2\varepsilon (1-\varepsilon )}\sigma'_{I}}
{\sigma_{T}+\varepsilon \sigma_{L}+\varepsilon \cos(2\varphi)\sigma_{P}+
\sqrt{2\varepsilon (1+\varepsilon )}\cos\varphi \sigma_{I}},
\label{eq:eq11}
\ea
which contains a $\varphi $--dependence. Therefore, this asymmetry has to be measured in
non coplanar geometry (out--of--plane kinematics).

We see that this asymmetry is determined by the structure function $\alpha_5$
which is defined by the interference of the reaction amplitudes that
characterize the absorption of virtual photons with nonzero longitudinal and
transverse components of the electromagnetic current corresponding to the
process $\gamma ^*+d\rightarrow \Delta +N$. One finds that
$\alpha_5\sim \sin\vartheta $ independently from the reaction mechanism. 
It vanishes when the $\Delta $--isobar emission angles are 
$\vartheta =0^{\circ}$ and $\vartheta =180^{\circ}$,  due to the conservation
of the total helicity of the interacting particles in the $\gamma ^*+d
\rightarrow \Delta +N$ reaction. The structure function $\alpha_5$ is nonzero
only if the complex amplitudes of the $\gamma ^*+d\rightarrow \Delta +N$
reaction have nonzero relative phases. This is a very specific observable,
which has no corresponding quantity in the $\Delta $--isobar excitation in
the deuteron photodisintegration process $\gamma +d\rightarrow \Delta +N$.

The study of the single--spin asymmetry $\Sigma_e$ was firstly suggested for 
pion production in electron--nucleon scattering, $e+N\to e+N+\pi $
\cite{G71}. Afterward, this asymmetry has been discussed for the hadron
production in the exclusive processes of the type $A({\vec e}, eh)X$,
where $A$ is a nucleus and $h$ is the detected hadron \cite{BP85,Pi85}. A number
of experiments to measure the asymmetry $\Sigma_e$ has already been done
\cite{MBDD}.
%%%%%%%%%%%%%%%%%%%%%%%%%%%%%%%%%%%%%%%%%%%%%%%%%%%%
\subsection{Vector--polarized deuteron target}
%%%%%%%%%%%%%%%%%%%%%%%%%%%%%%%%%%%%%%%%%%%%%%%%%%%%%%%
The part of the hadronic tensor depending on the deuteron vector polarization
has the following general structure:
\ba
H_{ij}(\xi )&=&{\vec\xi }{\vec n}(\beta_1\hat k_i\hat k_j+\beta_2m_im_j
+ \beta_3n_in_j+\beta_4\{\hat k,m\}_{ij}+i\beta_5[\hat k,m]_{ij})+
\nn \\
&&+{\vec\xi }\hat {\vec k}(\beta_6\{\hat k,n\}_{ij}+\beta_7\{m,n\}_{ij}+
i\beta_8[\hat k,n]_{ij}+i\beta_9[m,n]_{ij})+\nn\\
&&
+{\vec\xi }{\vec m}(\beta_{10}\{\hat k,n\}_{ij}+\beta_{11}\{m,n\}_{ij}+
i\beta_{12}[\hat k,n]_{ij}+i\beta_{13}[m,n]_{ij}), 
\label{eq:eq12}
\ea
where $\{a,b\}_{ij}=a_ib_j+a_jb_i, \ [a,b]_{ij}=a_ib_j-a_jb_i. $

Therefore, the dependence of the polarization observables on the deuteron
vector polarization is determined by thirteen structure functions. On the basis
of this formula one can make the following general conclusions:
\begin{itemize}

\item If the deuteron is vector--polarized and the  polarization vector 
is perpendicular to the $\gamma ^* +d\rightarrow \Delta +N$ reaction plane,
then the dependence of the differential cross section on the 
$\varepsilon$ and $\varphi$ variables is the same as in the case of unpolarized target, 
and the non vanishing components of the $H_{ij}(\xi )$ tensor are:
\ba
&&
H_{xx}(\xi ) \pm H_{yy}(\xi )=(\beta_2\pm \beta_3){\vec\xi }{\vec n},
~ H_{zz}(\xi )=\beta_1{\vec\xi }{\vec n},~ \nn \\
&&
H_{xz}(\xi )+H_{zx}(\xi )=2\beta_4{\vec\xi }{\vec n},~
H_{xz}(\xi )-H_{zx}(\xi )=-2i\beta_5{\vec\xi }{\vec n}. 
\ea
\item If the deuteron target is polarized in the $\gamma ^* +d\rightarrow
\Delta +N$ reaction plane (in direction of the vector ${\vec k}$ or ${\vec m}$),
then the dependence of the differential cross section of the $e^-+d\rightarrow
e^-+\Delta +N$ reaction on the $\varepsilon$ and $\varphi$ variables is:
\begin{itemize}
\item for unpolarized electron beam:
$$\varepsilon sin(2\varphi), ~ \sqrt{2\varepsilon(1+\varepsilon )}sin\varphi, $$
\item for longitudinally polarized electron beam:
$$\pm i\lambda\sqrt{1-\varepsilon ^2},~ \mp i\lambda\sqrt{2\varepsilon
(1-\varepsilon )}\\cos\varphi. $$
\end{itemize}
\end{itemize}
The differential cross section of the reaction ${\vec d}({\vec e},e'\Delta )N$,
where the electron beam is longitudinally polarized and the deuteron target
is vector--polarized, can be written as follows:
\ba
\frac{d^3\sigma}{dE'd\Omega_ed\Omega_{\Delta}}& =&
\sigma_0\biggl [1+\lambda \Sigma_{e}+(A_x^d+\lambda A_x^{ed})\xi_x+
(A_y^d+\lambda A_y^{ed})\xi_y+
\nn\\
&&
(A_z^d+\lambda A_z^{ed})\xi_z
\biggr ],
\label{eq:eq13}
\ea
where $\sigma_0$ is the unpolarized differential cross section, $\Sigma_{e}$
is the beam asymmetry (the asymmetry induced by the electron--beam
polarization), $A_i^d (i=x,y,z)$ are the analyzing powers due to the vector
polarization of the deuteron target, and $A_i^{ed} (i=x,y,z)$ are the spin--
correlation parameters. The direction of the deuteron polarization vector is
defined by the angles $\vartheta^*$, $\varphi^*$ in the reference frame where the $z$
axis is along the direction of the three--momentum transfer ${\vec k}$, and
the $y$ axis is defined by the vector product of the momenta of the detected $\Delta $--
isobar and the virtual photon (along the unit vector ${\vec n}$). The
target analyzing powers and the spin--correlation parameters depend on the
orientation of the deuteron polarization vector. The quantities $\Sigma_{e}$
and $A_i^d$ are T--odd observables and they are completely determined by the
reaction mechanism beyond the impulse approximation, for example, by 
final--state interaction effects. On the contrary, the quantities $A_i^{ed}$
are T--even observables and they do not vanish in absence of final--state 
interaction effects.

The expressions of the $A_i^{d}$ and $A_i^{ed}$ asymmetries can be explicitly
written as functions of the azimuthal angle $\varphi $, of the virtual--photon
linear polarization $\varepsilon $, and of the contributions of the longitudinal
(L) and transverse (T) components (relative to the virtual--photon momentum
${\vec k}$) of the hadron electromagnetic current of the $\gamma^*+d\to
\Delta +N$ reaction:
\ba
A_x^d\sigma_0&=&N\sin\varphi \biggl [\sqrt{2\varepsilon (1+\varepsilon )}
A_x^{(LT)}+\varepsilon \cos\varphi A_x^{(TT)}\biggr ], 
\nn\\ 
A_z^d\sigma_0&=&N \sin\varphi \biggl [\sqrt{2\varepsilon (1+\varepsilon )}
A_z^{(LT)}+\varepsilon \cos\varphi A_z^{(TT)}\biggr ], 
\nn\\ 
A_y^d\sigma_0&=&N\biggl [A_y^{(TT)}+\varepsilon A_y^{(LL)}+
\sqrt{2\varepsilon (1+\varepsilon )}\cos\varphi A_y^{(LT)}+
\varepsilon \cos(2\varphi )\bar A_y^{(TT)}\biggr ], 
\nn\\ 
A_x^{ed}\sigma_0&=&N\biggl [\sqrt{1-\varepsilon^2}
B_x^{(TT)}+\sqrt{2\varepsilon (1-\varepsilon )}
\cos\varphi B_x^{(LT)}\biggr ], 
\nn\\ 
A_z^{ed}\sigma_0&=&N\biggl [\sqrt{1-\varepsilon^2}
B_z^{(TT)}+\sqrt{2\varepsilon (1-\varepsilon )}
\cos\varphi B_z^{(LT)}\biggr ], 
\nn\\ 
A_y^{ed}\sigma_0&=&N\sqrt{2\varepsilon (1-\varepsilon )}
\sin\varphi B_y^{(LT)}, 
\label{eq:eq14}
\ea
where the individual contributions to the considered asymmetries in terms of
the structure functions $\beta_i$ are given by
\ba 
A_x^{(TT)}&=&4\beta_{11},~ A_y^{(TT)}=\beta_2+\beta_3,~
\bar A_y^{(TT)}=\beta_2-\beta_3, ~
A_z^{(TT)}=4\beta_{7}, \nn\\
A_x^{(LT)}&=&-2\frac{\sqrt{Q^2}}{k_0}\beta_{10}, ~
A_y^{(LT)}=-2\frac{\sqrt{Q^2}}{k_0}\beta_{4},~
A_z^{(LT)}=-2\frac{\sqrt{Q^2}}{k_0}\beta_{6}, \nn\\
A_y^{(LL)}&=&2\frac{Q^2}{k_0^2}\beta_{1},~
B_x^{(TT)}=2\beta_{13},~ B_z^{(TT)}=2\beta_{9}, \nn\\
B_x^{(LT)}&=&-2\frac{\sqrt{Q^2}}{k_0}\beta_{12},~
B_y^{(LT)}=2\frac{\sqrt{Q^2}}{k_0}\beta_{5}, ~
B_z^{(LT)}=-2\frac{\sqrt{Q^2}}{k_0}\beta_{8}. \nn
\ea
At this stage, the general model--independent analysis of the polarization
observables in the reactions ${\vec d}(e, e'\Delta )N$ and ${\vec d}({\vec e},
e'\Delta )N$ is completed. To proceed further in the calculation of the
observables, one needs a model for the reaction mechanism and for the deuteron
structure.
%%%%%%%%%%%%%%%%%%%%%%%%%%%%%%%%%%%%%%%%%%%%%%%%%%
\subsection{Tensor--polarized deuteron target}
%%%%%%%%%%%%%%%%%%%%%%%%%%%%%%%%%%%%%%%%%%%%%%%%%%
The part of the hadronic tensor $H_{ij}(S)$, which depends on the deuteron
tensor polarization, has the following general structure:
\ba
&&H_{ij}(S)=S_{ab}\hat k_a\hat k_b(\gamma_1\hat k_i\hat k_j+\gamma_2m_im_j
+ \gamma_3n_in_j+\gamma_4\{\hat k,m\}_{ij}+i\gamma_5[\hat k,m]_{ij})+
\nn\\ &&
S_{ab}m_am_b(\gamma_6\hat k_i\hat k_j+\gamma_7m_im_j
+ \gamma_8n_in_j+\gamma_9\{\hat k,m\}_{ij}+i\gamma_{10}[\hat k,m]_{ij})+ 
\nn\\ &&
S_{ab}\{\hat k,m\}_{ab}(\gamma_{11}\hat k_i\hat k_j+\gamma_{12}m_im_j
+ \gamma_{13}n_in_j+\gamma_{14}\{\hat k,m\}_{ij}+
i\gamma_{15}[\hat k,m]_{ij})+ 
\nn\\ &&
S_{ab}\{\hat k,n\}_{ab}(\gamma_{16}\{\hat k,n\}_{ij}+
\gamma_{17}\{m,n\}_{ij}+i\gamma_{18}[\hat k,n]_{ij}+
i\gamma_{19}[m,n]_{ij})+ 
\nn\\ &&
S_{ab}\{m,n\}_{ab}(\gamma_{20}\{\hat k,n\}_{ij}+
\gamma_{21}\{m,n\}_{ij}+i\gamma_{22}[\hat k,n]_{ij}+
i\gamma_{23}[m,n]_{ij}). 
\label{eq:eq15}
\ea
In this case, the dependence of the polarization observables on the deuteron
tensor polarization is determined by 23 structure functions.

From this equation one can conclude that:
\begin{itemize}
\item
If the deuteron is tensor polarized so that only $S_{zz},$ $S_{yy}$ and
$(S_{xz}+S_{zx})$ components of the quadrupole polarization tensor are
nonzero, then the dependence of the differential cross section of the
$e^-+d\rightarrow e^-+\Delta +N$ reaction on the parameter $\varepsilon$ and
on the azimuthal angle $\varphi$ must be the same as in the case of the
unpolarized target (more exactly, with similar $\varepsilon$-- and $\varphi$--
dependent terms).
\item If the deuteron is polarized so that only $(S_{xy}+S_{yx})$ and
$(S_{yz}+S_{zy})$ components of the quadrupole polarization tensor are nonzero,
then the typical terms follow $\sin\varphi$ and  $\sin(2\varphi )$ dependencies -
for deuteron disintegration by unpolarized electron beam, and terms which do
not depend on $\varepsilon$, $\varphi$, and $\cos\varphi$ - for deuteron
disintegration by longitudinally polarized electron beam.
\end{itemize}
In polarization experiments it is possible to prepare the deuteron target
with definite spin projection on some quantization axis. The corresponding
asymmetry is usually defined as
$$A=\frac{d\sigma (\lambda _d=+1)-d\sigma (\lambda _d=-1)}
{d\sigma (\lambda _d=+1)+d\sigma (\lambda _d=-1)}, $$
where $d\sigma (\lambda _d)$ is the differential cross section of the
$e^-+d\rightarrow e^-+\Delta +N$ reaction when the quantization axis for the
deuteron spin (in the $\Delta N$--pair CMS) coincides with its momentum, i.e.,
the deuteron has helicity $\lambda_d$. From an experimental point of view,
the measurement of an asymmetry is more convenient than a measurement of a
cross section, as most of systematic experimental errors and other
multiplicative factors cancel in the ratio.

The general form of the hadron tensor $H_{ij}(\lambda _d)$, which determines
the differential cross section of the process under consideration for the
case of the deuteron with helicity $\lambda _d$, can be written as
\ba
H_{ij}^{(\lambda _d=\pm 1)}&=&\delta_1\hat k_i\hat k_j+\delta_2m_im_j
+ \delta_3n_in_j+\delta_4\{\hat k,m\}_{ij}+i\delta_5[\hat k,m]_{ij}\pm
\nn\\
&&\pm \delta_6\{\hat k,n\}_{ij}\pm i\delta_7[\hat k,n]_{ij}\pm
\delta_8\{m,n\}_{ij}\pm i\delta_9[m,n]_{ij}. 
\label{eq:eq17}
\ea
The reaction amplitude is real in the Born (impulse) approximation. So,
assuming the T-invariance of the hadron electromagnetic interactions, we can
do the following statements, according to the deuteron polarization state:

\begin{itemize}
\item 
\underline {The deuteron is unpolarized}. Since the hadronic tensor
$H_{ij}(0)$ has to be symmetric (over i,j indexes) in this case, the
asymmetry in the scattering of longitudinally polarized electrons vanishes.
\item 
\underline {The deuteron is vector polarized}. Since the hadronic tensor
$H_{ij}(\xi )$ has to be antisymmetric in this case, then the deuteron vector
polarization can manifest itself in the scattering of longitudinally polarized
electrons. The perpendicular target polarization (normal to the
$\gamma^*+d\rightarrow \Delta +N$ reaction plane) leads to a correlation of
the following type: $\pm i\lambda\sqrt{2\varepsilon (1-\varepsilon )}
\sin\varphi $.  The longitudinal and transverse (along or perpendicular to the
virtual--photon momentum) target polarization (lying in the
$\gamma^*+d\rightarrow \Delta +N$ reaction plane) leads to two correlations
of the following type : $\mp i\lambda\sqrt{1-\varepsilon ^2}$ and
$\mp i\lambda\sqrt{2\varepsilon (1-\varepsilon )}\cos\varphi .$
\item
\underline{The deuteron is tensor polarized}. The hadronic tensor
$H_{ij}(S)$ is symmetric in this case. In the scattering of longitudinally
polarized electrons the contribution proportional to $\lambda S_{ab}$
vanishes. If the target is polarized so that only the $(S_{xy}+S_{yx})$ or
$(S_{yz}+S_{zy})$ components of the quadrupole polarization tensor are nonzero,
then in the differential cross section only the following two terms are present:
$\varepsilon \sin(2\varphi)$ and  $\sqrt{2\varepsilon (1+\varepsilon )}
\sin\varphi .$ For all other target polarizations the following structures are
present: a term which does not depend on $\varepsilon $ and $\varphi $
variables as well as terms with the following dependencies: $2\varepsilon $,
$\varepsilon \cos(2\varphi ) $, and $\sqrt{2\varepsilon (1+\varepsilon )}
\cos\varphi .$
\end{itemize}
The differential cross section of the $\Delta $--isobar excitation in the
scattering of longitudinally polarized electrons by tensor polarized deuteron
target (in the coincidence experimental setup) has the following general
structure
\ba
&&\frac{d^3\sigma}{dE'd\Omega_ed\Omega_{\Delta}} =
N\biggl \{\sigma_T+A_{xz}^TQ_{xz}+A_{xx}^T(Q_{xx}-Q_{yy})+A_{zz}^TQ_{zz}+
\nn\\
&&\varepsilon \biggl [\sigma_L+A_{xz}^LQ_{xz}+A_{xx}^L(Q_{xx}-Q_{yy})+
A_{zz}^LQ_{zz}\biggr ]+ 
\nn \\
&&
\sqrt{2\varepsilon (1+\varepsilon )}\cos\varphi \biggl [
\sigma_I+A_{xz}^IQ_{xz}+A_{xx}^I(Q_{xx}-Q_{yy})+A_{zz}^IQ_{zz}\biggr ]+ 
\nn \\
&&
\sqrt{2\varepsilon (1+\varepsilon )}\sin\varphi (
A_{xy}^IQ_{xy}+A_{yz}^IQ_{yz})+
\varepsilon \sin(2\varphi )(A_{xy}^PQ_{xy}+A_{yz}^PQ_{yz})+\nn \\
&&
\varepsilon \cos(2\varphi )
\biggl [\sigma_P+A_{xz}^PQ_{xz}+A_{xx}^P(Q_{xx}-Q_{yy})+
A_{zz}^PQ_{zz}\biggr ]+ 
\nn \\
&&
\lambda \sqrt{2\varepsilon (1-\varepsilon )}\sin\varphi \biggl [
\sigma_I'+\bar A_{xz}^IQ_{xz}+\bar A_{xx}^I(Q_{xx}-Q_{yy})+
\bar A_{zz}^IQ_{zz}\biggr ]+ 
\nn \\
&&\lambda \sqrt{2\varepsilon (1-\varepsilon )}\cos\varphi \biggl [
\bar A_{xy}^IQ_{xy}+\bar A_{yz}^IQ_{yz}\biggr ]+
\nn \\
&&
\lambda \sqrt{1-\varepsilon^2}\cos\varphi \biggl [
A_{xy}^TQ_{xy}+A_{yz}^TQ_{yz}\biggr ]\biggr \}, \nn
\label{eq:eq16}
\ea
where the quantities $Q_{ij} (i,j=x,y,z)$ are the components of the quadrupole
polarization tensor of the deuteron in its rest system (the coordinate system
is specified similarly to the case of the $\Delta N$--pair CMS). These
components satisfy the following conditions: $Q_{ij}=Q_{ji}$, $Q_{ii}=0$.
By writing this formula we took into account the following relation: $Q_{xx}+Q_{yy}+Q_{zz}=0$.

A general property of these tensor asymmetries is that they vanish in the region of the quasi--elastic scattering. This can be
explained as follows. All the asymmetries are determined by the convolution
$X_{\mu\nu}S_{\mu\nu}$, where the tensor $X_{\mu\nu}$ is built with the
four--momenta describing the $d\to np$ transition. Due to the condition
$P_{\mu}S_{\mu\nu}=0$, the most general form of this tensor is
$$X_{\mu\nu}=a_1g_{\mu\nu}+ia_2[\gamma_{\mu},\gamma_{\nu}]+
a_3\gamma_{\mu}p_{\nu}+a_4\gamma_{\nu}p_{\mu}+a_5p_{\mu}p_{\nu}, $$
where $p_{\mu}$ is the four--momentum of the nucleon--spectator. However, if
we take into account that $S_{\mu\nu}g_{\mu\nu}=0,$ $S_{\mu\nu}=S_{\nu\mu}$,
then the convolution $X_{\mu\nu}S_{\mu\nu}$ is determined by $a_3, a_4$ and
$a_5$. From the condition $P_{\mu}S_{\mu\nu}=0$, it follows that the time
components of the $S_{\mu\nu}$ tensor vanish in the Lab system. Therefore, the convolution $X_{\mu\nu}S_{\mu\nu}$ turns out to be proportional to the
nucleon--spectator three--momentum which is zero at the peak of the
quasi--elastic scattering.

Thus, in the general case, the number of independent asymmetries $A_{ij}^m(W,k^2,\vartheta )$, $i,j=x,y,z; m=T,P,L,I$, contributing to the exclusive cross section of the $\Delta $--isobar
excitation is 23 for the scattering of longitudinally polarized electrons by a tensor
polarized deuteron target, 16(7) for the scattering of unpolarized (longitudinally polarized) electrons $A_{ij}^m(W,k^2,\vartheta )$, where $i,j=x,y,z; m=T,P,L,I$. These asymmetries
can be related to the structure functions $\gamma_i$ which are the bilinear
combinations of the 36 independent scalar amplitudes describing the
$\gamma ^* +d\rightarrow \Delta +N$ reaction, by the following relations:
\ba
&&A_{xz}^T=2\frac{\omega}{M}(\gamma_{12}+\gamma_{13}),~
A_{xx}^T=\frac{1}{2}(\gamma_{7}+\gamma_{8}),~
\nn\\
&&
A_{zz}^T=\frac{\omega^2}{M^2}(\gamma_{2}+\gamma_{3})-
\frac{1}{2}(\gamma_{7}+\gamma_{8}),
A_{xz}^L =-4\frac{\omega}{M}\frac{k^2}{k_0^2}\gamma_{11},~
A_{xx}^L=-\frac{k^2}{k_0^2}\gamma_{6},~
\nn\\
&&A_{zz}^L=-\frac{k^2}{k_0^2}(2\frac{\omega^2}{M^2}\gamma_{1}-\gamma_{6}), 
A_{xz}^I=-4\frac{\omega}{M}\frac{\sqrt{-k^2}}{k_0}\gamma_{14},~
A_{xx}^I=-\frac{\sqrt{-k^2}}{k_0}\gamma_{9}, 
\nn\\
&&
A_{zz}^I=-\frac{\sqrt{-k^2}}{k_0}(2\frac{\omega^2}{M^2}\gamma_{4}-
\gamma_{9}), 
A_{xy}^I=-4\frac{\sqrt{-k^2}}{k_0}\gamma_{20},~
\nn\\
&&
A_{yz}^I=-4\frac{\omega}{M}\frac{\sqrt{-k^2}}{k_0}\gamma_{16},~
A_{xy}^P=4\gamma_{21},~ A_{yz}^P=4\frac{\omega}{M}\gamma_{17}, 
A_{xz}^P=2\frac{\omega}{M}(\gamma_{12}-\gamma_{13}),~
\nn\\
&&
A_{xx}^P=\frac{1}{2}(\gamma_{7}-\gamma_{8}),~
A_{zz}^P=\frac{\omega^2}{M^2}(\gamma_{2}-\gamma_{3})-
\frac{1}{2}(\gamma_{7}-\gamma_{8}), 
\nn\\
&&
\bar A_{xz}^I=4\frac{\omega}{M}\frac{\sqrt{-k^2}}{k_0}\gamma_{15},~
\bar A_{xx}^I=\frac{\sqrt{-k^2}}{k_0}\gamma_{10}, ~
\bar A_{zz}^I=\frac{\sqrt{-k^2}}{k_0}(2\frac{\omega^2}{M^2}\gamma_{5}-
\gamma_{10}), \nn\\
&&
\bar A_{xy}^I=-4\frac{\sqrt{-k^2}}{k_0}\gamma_{22},~
\bar A_{yz}^I=-4\frac{\omega}{M}\frac{\sqrt{-k^2}}{k_0}\gamma_{18},~
A_{xy}^T=4\gamma_{23},~ A_{yz}^T=4\frac{\omega}{M}\gamma_{19}. \nn
\ea
One can see from this formula that the scattering of unpolarized electrons
by a tensor polarized deuteron target with components $Q_{xy}=Q_{yz}=0$, is
characterized by the same $\varphi $-- and $\varepsilon $--dependences as
in the case of the scattering of unpolarized electrons by unpolarized
deuteron target. If $Q_{xy}\ne 0$, $Q_{yz}\ne 0$, then new terms of the type
$\sqrt{2\varepsilon (1+\varepsilon )}\sin\varphi $ and $\varepsilon
\sin(2\varphi )$ are present in the cross section. The asymmetries with upper
indexes $T, P (L)$ are determined only by the transverse (longitudinal)
components of the electromagnetic current for the $\gamma ^* +d\rightarrow
\Delta +N$ reaction, while the asymmetries with upper index $I$ are determined
by the interference of the longitudinal and transverse components of the
electromagnetic current.

Using the explicit form for the amplitude of the reaction under consideration
it is easy to obtain the expression for the hadronic tensor $H_{ij}$
in terms of the scalar amplitudes $f_i \ (i=1, ..., 36)$ .Appendix A
contains the formulas for the structure functions $\alpha _i, \beta _i,
\gamma _i$ in terms of the scalar amplitudes, which describe the polarization
effects in the $e^-+d\rightarrow e^-+\Delta +N$ reaction due the
deuteron polarization.

Let us stress again that the results listed above have a general nature and
are not related to a particular reaction mechanism. They have been derived assuming 
the one--photon--exchange mechanism, the spin one nature of the photon, the P-invariance of the hadron
electromagnetic interaction, and the hadron electromagnetic current
conservation. Other possible, model dependent contributions to the deuteron structure such as
meson--exchange current, the $D-$ wave
admixture in the deuteron ground state, a $\Delta\Delta-$
component, six--quark configuration etc., do not
affect the general results of this section.

%%%%%%%%%%%%%%%%%%%%%%%%%%%%%%%%%%
\section{Nucleon polarization}
%%%%%%%%%%%%%%%%%%%%%%%%%%%%%%%%%%

Let us write the matrix element of the reaction under consideration in
the following form (the vector indexes of the
virtual photon and $\Delta $--isobar are singled out)
$$M_{fi}=el_i\chi _{2k}^* F_{ki}\chi _1^c. $$
The polarization properties of the nucleon, produced in the
$\gamma^* +d\rightarrow \Delta +N$ reaction, are determined by the
${\vec P}_{ij} $ tensor
\be
{\vec P}_{ij}=Tr\rho_{lk}F_{ki}{\vec \sigma}F_{lj}^+, ~
i,j=x,y,z,
\label{eq:eq17a}
\ee
where $\rho_{lk} $ is the $\Delta $--isobar spin--density matrix. Let us consider
the case when the deuteron target and produced $\Delta $--isobar are
unpolarized. The nucleon polarization vector ${\vec P}$ (multiplied by the
unpolarized differential cross section $d^3\sigma /dE'd\Omega _ed
\Omega _{\Delta}$) is given by an expression obtained from Eq. (\ref{eq:eq3}),  
replacing the components of the hadronic tensor $H_{ij}$ by the corresponding
${\vec P}_{ij}$ tensor components. The tensor ${\vec P}_{ij}$ can be represented
in the following general form:
$${\vec P}_{ij}=\hat {\vec k}P_{ij}^{(k)}+{\vec m}P_{ij}^{(m)}+
{\vec n}P_{ij}^{(n)}. $$
Assuming the P-invariance of the hadron electromagnetic interaction, we can
write the tensor structure of the quantities $ P_{ij}^{(k)},$
$ P_{ij}^{(m)},$ and $ P_{ij}^{(n)}$, in terms of the structure functions
$P_i$, $i=1-13, $ which depend on three independent kinematical variables:
$k^2$, $W$, and $t$:
\ba
P_{ij}^{(k)}&=&P_1\{\hat k,n\}_{ij}+P_2\{m,n\}_{ij}+
iP_3[\hat k,n]_{ij}+iP_4[m,n]_{ij}, \nn\\
P_{ij}^{(m)}&=&P_5\{\hat k,n\}_{ij}+P_6\{m,n\}_{ij}+
iP_7[\hat k,n]_{ij}+iP_8[m,n]_{ij},\nn\\
P_{ij}^{(n)}&=&P_9\hat k_i\hat k_j+P_{10}m_im_j+
P_{11}n_in_j+P_{12}\{\hat k,m\}_{ij}+iP_{13}[\hat k,m]_{ij}.
\label{eq:eq18}
\ea
The expressions for the structure functions $P_i$, in terms of the scalar
amplitudes $f_i$, $i=1-36 $, are given in Appendix B. We can see that
the symmetric parts (with respect to the $i,j$ indexes) of the tensors in this equation
(which correspond to eight structure functions $P_i$, $i=1,2,5,6,9,10,11,12$)
determine the components of the polarization vector of the nucleon produced
in collisions of unpolarized electrons with an unpolarized deuteron target,
for the reaction $d(e,e'{\vec N})\Delta $. The antisymmetric parts of the
tensors in Eq. (\ref{eq:eq18}), (that is, the five structure functions $P_i$, 
$i=3,4,7,8,13$) determine the components of the polarization vector of the
nucleon produced in collisions of longitudinally polarized electrons with an
unpolarized deuteron target, for the reaction $d({\vec e},e'{\vec N})\Delta $.

Moreover, it can be shown that eight structure functions $P_1,$ $P_2,$ $P_5,$
$P_6,$ $P_{9-12}$ (in the symmetric parts of the corresponding tensors)
determine the T--odd contributions to the nucleon polarization vector
${\vec P}$ (for the scattering of unpolarized electrons), whereas the five
structure functions $P_3,$ $P_4,$ $P_7,$ $P_8,$ $P_{13}$ (in the antisymmetric
parts of the corresponding tensors) determine the T--even contributions to the
nucleon polarization vector ${\vec P}$ (for the scattering of longitudinally
polarized electrons).

These five T--even structure functions are nonzero even when the $\gamma^*+d
\rightarrow \Delta +N$ reaction amplitudes are real functions, which is true
in the framework of impulse approximation. In the scattering of the
longitudinally polarized electrons, they determine the nucleon polarization
induced by the absorption of circularly polarized virtual photons (by
unpolarized deuteron target) in the $\gamma ^*+d\rightarrow \Delta +N$
reaction: the polarization is transferred from the electron to the produced
nucleon by the virtual photon. The eight T-odd structure functions, defined
above, are nonzero only for complex $\gamma ^* +d\rightarrow \Delta +N$
reaction amplitudes (with different relative phases).

Due to the tensor structure of the quantities ${\vec P}_{ij}$ , in the
scattering of unpolarized electrons by unpolarized deuterons, the polarization
component of the nucleon which is orthogonal to the $\gamma ^* +d\rightarrow
\Delta +N$ reaction plane is characterized by the same $\varepsilon $ and
$\varphi $ dependences as in the unpolarized case. The polarization vector
of the nucleons polarized in the $\gamma ^*+d\rightarrow \Delta +N$ reaction
plane (components $P_x$ and $P_z$) is characterized by two dependences:
$\varepsilon \sin(2\varphi ) $ and $\sqrt{2\varepsilon(1+\varepsilon )}
\sin\varphi .$

To prove these statements, we explicitly single out the dependence of the
nucleon polarization on the kinematic variables $\varphi $ and $\varepsilon .$
In the general case, the vector of the nucleon polarization can be represented
as the sum of two terms: ${\vec P}^{(0)}$ and ${\vec P}^{(\lambda)}$, where
the polarization ${\vec P}^{(0)}$ corresponds to the unpolarized electron
beam (the so-called induced polarization) and the polarization
${\vec P}^{(\lambda)}$ corresponds to the longitudinally polarized electron
beam (polarization transfer). So, the components of the nucleon polarization
vector ${\vec P}$ in the reactions $d(e,e'{\vec N})\Delta $,
$d({\vec e},e'{\vec N})\Delta $ are given by:
$${\vec P}={\vec P}^{(0)}+\lambda {\vec P}^{(\lambda )}, $$
\ba
P_x^{(0)}\sigma _0&=&N\sin\varphi [\sqrt{2\varepsilon (1+\varepsilon
)}P_x^{(LT)}+\varepsilon \cos\varphi P_x^{(TT)}], \nn\\
P_z^{(0)}\sigma _0&=&N\sin\varphi
[\sqrt{2\varepsilon (1+\varepsilon )}P_z^{(LT)}+\varepsilon \cos\varphi
P_z^{(TT)}],  \nn\\
P_y^{(0)}\sigma _0&=&N[P_y^{(TT)}+\varepsilon
P_y^{(LL)}+\sqrt{2\varepsilon (1+\varepsilon )}\cos\varphi P_y^{(LT)}
+\varepsilon \cos(2\varphi )\bar P_y^{(TT)}], \nn\\
P_x^{(\lambda )}\sigma _0&=&N[\sqrt{1-\varepsilon
^2}R_x^{(TT)}+\sqrt{2\varepsilon (1-\varepsilon )} \cos\varphi R_x^{(LT)}],
\nn\\
P_z^{(\lambda )}\sigma _0&=&N[\sqrt{1-\varepsilon
^2}R_z^{(TT)}+\sqrt{2\varepsilon (1-\varepsilon )} \cos\varphi R_z^{(LT)}],
\nn\\
P_y^{(\lambda )}\sigma _0&=&N\sqrt{2\varepsilon (1-\varepsilon )} \sin\varphi
R_y^{(LT)}], 
\label{eq:eq19}
\ea
where $\sigma _0=d^3\sigma /dE'd\Omega_ed\Omega_{\Delta}$ is the unpolarized
differential cross section of the reaction under consideration, and the
individual contributions to the polarization vector in terms of the structure
functions $P_i$ are:
\ba
P_x^{(TT)}&=&4P_6,~
P_y^{(TT)}=P_{10}+P_{11},~\bar P_y^{(TT)}=P_{10}-P_{11},~
P_z^{(TT)}=4P_2, \nn\\
P_x^{(LT)}&=&-2\frac{\sqrt{Q^2}}{k_0}P_5,~
P_y^{(LT)}=-2\frac{\sqrt{Q^2}}{k_0}P_{12},~
P_z^{(LT)}=-2\frac{\sqrt{Q^2}}{k_0}P_1,\nn\\
P_y^{(LL)}&=&2\frac{Q^2}{k^2_0}P_9,~
R_x^{(TT)}=4P_8,~ R_z^{(TT)}=4P_4,\nn\\
R_x^{(LT)}&=&-2\frac{\sqrt{Q^2}}{k_0}P_7,~
R_y^{(LT)}=2\frac{\sqrt{Q^2}}{k_0}P_{13},~
R_z^{(LT)}=-2\frac{\sqrt{Q^2}}{k_0}P_3.
\nn
\ea
The expressions for the structure functions $P_i$ in terms of the reaction
amplitudes are general and do not depend on the details of the reaction
mechanism. As explicitly shown in Appendix B, each of the 13 structure
functions $P_i(W, k^2, t)$, $i=1-13,$ carries independent information about
the scalar amplitudes. Therefore, measurement of all these structure functions
is, in principle, necessary to perform the complete $\gamma ^* +d\rightarrow
\Delta +N$ experiment.
%%%%%%%%%%%%%%%%%%%%%%%%%%%%%%%
\section{Helicity amplitudes}
%%%%%%%%%%%%%%%%%%%%%%%%%%%%%%%

Since the spin structure of the matrix element of the reaction under
consideration is quite complicated, it is convenient to perform the
unitarization procedure (taking into account the final state interaction,
i.e., the $N\Delta \to N\Delta $ scattering effects) with the help of the
helicity amplitudes formalism. As it was shown above, the reaction
$\gamma ^* +d\rightarrow \Delta +N$ is described by 36 independent amplitudes.

Let us introduce the set of the helicity amplitudes $h_{\lambda\lambda '}
(k^2, W, \vartheta )$ (where $\lambda$ and $\lambda '$ are the helicities
of the initial ($\gamma ^*+d$) and final ($\Delta +N$) states) and consider
the amplitudes
$$h_{\lambda\lambda '}=< \lambda _{\Delta}, \lambda _N|T|\lambda _{\gamma},
\lambda _d>={\vec \chi}_{2}^*(\lambda _{\Delta}){\vec F}
(\lambda _{\gamma}, \lambda _d)\chi _1^c(\lambda _N), $$
where  $\lambda _{\gamma}, \lambda _d, \lambda _N $ and $\lambda _{\Delta}$
are the helicities of the virtual photon, deuteron, nucleon and $\Delta $--
isobar, respectively, with  $\lambda =\lambda _{\gamma}- \lambda _d$ and
$\lambda '= \lambda _{\Delta}-\lambda _N$. We choose the following convention:
\ba
&&h_1=< \frac{1}{2}\frac{1}{2}|T|11> ,~
h_2=< -\frac{1}{2}-\frac{1}{2}|T|11> ,~
h_3=< \frac{1}{2}\frac{1}{2}|T|10> ,~
\label{eq:eq19a}
\\
&&h_4=< -\frac{1}{2}-\frac{1}{2}|T|10> ,~
h_5=< \frac{1}{2}\frac{1}{2}|T|1-1> ,~
\nn\\
&&
h_6=< -\frac{1}{2}-\frac{1}{2}|T|1-1> ,~
h_7=< \frac{1}{2}-\frac{1}{2}|T|11> ,~
h_8=< -\frac{1}{2}\frac{1}{2}|T|11> ,~  
\nn\\
&&
h_9=< \frac{1}{2}-\frac{1}{2}|T|10> ,  
h_{10}=< -\frac{1}{2}\frac{1}{2}|T|10> ,~
h_{11}=< \frac{1}{2}-\frac{1}{2}|T|1-1> ,~
\nn\\
&&
h_{12}=< -\frac{1}{2}\frac{1}{2}|T|1-1> ,~  
h_{13}=< \frac{1}{2}\frac{1}{2}|T|01> ,~
h_{14}=< \frac{1}{2}\frac{1}{2}|T|00> ,~
\nn\\
&&
h_{15}=< \frac{1}{2}\frac{1}{2}|T|0-1> ,~
h_{16}=< -\frac{1}{2}\frac{1}{2}|T|01> ,~ 
h_{17}=< -\frac{1}{2}\frac{1}{2}|T|00> ,~
\nn\\
&&
h_{18}=< -\frac{1}{2}\frac{1}{2}|T|0-1> ,
h_{19}=< \frac{3}{2}\frac{1}{2}|T|11> ,~
h_{20}=< -\frac{3}{2}-\frac{1}{2}|T|11> ,~
\nn\\
&&
h_{21}=< \frac{3}{2}\frac{1}{2}|T|10> ,~
h_{22}=< -\frac{3}{2}-\frac{1}{2}|T|10> ,~ 
h_{23}=< \frac{3}{2}\frac{1}{2}|T|1-1> ,~
\nn\\
&&
h_{24}=< -\frac{3}{2}-\frac{1}{2}|T|1-1> ,~
h_{25}=< \frac{3}{2}-\frac{1}{2}|T|11> ,~
\nn\\
&&
h_{26}=< -\frac{3}{2}\frac{1}{2}|T|11> ,~ 
h_{27}=< \frac{3}{2}-\frac{1}{2}|T|10> ,~
h_{28}=< -\frac{3}{2}\frac{1}{2}|T|10> ,~
\nn\\
&&
h_{29}=< \frac{3}{2}-\frac{1}{2}|T|1-1> ,~
h_{30}=< -\frac{3}{2}\frac{1}{2}|T|1-1> ,~  
\nn\\ 
&&h_{31}=< \frac{3}{2}\frac{1}{2}|T|01> ,~
h_{32}=< \frac{3}{2}\frac{1}{2}|T|00> ,~
h_{33}=< \frac{3}{2}\frac{1}{2}|T|0-1> ,~
\nn\\
&&h_{34}=< -\frac{3}{2}\frac{1}{2}|T|01> ,~  
h_{35}=< -\frac{3}{2}\frac{1}{2}|T|00> ,~
h_{36}=< -\frac{3}{2}\frac{1}{2}|T|0-1>. 
\nn
\ea
We choose the helicity amplitudes in such a way that the first 18 helicity amplitudes
(corresponding to the $\Delta $--isobar helicities $\pm 1/2$) coincide with
the helicity amplitudes for the deuteron electrodisintegration reaction
$\gamma ^*+d\to n+p$ \cite{RTB}. As it was shown above, the matrix element
of the process under consideration can be described in terms of the scalar
amplitudes. The formulas relating the two sets of independent amplitudes
$f_i$ and $h_i$  are given in Appendix C.

%%%%%%%%%%%%%%%%%%%%%%%%%%%%%%%%%%%%%%%%%%%%%%%%%%%%%%%%%%%%%%%%%%%%%%%%
\section{$\Delta $--isobar production in deuteron photodisintegration
process}
%%%%%%%%%%%%%%%%%%%%%%%%%%%%%%%%%%%%%%%%%%%%%%%%%%%%%%%%%%%%%%%%%%%%%%%%%

Let us consider the particular case of the $\Delta $--isobar production, in the deuteron photodisintegration reaction
\be
\gamma (k)+d(P)\to \Delta (p_1)+N(p_2),
\label{eq:aq21}
\ee
where the four--momenta of the particles are given in the brackets. Of course, 
all observables for this reaction can be obtained using the formulas presented
above for the case of the virtual photon, but it is rather tedious procedure. 
So, it is worth to have the expressions for the differential cross section and 
various polarization observables which are suitable for the analysis of the 
future data on this reaction.

The matrix element of this reaction can be written as
\be
M=eA_{\mu}J_{\mu}=-eA_{i}J_{i},
\label{eq:eq21}
\ee 
where $A_{\mu}$ is the photon polarization four--vector and we use the
transverse gauge: ${\vec k}\cdot{\vec A}=0$ (${\vec k}$ is the photon momentum).

The differential cross section in CMS (not averaged over the spins of the 
initial particles) can be written as
\be
\frac{d\sigma}{d\Omega}=\frac{\alpha}{8\pi}\frac{p}{W}
\frac{1}{W^2-M^2}\rho_{ij}H_{ij},
\label{eq:eq22}
\ee
where $\rho_{ij}=A_iA_j^*$ and hadronic tensor is determined as $H_{ij}=
J_iJ_j^*$. The notation of the other quantities have been defined in previous sections.

In the reaction CMS, the quantity $J_i$ can be written as 
\be
J_{i}={\vec \chi}_2^+{\vec G}_i\chi _1^c, 
\label{eq:eq23}
\ee
where ${\vec \chi}_2^+ $ and $\chi _1^c$ are the $\Delta $--isobar vector
spinor and nucleon spinor, respectively. The quantity ${\vec G}_i$ can be 
chosen as
\be
{\vec G}_i={\vec m}G_i^{(m)}+\hat {\vec k}G_i^{(k)}, 
\label{eq:eq24}
\ee
with
\ba
G_i^{(m)}&=&m_i(ig_1{\vec U}\cdot {\vec m}
+ig_2{\vec U}\cdot \hat {\vec k}+
g_3{\vec \sigma }\cdot {\vec n}{\vec U}\cdot {\vec m}+
g_4{\vec \sigma}\cdot {\vec n}{\vec U}\cdot \hat {\vec k}+
\label{eq:eq26}
\\
&&
g_5{\vec \sigma }\cdot {\vec m}{\vec U}\cdot {\vec n}
+g_6{\vec \sigma}\cdot \hat {\vec k}{\vec U}\cdot {\vec n})+
n_i(ig_7{\vec U}\cdot {\vec n}+
g_{8}{\vec \sigma }\cdot {\vec n}{\vec U}\cdot {\vec n}+
\nn\\
&&
g_9{\vec \sigma}\cdot {\vec m}{\vec U}\cdot {\vec m}+
g_{10}{\vec \sigma }\cdot \hat {\vec k}{\vec U}\cdot {\vec m}+
g_{11}{\vec \sigma}\cdot {\vec m}{\vec U}\cdot \hat {\vec k}+
g_{12}{\vec \sigma }\cdot \hat {\vec k}{\vec U}\cdot \hat {\vec k}),  
\nn\\
G_i^{(k)}&=&m_i(ig_{13}{\vec U}\cdot {\vec m}
+ig_{14}{\vec U}\cdot \hat {\vec k}+
g_{15}{\vec \sigma }\cdot {\vec n}{\vec U}\cdot {\vec m}+
g_{16}{\vec \sigma}\cdot {\vec n}{\vec U}\cdot \hat {\vec k}+
\nn\\&&
g_{17}{\vec \sigma }\cdot {\vec m}{\vec U}\cdot {\vec n}+
g_{18}{\vec \sigma}\cdot \hat {\vec k}{\vec U}\cdot {\vec n})+
n_i(ig_{19}{\vec U}\cdot {\vec n}+
g_{20}{\vec \sigma }\cdot {\vec n}{\vec U}\cdot {\vec n}+
\nn\\&&
g_{21}{\vec \sigma}\cdot {\vec m}{\vec U}\cdot {\vec m}+
g_{22}{\vec \sigma }\cdot \hat {\vec k}{\vec U}\cdot {\vec m}+
g_{23}{\vec \sigma}\cdot {\vec m}{\vec U}\cdot \hat {\vec k}+
g_{24}{\vec \sigma }\cdot \hat {\vec k}{\vec U}\cdot \hat {\vec k}),
\nn
\ea
where $g_i (i=1-24)$ are the scalar amplitudes, depending on two variables 
(energy and scattering angle), which completely determine the reaction dynamics,
and ${\vec U}$ is the deuteron polarization vector.
 
The hadronic tensor $H_{ij} (i,j=x,y,z)$ depends linearly on the target
polarization and it can be represented as follows
\be
H_{ij}=H_{ij}(0)+H_{ij}(\xi)+H_{ij}(S),
\label{eq:eqtensor}
\ee
where the term $H_{ij}(0)$ corresponds to the case of the unpolarized deuteron
target, and the term $H_{ij}(\xi) (H_{ij}(S))$ corresponds to the case of the
vector (tensor-)-polarized target. Let us consider the polarization observables of the $\gamma +d\to \Delta +N$ reaction for each contribution to the hadronic tensor $H_{ij}.$

%%%%%%%%%%%%%%%%%%%%%%%%%%%%%%%%%%%%%%%%%
\subsection{Unpolarized deuteron target}
%%%%%%%%%%%%%%%%%%%%%%%%%%%%%%%%%%%%%%%%%

The general structure of the hadronic tensor for unpolarized 
deuteron target has following form
\be
H_{ij}(0)=a_1m_im_j+a_2n_in_j,
\label{eq:eq28}
\ee
where $a_1$ and $a_2$ are the structure functions which can be expressed in
terms of the reaction scalar amplitudes. The expressions of these structure
functions can be found in Appendix D.

The differential cross section of this reaction for the case of 
unpolarized particles can be written as
\be
\frac{d\sigma_{un}}{d\Omega}=N(a_1+a_2),~ 
N=\frac{\alpha}{16\pi}\frac{p}{W}\frac{1}{W^2-M^2}.
\label{eq:eq29}
\ee
Let us consider the case when photon is polarized. The general expression of the
photon polarization vector is determined by two real parameters $\beta $ and 
$\delta $ and it can be written as \cite{AB65}
\be
{\vec A}=\cos\beta {\vec m}+\sin\beta exp(i\delta ){\vec n}.
\label{eq:eq30}
\ee
If the parameter $\delta =0$ then the photon polarization vector describe the 
linear polarization state of the photon, directed at an angle $\beta $ with respect 
to the $x$ axis. The parameters $\beta =\pi /4$ and $\delta =\pm \pi /2$ denote the 
circular polarization of the photon. Arbitrary $\beta $ and $\delta $ correspond to 
elliptic photon polarization.

The differential cross section in the case of polarized photon has the 
following form
\be
\frac{d\sigma}{d\Omega}=\frac{d\sigma_{un}}{d\Omega}
(1+A_{\perp}\cos2\beta ), 
\label{eq:eq31}
\ee
where $A_{\perp}$ is the asymmetry due to the linear polarization of 
the photon and it can be written as
\be
A_{\perp}=
\frac{d\sigma /d\Omega (\beta =0^0)-d\sigma /d\Omega (\beta =90^0)}
{d\sigma /d\Omega (\beta =0^0)+d\sigma /d\Omega (\beta =90^0)}.
\label{eq:eq32}
\ee
This asymmetry has following form in terms of the structure functions
\be
\frac{d\sigma_{un}}{d\Omega}A_{\perp}=N(a_1-a_2)~ or~ 
A_{\perp}=\frac{a_1-a_2}{a_1+a_2}. 
\label{eq:eq33}
\ee
Note that circular polarization of the photon does not contribute to the 
differential cross section due to the P--invariance of the hadron
electromagnetic interaction.
%%%%%%%%%%%%%%%%%%%%%%%%%%%%%%%%%%%%%%%%%%%%%%%%%%
\subsection{Vector polarized deuteron target}
%%%%%%%%%%%%%%%%%%%%%%%%%%%%%%%%%%%%%%%%%%%%%%%%%%%%%%
For $\gamma +d\to \Delta +N$, the 
dependence of the polarization observables on the deuteron vector polarization is determined by six structure 
functions. The part of the hadronic tensor which depends on the deuteron vector 
polarization has the following general structure:
\ba
H_{ij}(\xi )&=&{\vec\xi }{\vec n}(b_1m_im_j+b_2n_in_j)+
{\vec\xi }\hat {\vec k}(b_3\{m,n\}_{ij}+ib_4[m,n]_{ij})+
\nn\\
&&+{\vec\xi }{\vec m}(b_{5}\{m,n\}_{ij}+ib_{6}[m,n]_{ij}), 
\label{eq:eq34}
\ea
where $b_{i}$, $(i=1-6)$ are the structure functions, depending on two variables, 
which can be expressed in terms of the reaction scalar amplitudes. The 
expressions of these structure functions are given in Appendix D. 

The part of the differential cross section of the $\gamma +d\to \Delta +N$ 
reaction which depends on the deuteron vector polarization, for the case of 
arbitrarily polarized photon, can be written as 
\ba
\frac{d\sigma_v}{d\Omega}&=&\frac{d\sigma_{un}}{d\Omega}
\biggl [A_{y}\xi_y+C_y^l\cos2\beta \xi_y+\sin2\beta \cos\delta (C_x^l\xi_x+
C_z^l\xi_z)+ 
\nn \\
&&+\sin2\beta \sin\delta (C_x^c\xi_x+C_z^c\xi_z)\biggr ], 
\label{eq:eq35}
\ea
where $A_{y}$ is the asymmetry due to the vector polarization of the deuteron
target when the photon is unpolarized (the so--called single target asymmetry). 
This asymmetry is due to the component of the 
polarization vector ${\vec \xi}$ describing the vector polarization of the
target, which is normal to the reaction plane. If the reaction amplitudes are real functions (as, for example, in  
the impulse approximation) then this asymmetry is equal to zero. The quantities
$C^l_{x,y,z} (C^c_{x,z})$ are the correlation coefficients due to the vector 
polarization of the deuteron target when the photon is linearly (circularly) 
polarized. The correlation coefficients $C^l_{x,y,z}$ are zero when the  
amplitudes are real. The correlation coefficients $C^c_{x,z}$ are determined by the 
components of the polarization vector lying in the reaction plane and they 
are non--zero, in general, for the real amplitudes. All these 
polarization observables can be expressed in terms of the structure functions 
$b_i (i=1-6)$ as:
\ba
&&\frac{d\sigma_{un}}{d\Omega}A_{y}=N(b_1+b_2),~  
\frac{d\sigma_{un}}{d\Omega}C_{y}^l=N(b_1-b_2), 
\nn \\
&&\frac{d\sigma_{un}}{d\Omega}C_{x}^l=2Nb_5,~  
\frac{d\sigma_{un}}{d\Omega}C_{z}^l=2Nb_3, \nn \\
&&\frac{d\sigma_{un}}{d\Omega}C_{x}^c=2Nb_6,~  
\frac{d\sigma_{un}}{d\Omega}C_{z}^c=2Nb_4. \nn 
\label{eq:eq36}
\ea
%%%%%%%%%%%%%%%%%%%%%%%%%%%%%%%%%%%%%%%%%%%%%
\subsection{Tensor polarized deuteron target}
%%%%%%%%%%%%%%%%%%%%%%%%%%%%%%%%%%%%%%%%%%%%%
For $\gamma +d\to \Delta +N$, the dependence of the polarization observables,  on the deuteron tensor (quadrupole) polarization 
is completely determined by ten structure functions.
The part of the hadronic tensor which depends on the tensor (quadrupole)  
polarization of the deuteron target has the following general structure:
\ba
H_{ij}(S)&=&S_{ab}\hat k_a\hat k_b(c_1m_im_j+c_2n_in_j)+
S_{ab}m_am_b(c_3m_im_j+c_4n_in_j)+ 
 \nn \\
&&
S_{ab}\{\hat k,m\}_{ab}(c_{5}m_im_j+c_{6}n_in_j)+
S_{ab}\{\hat k,n\}_{ab}(c_{7}\{m,n\}_{ij}+ \nn \\
&&
ic_{8}[m,n]_{ij})+ 
S_{ab}\{m,n\}_{ab}(c_{9}\{m,n\}_{ij}+ic_{10}[m,n]_{ij}), 
\label{eq:eq36a}
\ea
where $c_i (i=1=10)$ are structure functions, which depend on two variables. 
Their expressions in terms of the reaction scalar amplitudes are given in Appendix D. 

The part of the differential cross section of the $\gamma +d\to \Delta +N$ 
reaction which depends on the deuteron tensor polarization, for the case of 
arbitrarily polarized photon, can be written as 
\ba
\frac{d\sigma_t}{d\Omega}&=&\frac{d\sigma_{un}}{d\Omega}
\biggl \{A_{zz}Q_{zz}+A_{xx}(Q_{xx}-Q_{yy})+A_{xz}Q_{xz}+\cos2\beta 
\biggl [C^l_{zz}Q_{zz}+ \nn\\
&&
C^l_{xx}(Q_{xx}-Q_{yy})+C^l_{xz}Q_{xz}\biggr ]+  
\sin2\beta \cos\delta (C^l_{xy}Q_{xy}+C^l_{yz}Q_{yz})+ \nn\\
&&
\sin2\beta \sin\delta (C^c_{xy}Q_{xy}+C^c_{yz}Q_{yz})\biggr \},
\label{eq:eq37}
\ea
where $A_{zz}$, $A_{xx}$ and $A_{xz}$ are the asymmetries due to the tensor 
polarization of the deuteron target when the photon is unpolarized. These 
asymmetries are non--zero, in the general case, if the reaction amplitudes are 
real, in contrast to the $A_y$ asymmetry. 
The quantities $C^l_{zz}$, $C^l_{xx}$, $C^l_{xz}$, $C^l_{xy}$ 
and $C^l_{yz}$ are the correlation coefficients due to the tensor polarization 
of the deuteron target, when the photon is linearly polarized (they
can be non--zero even if the reaction amplitudes are real). The quantities 
$C^c_{xy}$ and $C^c_{yz}$ are the correlation coefficients which are determined 
by the tensor polarization of the deuteron target and the circular polarization of 
the photon (they are completely determined by the reaction mechanism beyond the 
impulse approximation, for example, by the final--state interaction effects).
All these polarization observables can be expressed in terms of the structure 
functions $c_i$, $(i=1-10)$ as
\ba
\frac{d\sigma_{un}}{d\Omega}A_{zz}&=&\frac{N}{2}[2\gamma_1^2(c_1+c_2)-c_3-c_4], ~
\frac{d\sigma_{un}}{d\Omega}A_{xx}=\frac{N}{2}(c_3+c_4), \nn\\
\frac{d\sigma_{un}}{d\Omega}A_{xz}&=&2N\gamma_1(c_5+c_6), 
\frac{d\sigma_{un}}{d\Omega}C^l_{zz}=\frac{N}{2}[2\gamma_1^2(c_1-c_2)-c_3+c_4],\nn\\
\frac{d\sigma_{un}}{d\Omega}C^l_{xx}&=&\frac{N}{2}(c_3-c_4), \ \
\frac{d\sigma_{un}}{d\Omega}C^l_{xz}=2N\gamma_1(c_5-c_6), \ \
\frac{d\sigma_{un}}{d\Omega}C^l_{xy}=4Nc_9, \nn\\
\frac{d\sigma_{un}}{d\Omega}C^l_{yz}&=&4N\gamma_1c_7, \ \
\frac{d\sigma_{un}}{d\Omega}C^c_{xy}=4Nc_{10}, \ \
\frac{d\sigma_{un}}{d\Omega}C^c_{yz}=4N\gamma_1c_8, \nn\\
\gamma_1&=&\frac{W^2+M^2}{2MW}.
\label{eq:eq41n}
\ea
%%%%%%%%%%%%%%%%%%%%%%%%%%%%%%%%%%%%%
\subsection{Polarization of the nucleon}
%%%%%%%%%%%%%%%%%%%%%%%%%%%%%%%%%%%%%
Taking into account the expression of the quantity $J_i$ we may write the nucleon polarization, in  the $\gamma +d\to \Delta +N$ 
reaction, in the following form:
\be
\frac{d\sigma_{un}}{d\Omega}{\vec P}=N\rho_{ij}{\vec P}_{ij},
\label{eq:eq42n}
\ee
where ${\vec P}_{ij}=Tr\rho^{\Delta}_{lk}G_{ki}{\vec \sigma}G^+_{lj}$ 
($\rho^{\Delta}_{lk}$ is the $\Delta $--isobar spin--density matrix). The
general structure of this tensor, for the case of unpolarized 
$\Delta $--isobar and deuteron target, can be represented in the following 
form
\ba
{\vec P}_{ij}&=&\hat {\vec k}(d_1\{m,n\}_{ij}+id_2[m,n]_{ij})+
{\vec m}(d_3\{m,n\}_{ij}+id_4[m,n]_{ij})+\nn\\
&&+{\vec n}(d_{5}m_im_j+d_{6}n_in_j), 
\label{eq:eq40}
\ea
where $d_i \ (i=1-6)$ are the structure functions and their expressions in terms
of the reaction scalar amplitudes are given in Appendix E. The 
nucleon polarization in the $\gamma +d\to \Delta +N$ reaction, is completely 
determined by six structure functions, when the photon 
is arbitrarily polarized and the other particles are unpolarized.

The vector components of the nucleon polarization are
\ba
\frac{d\sigma_{un}}{d\Omega}P_y&=&P_y^0+\cos2\beta P_y^l,~ \nn\\
\frac{d\sigma_{un}}{d\Omega}P_x&=&\sin2\beta (\cos\delta P_x^l+\sin\delta P_x^c),~\nn\\
\frac{d\sigma_{un}}{d\Omega}P_z&=&\sin2\beta (\cos\delta P_z^l+\sin\delta P_z^c), 
\label{eq:eq45n}
\ea
where $P_y^0$ is the $y$--component of the nucleon polarization when all 
other particles are unpolarized, whereas $P_y^l$, $P_x^l$ and $P_z^l$ are the components
of the nucleon polarization when the photon is linear polarized. All these 
observables arise due to reaction mechanisms beyond the 
impulse approximation. The quantities $P_x^c$ and $P_z^c$ are the $x$-- and 
$z$--components of the nucleon polarization when the photon is circularly 
polarized and in general, they can be non--zero in the impulse 
approximation. The expressions of these observables in terms of the structure 
functions $d_i$ are
\ba
&&P_y^0=\frac{N}{2}(d_5+d_6),~ P_y^l=\frac{N}{2}(d_5-d_6), P_x^l=Nd_3,~ P_z^l=Nd_1,~ ~\nn\\
&&P_x^c=Nd_4,~ P_z^c=Nd_2. 
\label{eq:eq46n}
\ea
%%%%%%%%%%%%%%%%%%%%%%%%%%%
\section{Conclusions}
%%%%%%%%%%%%%%%%%%%%%%%%%%%

We developed a relativistic approach to the calculation of the differential 
cross section and various polarization observables for the $\Delta $--isobar
production in deuteron photo-- and electrodisintegration processes,
$\gamma +d\to \Delta +N$ and $e^-+d\to e^-+\Delta +N$.

A general analysis of the structure of the differential cross
section and polarization observables for the $\Delta $--isobar excitation
in the scattering of the electrons by the deuteron target, $\gamma ^*+d\to
\Delta +N$ was derived. Our formalism is based on the most general symmetry properties
of the hadron electromagnetic interaction, such as gauge invariance (the
conservation of the hadronic and leptonic electromagnetic currents) and
$P$--invariance (invariance with respect to the space reflections) and does
not depend on the deuteron structure and on the details of the reaction
mechanism for $\gamma ^*+d\to \Delta +N$. This general analysis was done with
the help of the structure function formalism which is especially convenient
for the investigation of the polarization phenomena in this reaction.

The observables related to the cases of an arbitrary polarized deuteron target,
longitudinally polarized electron beam, polarization of the outgoing nucleon,
as well as the polarization transfer from electron to final nucleon, and
the correlation of the electron and deuteron polarizations were considered
in detail. We derived the expressions for polarization effects which are absent in the 
impulse approximation and due to the strong $\Delta N-$ interaction in the 
final state. 

A particular case of the process of the photoproduction of the $\Delta $--isobar 
on the deuteron target has been considered in details. The differential cross 
section and various polarization observables have been derived in terms of the 
reaction amplitudes. The polarization observables due to the linear and circular 
polarizations of the photon provided the deuteron target is arbitrarily 
polarized have been derived in terms of the reaction amplitudes. The 
polarization of the final nucleon is also considered.

General properties of these observables have been derived and underlined.
Such properties should be fulfilled by any model calculation. In this respect, the present approach is important, as it gives on one side, guidelines for models and, on the other side, defines the strategy (the observables and the kinematical conditions) for experiments.

%%%%%%%%%%%%%%%%%%%%%%%%%%
\section{Acknowledgment}
%%%%%%%%%%%%%%%%%%%%%%%%%%

One of us (G.I.G.) acknowledges the hospitality of CEA, Saclay, where part of 
this work was done. This work is supported in part by grant INTAS Ref. 
No. 05-1000008-8328.

%%%%%%%%%%%%%%%%%%%%%%%
\section{Appendix A}
%%%%%%%%%%%%%%%%%%%%%%%

In this Appendix, we present the formulas for the structure functions which
determine the hadronic tensor $H_{ij}$ for various polarization states of the
deuteron target. The functions are written in terms of the scalar amplitudes
$f_i \ (i=1, ..., 36)$ determining the $\gamma^*+d\rightarrow \Delta +N$
reaction.

$\bullet $ \underline{Unpolarized deuteron target}.

The hadronic tensor $H_{ij}(0)$ is determined by the structure functions
$\alpha _i$, $(i=1, .., 5)$

\ba
\alpha _1&=&\frac{2}{3}\biggl \{Ax_1\biggl[|f_{13}|^2+|f_{15}|^2+
|f_{17}|^2+|f_{18}|^2+z|f_{14}|^2+z|f_{16}|^2\biggr]+ 
\nn \\
&&
Ax_2\biggl[|f_{31}|^2+|f_{33}|^2+
|f_{35}|^2+|f_{36}|^2+z|f_{32}|^2+z|f_{34}|^2\biggr]+
 \nn \\
&&
2Ax_3Re(f_{13}f_{31}^*+f_{18}f_{36}^*+f_{17}f_{35}^*+f_{15}f_{33}^*+
zf_{14}f_{32}^*+zf_{16}f_{34}^*)+  \nn \\
&&
2BRe(f_{18}f_{35}^*-f_{17}f_{36}^*+f_{33}f_{13}^*-f_{31}f_{15}^*+
zf_{34}f_{14}^*-zf_{32}f_{16}^*)\biggr \},  \nn \\
\alpha _2&=&\frac{2}{3}\biggl \{Ax_1\biggl[|f_{7}|^2+|f_{8}|^2+
|f_{9}|^2+|f_{10}|^2+z|f_{11}|^2+z|f_{12}|^2\biggr]+ \nn \\
&&
Ax_2\biggl[|f_{25}|^2+|f_{26}|^2+
|f_{27}|^2+|f_{28}|^2+z|f_{29}|^2+z|f_{30}|^2\biggr]+ \nn \\
&&
2Ax_3Re(f_{7}f_{25}^*+f_{10}f_{28}^*+f_{9}f_{27}^*+f_{8}f_{26}^*+
zf_{11}f_{29}^*+zf_{12}f_{30}^*)+ \nn \\
&&
2BRe(f_{10}f_{27}^*-f_{9}f_{28}^*+f_{26}f_{7}^*-f_{25}f_{8}^*+
zf_{12}f_{29}^*-zf_{11}f_{30}^*)\biggr \}, \nn \\
\alpha _3&=&\frac{2}{3}\biggl \{Ax_1\biggl[|f_{1}|^2+|f_{3}|^2+
|f_{5}|^2+|f_{6}|^2+z|f_{2}|^2+z|f_{4}|^2\biggr]+ \nn \\
&&
Ax_2\biggl[|f_{19}|^2+|f_{21}|^2+
|f_{23}|^2+|f_{24}|^2+z|f_{20}|^2+z|f_{22}|^2\biggr]+ \nn \\
&&
2Ax_3Re(f_{1}f_{19}^*+f_{3}f_{21}^*+f_{5}f_{23}^*+f_{6}f_{24}^*+
zf_{2}f_{20}^*+zf_{4}f_{22}^*)+
\nn \\
&&
2BRe(f_{6}f_{23}^*-f_{5}f_{24}^*+f_{21}f_{1}^*-f_{19}f_{3}^*+
zf_{22}f_{2}^*-zf_{20}f_{4}^*)\biggr \}, \nn \\
\alpha _4&=&\frac{2}{3}ReC,~ \alpha _5=-\frac{2}{3}ImC, \nn \\
C&=&Ax_1\biggl [f_{1}f_{13}^*+f_{3}f_{15}^*+f_{5}f_{17}^*+f_{6}f_{18}^*+
zf_{2}f_{14}^*+zf_{4}f_{16}^*\biggr ]+ \nn \\
&&
Ax_2\biggl [f_{19}f_{31}^*+f_{21}f_{33}^*+f_{23}f_{35}^*+f_{24}f_{36}^*+
zf_{20}f_{32}^*+zf_{22}f_{34}^*\biggr ]+ \nn \\
&&
Ax_3\biggl [f_{1}f_{31}^*+f_{3}f_{33}^*+f_{5}f_{35}^*+f_{6}f_{36}^*+
f_{19}f_{13}^*+f_{21}f_{15}^*+f_{23}f_{17}^*+ \nn \\
&&
f_{24}f_{18}^*+zf_{2}f_{32}^*+zf_{4}f_{34}^*+zf_{20}f_{14}^*+
zf_{22}f_{16}^*\biggr ]+ \nn \\
&&
B\biggl [f_{6}f_{35}^*+f_{23}f_{18}^*-f_{5}f_{36}^*-f_{14}f_{17}^*+
f_{21}f_{13}^*-f_{19}f_{15}^*+f_{1}f_{33}^*-\nn \\
&& 
f_{3}f_{31}^*-zf_{20}f_{16}^*+zf_{22}f_{14}^*+zf_{2}f_{34}^*-
zf_{4}f_{32}^*\biggr ]. 
\nn 
\ea
We use here the notation
\ba
x_1&=&1+\frac{({\vec m}\cdot {\vec p})^2}{M_{\Delta}^2},~
x_2=1+\frac{(\hat {\vec k}\cdot {\vec p})^2}{M_{\Delta}^2},~
x_3=\frac{{\vec m}\cdot {\vec p}\hat {\vec k}\cdot {\vec p}}{M_{\Delta}^2}, 
\nn\\
A&=&\frac{2M_{\Delta}^2}{3M_{\Delta}^2+{\vec p}^2},~
B=\frac{M_{\Delta}E_{\Delta}}{3M_{\Delta}^2+{\vec p}^2},~
z=\frac{\omega^2}{M^2},\nn 
\ea
where $M_{\Delta}$ is the $\Delta $--isobar mass, ${\vec p} \ (E_{\Delta})$
and $\omega $ are the momentum (energy) and energy of the $\Delta $--isobar
and deuteron in CMS of the $\gamma^*+d\rightarrow \Delta +N$ reaction, which are expressed in term of the total energy and of the masses of the particles as:
\ba
\omega &=&\frac{W^2+M^2-k^2}{2W},~
E_{\Delta} =\frac{W^2+M^2_{\Delta}-m^2}{2W},~
\nn\\
|{\vec p}|&=&\frac{1}{2W}\sqrt{(W^2+M^2_{\Delta}-m^2)^2-4W^2M^2_{\Delta}}. 
\nn 
\ea

$\bullet $ \underline{Vector polarized deuteron target.}

The hadronic tensor $H_{ij}(\xi )$ is determined by the structure functions
$\beta _i \ (i=1, .., 13)$

\ba
\beta_1&=&-2\frac{\omega}{M}ImD_1,~
\beta_2=-2\frac{\omega}{M}ImD_2, \nn \\
D_1&=&Ax_1(f_{14}f_{13}^*+f_{16}f_{15}^*)+Ax_2(f_{32}f_{31}^*+
f_{34}f_{33}^*)+Ax_3(f_{14}f_{31}^*-f_{13}f_{32}^*+ 
\nn \\
&&
+f_{16}f_{33}^*-f_{15}f_{34}^*)-B(f_{13}f_{34}^*+
f_{16}f_{31}^*-f_{14}f_{33}^*-f_{15}f_{32}^*), \nn \\
D_2&=&Ax_1(f_{2}f_{1}^*+f_{4}f_{3}^*)+Ax_2(f_{20}f_{19}^*+
f_{22}f_{21}^*)+Ax_3(f_{2}f_{19}^*-f_{1}f_{20}^*+ \nn \\
&&
f_{4}f_{21}^*-f_{3}f_{22}^*)-B(f_{1}f_{22}^*+
f_{4}f_{19}^*-f_{2}f_{21}^*-f_{3}f_{20}^*), \nn \\
\beta_3&=&-2\frac{\omega}{M}ImD_3, \nn \\
D_3&=&Ax_1(f_{11}f_{9}^*+f_{12}f_{10}^*)+Ax_2(f_{29}f_{27}^*+
f_{30}f_{28}^*)+Ax_3(f_{12}f_{28}^*+\nn \\
&&
f_{30}f_{10}^*+ f_{11}f_{27}^*-f_{9}f_{29}^*)+B(f_{12}f_{27}^*+
f_{9}f_{30}^*-f_{10}f_{29}^*-f_{11}f_{28}^*), \nn \\
\beta_4&=&-\frac{\omega}{M}ImD_4,~
\beta_5=-\frac{\omega}{M}ReD_4, \nn \\
\beta_6&=&-ImD_5,~  \beta_8=-ReD_5, \nn \\
\beta_7&=&-ImD_6,~  \beta_9=-ReD_6, \nn \\
D_4&=&Ax_1(f_{2}f_{13}^*+f_{4}f_{15}^*-f_{1}f_{14}^*-f_{3}f_{16}^*)+
\nn \\
&&
Ax_2(f_{20}f_{31}^*+f_{22}f_{33}^*-f_{19}f_{32}^*-
f_{21}f_{34}^*)+\nn \\
&&
Ax_3(f_{2}f_{31}^*+f_{20}f_{13}^*-f_{1}f_{32}^*-f_{19}f_{14}^*+
f_{4}f_{33}^*-f_{3}f_{34}^*+f_{22}f_{15}^*-f_{21}f_{16}^*)- \nn \\
&&
B(f_{21}f_{14}^*+f_{1}f_{34}^*-f_{22}f_{13}^*-f_{2}f_{33}^*+
f_{20}f_{15}^*+f_{4}f_{31}^*-f_{19}f_{16}^*-f_{3}f_{22}^*), \nn \\
D_5&=&Ax_1(f_{9}f_{17}^*+f_{10}f_{18}^*-f_{7}f_{13}^*-f_{8}f_{15}^*)+
\nn \\
&&
Ax_2(f_{27}f_{35}^*+f_{28}f_{36}^*-f_{25}f_{31}^*-f_{26}f_{33}^*)+ \nn \\
&&
Ax_3(f_{10}f_{36}^*+f_{28}f_{18}^*-f_{7}f_{31}^*-f_{25}f_{13}^*+
f_{27}f_{17}^*-f_{8}f_{33}^*+f_{9}f_{35}^*-f_{26}f_{15}^*)+ \nn \\
&&
B(f_{10}f_{35}^*+f_{27}f_{18}^*-f_{9}f_{36}^*-f_{28}f_{17}^*+
f_{8}f_{31}^*+f_{25}f_{15}^*-f_{7}f_{33}^*-f_{26}f_{13}^*),\nn \\
D_6&=&Ax_1(f_{9}f_{5}^*+f_{10}f_{6}^*-f_{7}f_{1}^*-f_{8}f_{3}^*)+\nn \\
&&
Ax_2(f_{27}f_{23}^*+f_{28}f_{24}^*-f_{25}f_{19}^*-f_{26}f_{21}^*)+ \nn \\
&&
Ax_3(f_{10}f_{24}^*+f_{28}f_{6}^*-f_{7}f_{19}^*-f_{25}f_{1}^*+
f_{27}f_{5}^*-f_{8}f_{21}^*+f_{9}f_{23}^*-f_{26}f_{3}^*)+ \nn \\
&&
B(f_{10}f_{23}^*+f_{27}f_{6}^*-f_{9}f_{24}^*-f_{28}f_{5}^*+
f_{8}f_{19}^*+f_{25}f_{3}^*-f_{7}f_{21}^*-f_{26}f_{1}^*), \nn \\
\beta_{10}&=&-\frac{\omega}{M}ImD_7,~
\beta_{12}=-\frac{\omega}{M}ReD_7, \nn \\
\beta_{11}&=&-\frac{\omega}{M}ImD_8,~
\beta_{13}=-\frac{\omega}{M}ReD_8, \nn \\
D_7&=&Ax_1(f_{7}f_{14}^*+f_{8}f_{16}^*-f_{11}f_{17}^*-f_{12}f_{18}^*)+
Ax_2(f_{25}f_{32}^*+f_{26}f_{34}^*-
\nn \\
&&
f_{29}f_{35}^*-f_{36}f_{30}^*)+Ax_3(f_{7}f_{32}^*+f_{25}f_{14}^*- \nn \\
&&
f_{12}f_{36}^*-f_{30}f_{18}^*+
f_{8}f_{34}^*-f_{29}f_{17}^*+f_{26}f_{16}^*-f_{11}f_{35}^*)+ \nn \\
&&
B(f_{30}f_{17}^*+f_{11}f_{36}^*-f_{29}f_{18}^*-f_{12}f_{35}^*+
f_{26}f_{14}^*+f_{7}f_{34}^*-f_{25}f_{16}^*-f_{8}f_{32}^*), \nn \\
D_8&=&Ax_1(f_{7}f_{2}^*+f_{8}f_{4}^*-f_{11}f_{5}^*-f_{12}f_{6}^*)+
\nn \\
&&
Ax_2(f_{25}f_{20}^*+f_{26}f_{22}^*-f_{29}f_{23}^*-f_{30}f_{24}^*)+ \nn \\
&&
Ax_3(f_{7}f_{20}^*+f_{25}f_{2}^*-f_{30}f_{6}^*-f_{12}f_{24}^*+
f_{8}f_{22}^*-f_{29}f_{5}^*+f_{26}f_{4}^*-f_{11}f_{23}^*)+ \nn \\
&&
B(f_{30}f_{5}^*+f_{11}f_{24}^*-f_{29}f_{6}^*-f_{12}f_{23}^*+
f_{26}f_{2}^*+f_{7}f_{22}^*-f_{25}f_{4}^*-f_{8}f_{20}^*). \nn \\
\nn 
\ea

$\bullet $ \underline{Tensor polarized deuteron target.}

The hadronic tensor $H_{ij}(S)$ is determined by the structure functions
$\gamma _i \ (i=1, .., 23)$
\ba
\gamma_1&=&2Ax_1\biggl [|f_{14}|^2+|f_{16}|^2-\frac{M^2}{\omega^2}
(|f_{17}|^2+|f_{18}|^2)\biggr ]+
2Ax_2\biggl [|f_{32}|^2+|f_{34}|^2-
\nn \\
&&
\frac{M^2}{\omega^2}(|f_{35}|^2+|f_{36}|^2)\biggr ]+\nn 
4Ax_3Re\biggl [f_{16}f_{34}^*+f_{14}f_{32}^*-
\frac{M^2}{\omega^2}
(f_{17}f_{35}^*+ \nn \\
&&
f_{18}f_{36}^*)\biggr ]-4BRe\biggl [f_{32}f_{16}^*-f_{34}f_{14}^*-\frac{M^2}{\omega^2}
(f_{17}f_{36}^*-f_{18}f_{35}^*)\biggr ], \nn \\
\gamma_2&=&2Ax_1\biggl [|f_{2}|^2+|f_{4}|^2-\frac{M^2}{\omega^2}
(|f_{5}|^2+|f_{6}|^2)\biggr ]+
2Ax_2\biggl [|f_{20}|^2+|f_{22}|^2- \nn \\
&&
\frac{M^2}{\omega^2}(|f_{23}|^2+|f_{24}|^2)\biggr ]+ 
4Ax_3Re\biggl [f_{4}f_{22}^*+f_{2}f_{20}^*-
\frac{M^2}{\omega^2}
(f_{5}f_{23}^*+ \nn \\
&&
f_{6}f_{24}^*)\biggr ]-4BRe\biggl [f_{20}f_{4}^*-f_{22}f_{2}^*-\frac{M^2}{\omega^2}
(f_{5}f_{24}^*-f_{6}f_{23}^*)\biggr ], \nn \\
\gamma_3&=&2Ax_1\biggl [|f_{11}|^2+|f_{12}|^2-\frac{M^2}{\omega^2}
(|f_{7}|^2+|f_{8}|^2)\biggr ]+
2Ax_2\biggl [|f_{29}|^2+|f_{30}|^2-\nn \\
&&
\frac{M^2}{\omega^2}(|f_{25}|^2+|f_{26}|^2)\biggr ]+
4Ax_3Re\biggl [f_{11}f_{29}^*+f_{12}f_{30}^*-\frac{M^2}{\omega^2}
(f_{8}f_{26}^*+ \nn \\
&&
+f_{7}f_{25}^*)\biggr ]+4BRe\biggl [f_{12}f_{29}^*-f_{11}f_{30}^*-\frac{M^2}{\omega^2}
(f_{26}f_{7}^*-f_{25}f_{8}^*)\biggr ], \nn \\
\gamma_4&=&2Ax_1Re\biggl [f_{2}f_{14}^*+f_{4}f_{16}^*-\frac{M^2}{\omega^2}
(f_{6}f_{18}^*+f_{5}f_{17}^*)\biggr ]+
2Ax_2Re\biggl [ f_{20}f_{32}^*+ \nn \\
&&
f_{22}f_{34}^*-\frac{M^2}{\omega^2}(f_{24}f_{36}^*+f_{23}f_{35}^*)\biggr ]+
2Ax_3Re\biggl [f_{4}f_{34}^*+f_{22}f_{16}^*+f_{2}f_{32}^*+ 
\nn \\
&&
f_{20}f_{14}^*-\frac{M^2}{\omega^2}(f_{5}f_{35}^*+f_{23}f_{17}^*+f_{6}f_{36}^*+
f_{24}f_{18}^*)\biggr ]+
2BRe\biggl [f_{34}f_{2}^*+
\nn \\
&&
f_{14}f_{22}^*-f_{32}f_{4}^*- f_{16}f_{20}^*-
\frac{M^2}{\omega^2}(f_{6}f_{35}^*+f_{23}f_{18}^*
-f_{5}f_{36}^*-f_{24}f_{17}^*)\biggr ], \nn \\
%%%%%%%%
\gamma_5&=&-2Ax_1Im\biggl [f_{2}f_{14}^*+f_{4}f_{16}^*-\frac{M^2}{\omega^2}
(f_{6}f_{18}^*+f_{5}f_{17}^*)\biggr ]-
\nn \\
&&
2Ax_2Im\biggl [f_{20}f_{32}^*+ f_{22}f_{34}^*- 
\frac{M^2}{\omega^2}(f_{24}f_{36}^*+f_{23}f_{35}^*)\biggr ]-
\nn \\
&&
2Ax_3Im\biggl [f_{4}f_{34}^*+f_{22}f_{16}^*+f_{2}f_{32}^*+
f_{20}f_{14}^*-
\nn \\
&&
\frac{M^2}{\omega^2}(f_{5}f_{35}^*+f_{23}f_{17}^*+f_{6}f_{36}^*+
f_{24}f_{18}^*)\biggr ]-
\nn \\
&&
2BIm\biggl [-f_{34}f_{2}^*-f_{14}f_{22}^*+
f_{32}f_{4}^*+ 
f_{16}f_{20}^*-
\nn \\
&&
\frac{M^2}{\omega^2}(f_{6}f_{35}^*+f_{23}f_{18}^*
-f_{5}f_{36}^*-f_{24}f_{17}^*)\biggr ],\nn \\
%%%%%%%%%%%%%%%%%%%%%
\gamma_6&=&2Ax_1\biggl [|f_{13}|^2+|f_{15}|^2-|f_{17}|^2-|f_{18}|^2\biggr ]+
2Ax_2\biggl [|f_{31}|^2+|f_{33}|^2-
\nn \\
&&
|f_{35}|^2-|f_{36}|^2\biggr ]+ 
4Ax_3Re\biggl [f_{15}f_{33}^*+f_{13}f_{31}^*-
f_{17}f_{35}^*-f_{18}f_{36}^*\biggr ]- \nn \\
&&
4BRe\biggl [f_{18}f_{35}^*-f_{17}f_{36}^*+
f_{31}f_{15}^*-f_{33}f_{13}^*\biggr ], \nn \\
%%%%%%%%%%%%%%%%%%%
\gamma_7&=&2Ax_1\biggl [|f_{1}|^2+|f_{3}|^2-|f_{5}|^2-|f_{6}|^2\biggr ]+
2Ax_2\biggl [|f_{19}|^2+|f_{21}|^2-|f_{23}|^2-
\nn \\
&&
|f_{24}|^2\biggr ]+ 
4Ax_3Re\biggl [f_{3}f_{21}^*+f_{1}f_{19}^*-
f_{5}f_{23}^*-f_{6}f_{24}^*\biggr ]- \nn \\
&&
4BRe\biggl [f_{6}f_{23}^*-f_{5}f_{24}^*+
f_{19}f_{3}^*-f_{21}f_{1}^*\biggr ], \nn \\
%%%%%%%%%%%%%%%%
\gamma_8&=&2Ax_1\biggl [|f_{9}|^2+|f_{10}|^2-|f_{7}|^2-|f_{8}|^2\biggr ]+
2Ax_2\biggl [|f_{27}|^2+|f_{28}|^2-|f_{25}|^2-
\nn \\
&&
|f_{26}|^2\biggr ]+ 
4Ax_3Re\biggl [f_{9}f_{27}^*+f_{10}f_{28}^*-
f_{8}f_{26}^*-f_{7}f_{25}^*\biggr ]+ \nn \\
&& 4BRe\biggl [f_{10}f_{27}^*-f_{9}f_{28}^*+
f_{25}f_{8}^*-f_{26}f_{7}^* \biggr ], \nn \\
%%%%%%%%%%%%%%%%%%%
\gamma_9&=&2Ax_1Re\biggl [f_{1}f_{13}^*+f_{3}f_{15}^*-
f_{6}f_{18}^*-f_{5}f_{17}^*\biggr ]+
2Ax_2Re\biggl [f_{19}f_{31}^*+f_{21}f_{33}^*- \nn \\
&&
f_{24}f_{36}^*-f_{23}f_{35}^*\biggr ]+
2Ax_3Re\biggl [
f_{3}f_{33}^*+f_{21}f_{15}^*+f_{1}f_{31}^*+f_{19}f_{13}^*- 
f_{5}f_{35}^*-
\nn \\
&&
f_{23}f_{17}^*-f_{6}f_{36}^*-f_{24}f_{18}^*\biggr ]-
2BRe\biggl [f_{6}f_{35}^*+f_{23}f_{18}^*-
f_{5}f_{36}^*- 
f_{24}f_{17}^*+
\nn \\
&&
f_{15}f_{19}^*+f_{31}f_{3}^*-f_{13}f_{21}^*-
f_{33}f_{1}^*\biggr ], \nn \\
%%%%%%%%%%%%%%%%%%%
\gamma_{10}&=&-2Ax_1Im\biggl [f_{1}f_{13}^*+f_{3}f_{15}^*-
f_{6}f_{18}^*-f_{5}f_{17}^*\biggr ]-
2Ax_2Im\biggl [f_{19}f_{31}^*+ \nn \\
&&
f_{21}f_{33}^*-f_{24}f_{36}^*-f_{23}f_{35}^*\biggr ]-2Ax_3Im\biggl [
f_{3}f_{33}^*+f_{21}f_{15}^*+f_{1}f_{31}^*+f_{19}f_{13}^*- \nn \\
&&
f_{5}f_{35}^*-f_{23}f_{17}^*-f_{6}f_{36}^*-f_{24}f_{18}^*\biggr ]+
2BRe\biggl [f_{6}f_{35}^*+f_{23}f_{18}^*-f_{5}f_{36}^*- \nn \\
&&
f_{24}f_{17}^*-f_{15}f_{19}^*-f_{31}f_{3}^*+f_{13}f_{21}^*+
f_{33}f_{1}^*\biggr ], \nn \\
%%%%%%%%%%%%%%%%%%%
\gamma_{11}&=&2Ax_1Re\biggl [f_{13}f_{14}^*+f_{15}f_{16}^*\biggr ]+
2Ax_2Re\biggl [f_{31}f_{32}^*+f_{33}f_{34}^*\biggr ]+
\nn \\
&&
2Ax_3Re\biggl [f_{15}f_{34}^*+  f_{16}f_{33}^*+f_{13}f_{32}^*+f_{14}f_{31}^*\biggr ]-\nn \\
&&
2BRe\biggl [f_{31}f_{16}^*+f_{32}f_{15}^*-f_{33}f_{14}^*-
f_{34}f_{13}^*\biggr ], \nn \\
%%%%%%%%%%%%%%%%%%%%%%%%%%%%%%%%%%%%%
\gamma_{12}&=&2Ax_1Re\biggl [f_{1}f_{2}^*+f_{3}f_{4}^*\biggr ]+
2Ax_2Re\biggl [f_{19}f_{20}^*+f_{21}f_{22}^*\biggr ]+
2Ax_3Re\biggl [f_{3}f_{22}^*+ \nn \\
&&
f_{4}f_{21}^*+f_{2}f_{19}^*+f_{1}f_{20}^*\biggr ]-
2BRe\biggl [f_{19}f_{4}^*+f_{20}f_{3}^*-f_{21}f_{2}^*-
f_{22}f_{1}^*\biggr ], \nn \\
%%%%%%%%%%%%%%%%%%%%%%%%%%%%%%%
\gamma_{13}&=&2Ax_1Re\biggl [f_{9}f_{11}^*+f_{10}f_{12}^*\biggr ]+
2Ax_2Re\biggl [f_{27}f_{29}^*+f_{28}f_{30}^*\biggr ]+\nn \\
&&
2Ax_3Re\biggl [f_{9}f_{29}^*+  
f_{10}f_{30}^*+f_{11}f_{27}^*+f_{12}f_{28}^*\biggr ]+\nn \\
&&
2BRe\biggl [f_{10}f_{29}^*+f_{12}f_{27}^*-f_{9}f_{30}^*-
f_{11}f_{28}^*\biggr ], \nn \\
\gamma_{14}&=&Ax_1Re\biggl [f_{1}f_{14}^*+f_{2}f_{13}^*+
f_{3}f_{16}^*+f_{4}f_{15}^*\biggr ]+
Ax_2Re\biggl [f_{19}f_{32}^*+f_{20}f_{31}^*+ \nn \\
&&
f_{21}f_{34}^*+f_{22}f_{33}^*\biggr ]+Ax_3Re\biggl [
f_{1}f_{32}^*+f_{2}f_{31}^*+f_{22}f_{15}^*+f_{3}f_{34}^*+ \nn \\
&&
f_{4}f_{33}^*+f_{19}f_{14}^*+f_{20}f_{13}^*+f_{21}f_{16}^*\biggr ]-
BRe\biggl [f_{15}f_{20}^*+f_{16}f_{19}^*+f_{31}f_{4}^*+\nn \\
&&
f_{32}f_{3}^*-f_{13}f_{22}^*-f_{14}f_{21}^*-f_{33}f_{2}^*-
f_{34}f_{1}^*\biggr ], \nn \\
\gamma_{15}&=&-Ax_1Im\biggl [f_{1}f_{14}^*+f_{2}f_{13}^*+
f_{3}f_{16}^*+f_{4}f_{15}^*\biggr ]-
Ax_2Im\biggl [f_{19}f_{32}^*+f_{20}f_{31}^*+ \nn \\
&&
f_{21}f_{34}^*+f_{22}f_{33}^*\biggr ]-Ax_3Im\biggl [
f_{1}f_{32}^*+f_{2}f_{31}^*+f_{22}f_{15}^*+f_{3}f_{34}^*+ \nn \\
&&
f_{4}f_{33}^*+f_{19}f_{14}^*+f_{20}f_{13}^*+f_{21}f_{16}^*\biggr ]-
BIm\biggl [f_{15}f_{20}^*+f_{16}f_{19}^*+f_{31}f_{4}^*+ \nn \\
&&
f_{32}f_{3}^*-f_{13}f_{22}^*-f_{14}f_{21}^*-f_{33}f_{2}^*-
f_{34}f_{1}^*\biggr ], \nn \\
\gamma_{16}&=&Ax_1Re\biggl [f_{7}f_{14}^*+f_{11}f_{17}^*+
f_{8}f_{16}^*+f_{12}f_{18}^*\biggr ]+
Ax_2Re\biggl [f_{25}f_{32}^*+f_{26}f_{34}^*+ \nn \\
&&
f_{29}f_{35}^*+f_{30}f_{36}^*\biggr ]+Ax_3Re\biggl [
f_{7}f_{32}^*+f_{25}f_{14}^*+f_{18}f_{30}^*+f_{36}f_{12}^*+ \nn \\
&&
f_{17}f_{29}^*+f_{35}f_{11}^*+f_{8}f_{34}^*+f_{26}f_{16}^*\biggr ]+
BRe\biggl [f_{12}f_{35}^*+f_{29}f_{18}^*-f_{11}f_{36}^*- \nn \\
&&
f_{30}f_{17}^*+f_{14}f_{26}^*+f_{34}f_{7}^*-f_{16}f_{25}^*-
f_{32}f_{8}^*\biggr ], \nn \\
\gamma_{17}&=&Ax_1Re\biggl [f_{2}f_{7}^*+f_{4}f_{8}^*+
f_{5}f_{11}^*+f_{6}f_{12}^*\biggr ]+
Ax_2Re\biggl [f_{20}f_{25}^*+f_{22}f_{26}^*+ \nn \\
&&
f_{23}f_{29}^*+f_{24}f_{30}^*\biggr ]+Ax_3Re\biggl [
f_{5}f_{29}^*+f_{23}f_{11}^*+f_{4}f_{26}^*+f_{22}f_{8}^*+ \nn \\
&&
f_{2}f_{25}^*+f_{20}f_{7}^*+f_{6}f_{30}^*+f_{24}f_{12}^*\biggr ]+
BRe\biggl [f_{6}f_{29}^*+f_{23}f_{12}^*-f_{5}f_{30}^*- \nn \\
&&
f_{24}f_{11}^*+f_{26}f_{2}^*+f_{7}f_{22}^*-f_{25}f_{4}^*-
f_{8}f_{20}^*\biggr ], \nn \\
\gamma_{18}&=&-Ax_1Im\biggl [f_{7}f_{14}^*+f_{11}f_{17}^*+
f_{8}f_{16}^*+f_{12}f_{18}^*\biggr ]-
Ax_2Im\biggl [f_{25}f_{32}^*+f_{26}f_{34}^*+ \nn \\
&&
f_{29}f_{35}^*+f_{30}f_{36}^*\biggr ]+Ax_3Im\biggl [
-f_{7}f_{32}^*-f_{25}f_{14}^*+f_{18}f_{30}^*+f_{36}f_{12}^*+ 
\nn \\
&&
f_{17}f_{29}^*+f_{35}f_{11}^*-f_{8}f_{34}^*-f_{26}f_{16}^*\biggr ]+
BIm\biggl [-f_{12}f_{35}^*-f_{29}f_{18}^*+ \nn \\
&&
f_{11}f_{36}^*+f_{30}f_{17}^*+f_{14}f_{26}^*+f_{34}f_{7}^*-f_{16}f_{25}^*-
f_{32}f_{8}^*\biggr ],\nn \\
\gamma_{19}&=&Ax_1Im\biggl [f_{2}f_{7}^*+f_{4}f_{8}^*+
f_{5}f_{11}^*+f_{6}f_{12}^*\biggr ]+
Ax_2Im\biggl [f_{20}f_{25}^*+\nn \\
&&
f_{22}f_{26}^*+f_{23}f_{29}^*+f_{24}f_{30}^*\biggr ]+Ax_3Im\biggl [
f_{5}f_{29}^*+f_{23}f_{11}^*+f_{4}f_{26}^*+ \nn \\
&&
f_{22}f_{8}^*+f_{2}f_{25}^*+f_{20}f_{7}^*+f_{6}f_{30}^*+f_{24}f_{12}^*\biggr ]+
BIm\biggl [f_{6}f_{29}^*+\nn \\
&&
f_{23}f_{12}^*-f_{5}f_{30}^*- f_{24}f_{11}^*-f_{26}f_{2}^*-f_{7}f_{22}^*+f_{25}f_{4}^*+
f_{8}f_{20}^*\biggr ], \nn \\
\gamma_{20}&=&Ax_1Re\biggl [f_{7}f_{13}^*+f_{10}f_{18}^*+
f_{8}f_{15}^*+f_{9}f_{17}^*\biggr ]+
Ax_2Re\biggl [f_{25}f_{31}^*+ \nn \\
&&
f_{26}f_{33}^*+f_{28}f_{36}^*+f_{27}f_{35}^*\biggr ]+Ax_3Re\biggl [
f_{7}f_{31}^*+f_{25}f_{13}^*+f_{18}f_{28}^*+ \nn \\
&&
f_{36}f_{10}^*+f_{17}f_{27}^*+f_{35}f_{9}^*+f_{8}f_{33}^*+f_{26}f_{15}^*\biggr ]+
BRe\biggl [f_{10}f_{35}^*+ \nn \\
&&
f_{27}f_{18}^*-f_{9}f_{36}^*-f_{28}f_{17}^*+f_{13}f_{26}^*+f_{33}f_{7}^*-f_{15}f_{25}^*-
f_{31}f_{8}^*\biggr ], \nn \\
\gamma_{21}&=&Ax_1Re\biggl [f_{1}f_{7}^*+f_{3}f_{8}^*+
f_{5}f_{9}^*+f_{6}f_{10}^*\biggr ]+
Ax_2Re\biggl [f_{19}f_{25}^*+f_{21}f_{26}^*+ \nn \\
&&
f_{23}f_{27}^*+f_{24}f_{28}^*\biggr ]+Ax_3Re\biggl [
f_{5}f_{27}^*+f_{23}f_{9}^*+f_{3}f_{26}^*+f_{21}f_{8}^*+ \nn \\
&&
f_{1}f_{25}^*+f_{19}f_{7}^*+f_{6}f_{28}^*+f_{24}f_{10}^*\biggr ]+
BRe\biggl [f_{6}f_{27}^*+f_{23}f_{10}^*-f_{5}f_{28}^*- \nn \\
&&
f_{24}f_{9}^*+f_{26}f_{1}^*+f_{7}f_{21}^*-f_{25}f_{3}^*-
f_{8}f_{19}^*\biggr ], \nn \\
\gamma_{22}&=&-Ax_1Im\biggl [f_{7}f_{13}^*+f_{10}f_{18}^*+
f_{8}f_{25}^*+f_{9}f_{17}^*\biggr ]-
Ax_2Im\biggl [f_{25}f_{31}^*+ \nn \\
&&
f_{26}f_{33}^*+f_{28}f_{36}^*+f_{27}f_{35}^*\biggr ]+Ax_3Im\biggl [
-f_{7}f_{31}^*-f_{25}f_{13}^*+f_{18}f_{28}^*+ \nn \\
&&
f_{36}f_{10}^*+f_{17}f_{27}^*+f_{35}f_{9}^*-f_{8}f_{33}^*-f_{26}f_{15}^*\biggr ]+
BIm\biggl [-f_{10}f_{35}^*-+ \nn \\
&&
f_{27}f_{18}^*+f_{9}f_{36}^*+f_{28}f_{17}^*+f_{13}f_{26}^*+f_{33}f_{7}^*-f_{15}f_{25}^*-
f_{31}f_{8}^*\biggr ],\nn \\
\gamma_{23}&=&Ax_1Im\biggl [f_{1}f_{7}^*+f_{3}f_{8}^*+
f_{5}f_{9}^*+f_{6}f_{10}^*\biggr ]+
Ax_2Im\biggl [f_{19}f_{25}^*+f_{21}f_{26}^*+ \nn \\
&&
f_{23}f_{27}^*+f_{24}f_{28}^*\biggr ]+Ax_3Im\biggl [
f_{5}f_{27}^*+f_{23}f_{9}^*+f_{3}f_{26}^*+f_{21}f_{8}^*+ \nn \\
&&
f_{1}f_{25}^*+f_{19}f_{7}^*+f_{6}f_{28}^*+f_{24}f_{10}^*\biggr ]+
BIm\biggl [f_{6}f_{27}^*+f_{23}f_{10}^*-f_{5}f_{28}^*- \nn \\
&&
f_{24}f_{9}^*-f_{26}f_{1}^*-f_{7}f_{21}^*+f_{25}f_{3}^*+
f_{8}f_{19}^*\biggr ]. \nn 
\ea
%%%%%%%%%%%%%%%%%%%%%%%%%%
\section{Appendix B}
%%%%%%%%%%%%%%%%%%%%%%%%%

Here we present the expressions for the structure functions $P_i, \ (i=1-13),$
which determine the tensor ${\vec P}_{ij}$. These structure functions, defining
the nucleon polarization vector ${\vec P}$, are written in terms of the scalar
amplitudes $f_i \ (i=1, ..., 36)$ determining the $\gamma^*+d\rightarrow \Delta +N$
reaction.
\ba
P_1&=&ImQ_1,~ P_2=ImQ_2,~ P_3=ReQ_1,~ P_4=ReQ_2,\nn\\
Q_{1}&=&\frac{2}{3}Ax_1\biggl [f_{10}f_{13}^*+f_{9}f_{15}^*-
f_{7}f_{18}^*-f_{8}f_{17}^*+zf_{11}f_{16}^*+zf_{12}f_{14}^*\biggr ]+
\nn\\
&&
\frac{2}{3}Ax_2\biggl [f_{27}f_{33}^*+ +f_{28}f_{31}^*-f_{25}f_{36}^*-f_{26}f_{35}^*+zf_{29}f_{34}^*+
zf_{30}f_{32}^*\biggr ]+\nn\\
&&
\frac{2}{3}Ax_3\biggl [
f_{9}f_{33}^*+f_{10}f_{31}^*+f_{27}f_{15}^*+  +f_{28}f_{13}^*-f_{7}f_{36}^*-f_{8}f_{35}^*-f_{25}f_{18}^*-\nn\\
&&
f_{26}f_{17}^*+zf_{11}f_{34}^*+zf_{12}f_{32}^*+zf_{29}f_{16}^*+zf_{30}f_{14}^*
\biggr ]+  
\nn\\
&&
\frac{2}{3}B\biggl [f_{8}f_{36}^*+f_{10}f_{33}^*+f_{25}f_{17}^*+
f_{27}f_{13}^*-f_{7}f_{35}^*-f_{9}f_{31}^*-f_{26}f_{18}^*-
\nn\\
&&
f_{28}f_{15}^*+ zf_{12}f_{34}^*+
+zf_{29}f_{14}^*-zf_{11}f_{32}^*-zf_{30}f_{16}^*\biggr ], \nn\\
%%%%%%%%%%%%%
Q_{2}&=&\frac{2}{3}Ax_1\biggl [f_{9}f_{3}^*+f_{10}f_{1}^*-
f_{7}f_{6}^*-f_{8}f_{5}^*+zf_{11}f_{4}^*+zf_{12}f_{2}^*\biggr ]+\nn\\
&&
\frac{2}{3}Ax_2\biggl [f_{27}f_{21}^*+ f_{28}f_{19}^*-f_{25}f_{24}^*-f_{26}f_{23}^*+zf_{29}f_{22}^*+
zf_{30}f_{20}^*\biggr ]+
\nn\\
&&
\frac{2}{3}Ax_3\biggl [
f_{9}f_{21}^*+f_{10}f_{19}^*+f_{27}f_{3}^*+  +f_{28}f_{1}^*-f_{7}f_{24}^*-f_{8}f_{23}^*-f_{25}f_{6}^*-\nn\\
&&
f_{26}f_{5}^*+zf_{11}f_{22}^*+zf_{12}f_{20}^*+zf_{29}f_{4}^*+zf_{30}f_{2}^*\biggr ]+  
\nn\\
&&
\frac{2}{3}B\biggl [f_{8}f_{24}^*+f_{10}f_{21}^*+f_{25}f_{5}^*+
f_{27}f_{1}^*-f_{7}f_{23}^*-f_{9}f_{19}^*-f_{26}f_{6}^*-\nn\\
&&
f_{28}f_{3}^*+zf_{12}f_{22}^*+ zf_{29}f_{2}^*-zf_{11}f_{20}^*-zf_{30}f_{4}^*\biggr ], \nn\\
P_5&=&ImQ_3,~ P_6=ImQ_4,~ P_7=ReQ_3,~ P_8=ReQ_4, \nn\\
%%%%%%%%%%%%
Q_{3}&=&\frac{2}{3}Ax_1\biggl [f_{9}f_{13}^*+f_{8}f_{18}^*-
f_{7}f_{17}^*-f_{10}f_{15}^*+zf_{11}f_{14}^*-zf_{12}f_{16}^*\biggr ]+\nn\\
&&
\frac{2}{3}Ax_2\biggl [f_{27}f_{31}^*+ f_{26}f_{36}^*-f_{25}f_{35}^*-f_{28}f_{33}^*+zf_{29}f_{32}^*+
zf_{30}f_{34}^*\biggr ]+
\nn\\
&&
\frac{2}{3}Ax_3\biggl [
f_{9}f_{31}^*+f_{8}f_{36}^*+f_{26}f_{18}^*+  f_{27}f_{13}^*-f_{7}f_{35}^*-f_{10}f_{33}^*-f_{25}f_{17}^*-
\nn\\
&&
f_{28}f_{15}^*+zf_{11}f_{32}^*-zf_{12}f_{34}^*+zf_{29}f_{14}^*-zf_{30}f_{16}^*\biggr ]+  
\nn\\
&&
\frac{2}{3}B\biggl [f_{7}f_{36}^*+f_{8}f_{35}^*+f_{9}f_{33}^*+
f_{10}f_{31}^*-f_{25}f_{18}^*-f_{27}f_{15}^*-f_{26}f_{17}^*-\nn\\
&&
f_{28}f_{13}^*+zf_{12}f_{32}^*- zf_{29}f_{16}^*+zf_{11}f_{34}^*-zf_{30}f_{14}^*\biggr ], \nn\\
%%%%%%%%%%%%
Q_{4}&=&\frac{2}{3}Ax_1\biggl [f_{9}f_{1}^*+f_{8}f_{6}^*-
f_{7}f_{5}^*-f_{10}f_{3}^*+zf_{11}f_{2}^*-zf_{12}f_{4}^*\biggr ]+
\nn\\
&&
\frac{2}{3}Ax_2\biggl [f_{26}f_{24}^*+ f_{27}f_{19}^*-f_{25}f_{23}^*-f_{28}f_{21}^*+zf_{29}f_{20}^*-
zf_{30}f_{22}^*\biggr ]+
\nn\\
&&
\frac{2}{3}Ax_3\biggl [
f_{9}f_{19}^*+f_{8}f_{24}^*+f_{26}f_{6}^*+  f_{27}f_{1}^*-f_{7}f_{23}^*-f_{10}f_{21}^*-f_{25}f_{5}^*-\nn\\
&&
f_{28}f_{3}^*+zf_{11}f_{20}^*-zf_{12}f_{22}^*+zf_{29}f_{2}^*-zf_{30}f_{4}^*\biggr ]+  \nn\\
&&\frac{2}{3}B\biggl [f_{7}f_{24}^*+f_{8}f_{23}^*+f_{9}f_{21}^*+
f_{10}f_{19}^*-f_{25}f_{6}^*-f_{26}f_{5}^*-f_{27}f_{3}^*-
\nn\\
&&
f_{28}f_{1}^*+
zf_{12}f_{20}^*- zf_{29}f_{4}^*+zf_{11}f_{22}^*-zf_{30}f_{2}^*\biggr ], 
%%%%%%%%%%%%%%%%%%%%%%%%
\nn\\
P_{9}&=&-\frac{4}{3}Im\biggl \{Ax_1\biggl [f_{13}f_{15}^*+f_{17}f_{18}^*
+zf_{14}f_{16}^*\biggr ]+
\nn\\
&&
Ax_2\biggl [f_{31}f_{33}^*+
f_{35}f_{36}^*+zf_{32}f_{34}^*\biggr ]+
\nn\\
&&Ax_3\biggl [f_{13}f_{33}^*+f_{17}f_{36}^*+f_{31}f_{15}^*+
f_{35}f_{18}^*+zf_{14}f_{34}^*+zf_{32}f_{16}^*\biggr ]+
\nn\\
&&
B\biggl [f_{18}f_{36}^*+f_{17}f_{35}^*+ +f_{31}f_{13}^*+f_{33}f_{15}^*+zf_{32}f_{14}^*+
zf_{34}f_{16}^*\biggr ]\biggr \}, \nn\\
%%%%%%%%%%%%%%%%%%%%%%%%%%%%%%%%%%%
P_{10}&=&-\frac{4}{3}Im\biggl \{Ax_1\biggl [f_{1}f_{3}^*+f_{5}f_{6}^*
+zf_{2}f_{4}^*\biggr ]+Ax_2\biggl [f_{19}f_{21}^*+
f_{25}f_{24}^*+zf_{20}f_{22}^*\biggr ]+ \nn\\
&&Ax_3\biggl [f_{1}f_{21}^*+f_{5}f_{24}^*+f_{19}f_{3}^*+
f_{23}f_{6}^*+zf_{2}f_{22}^*+zf_{20}f_{4}^*\biggr ]+
\nn\\
&&
B\biggl [f_{6}f_{24}^*+f_{5}f_{23}^*+ f_{21}f_{3}^*+f_{19}f_{1}^*+zf_{20}f_{2}^*+
zf_{2}f_{4}^*\biggr ]\biggr \}, \nn\\
%%%%%%%%%%%%%%%%%%%%%%%%%%%%%%%%
P_{11}&=&-\frac{4}{3}Im\biggl \{Ax_1\biggl [f_{7}f_{8}^*+f_{9}f_{10}^*
+zf_{11}f_{12}^*\biggr ]+
\nn\\
&&
Ax_2\biggl [f_{25}f_{26}^*+
f_{27}f_{28}^*+zf_{29}f_{30}^*\biggr ]+ 
\nn\\
&&Ax_3\biggl [f_{7}f_{26}^*+f_{9}f_{28}^*+f_{25}f_{8}^*+
f_{27}f_{10}^*+zf_{11}f_{30}^*+zf_{29}f_{12}^*\biggr ]+
\nn\\
&&
B\biggl [f_{9}f_{27}^*+f_{10}f_{28}^*+ f_{23}f_{7}^*+f_{26}f_{8}^*+zf_{12}f_{30}^*+
zf_{11}f_{29}^*\biggr ]\biggr \}, \nn\\
P_{12}&=&-ImQ_5,~ P_{13}=-ReQ_5, \nn\\
%%%%%%%%%%%%
Q_{5}&=&\frac{2}{3}\biggl \{Ax_1\biggl [f_{1}f_{15}^*+f_{5}f_{18}^*-
f_{3}f_{13}^*-f_{6}f_{17}^*+zf_{2}f_{16}^*-zf_{4}f_{14}^*\biggr ]+\nn\\
&&
Ax_2\biggl [f_{19}f_{33}^*+f_{23}f_{36}^*-f_{21}f_{31}^*-f_{24}f_{35}^*+zf_{20}f_{34}^*-
zf_{22}f_{32}^*\biggr ]+
\nn\\
&&
Ax_3\biggl [
f_{1}f_{33}^*+f_{5}f_{36}^*+f_{19}f_{15}^*+  f_{23}f_{18}^*-f_{3}f_{31}^*-f_{6}f_{35}^*-f_{21}f_{13}^*-\nn\\
&&
f_{24}f_{17}^*+zf_{2}f_{34}^*-zf_{4}f_{32}^*-zf_{22}f_{14}^*+zf_{20}f_{16}^*\biggr ]+  \nn\\
&&B\biggl [f_{5}f_{35}^*+f_{6}f_{36}^*+f_{19}f_{13}^*+
f_{21}f_{15}^*-f_{1}f_{31}^*-f_{3}f_{33}^*-f_{23}f_{17}^*-
\nn\\
&&
f_{24}f_{18}^*+zf_{20}f_{14}^*- zf_{4}f_{34}^*+zf_{22}f_{16}^*-zf_{2}f_{32}^*\biggr ]. \nn
\ea

%%%%%%%%%%%%%%%%%%%%%%
\section{Appendix C}
%%%%%%%%%%%%%%%%%%%%%%%%
The relations between the helicity amplitudes and the scalar
amplitudes are given here:
\ba
h_1&=&-\frac{1}{2\sqrt{6}}\biggl \{\cos\vartheta (f_3+f_8)-
\sin\vartheta (f_{21}+f_{26})+\cos^2\vartheta (f_9-f_5)+
\nn \\
&&
\sin^2\vartheta (f_{28}-f_{24})- 
\sin\vartheta \cos\vartheta (f_{10}-f_6+f_{27}-f_{23})-
\nn \\
&&
2\frac{E_1}{M_{\Delta}}\biggl [\cos\vartheta (f_{19}+f_{25})+
\sin\vartheta (f_1+f_7)+\cos^2\vartheta (f_{28}-f_{24})+
\nn\\
&&
\sin^2\vartheta (f_{9}-f_{5})+
\sin\vartheta \cos\vartheta (f_{10}-f_6+f_{27}-f_{23})\biggr ]\biggr \}, 
\nn \\
%%%%%%%%%%
h_2&=&-\frac{1}{2\sqrt{6}}\biggl \{-\cos\vartheta (f_3+f_8)+
\sin\vartheta (f_{21}+f_{26})+\cos^2\vartheta (f_9-f_5)+
\nn \\
&&
\sin^2\vartheta (f_{28}-f_{24})- 
\sin\vartheta \cos\vartheta (f_{10}-f_6+f_{27}-f_{23})+
\nn \\
&&
2\frac{E_1}{M_{\Delta}}\biggl [\cos\vartheta (f_{19}+f_{25})+
\sin\vartheta (f_1+f_7)- \cos^2\vartheta (f_{28}-f_{24})-
\nn \\
&&
\sin^2\vartheta (f_{9}-f_{5})-
\sin\vartheta \cos\vartheta (f_{10}-f_6+f_{27}-f_{23})\biggr ]\biggr \}, 
\nn \\
%%%%%%%%%%%%%
h_3&=&-\frac{1}{2\sqrt{3}}\frac{\omega}{M}
\biggl \{\cos\vartheta (f_4+\cos\vartheta f_{11})-
\sin\vartheta (f_{22}-\sin\vartheta f_{30})-
\nn \\
&&
\sin\vartheta \cos\vartheta (f_{12}+f_{29})- 
2\frac{E_1}{M_{\Delta}}\biggl [\cos\vartheta (f_{20}+\cos\vartheta f_{30})+
\nn \\
&&
\sin\vartheta (f_2+\sin\vartheta f_{11})+
\sin\vartheta \cos\vartheta (f_{12}+f_{29})\biggr ]\biggr \}, 
\nn \\
%%%%%%%%%
h_4&=&-\frac{1}{2\sqrt{3}}\frac{\omega}{M}
\biggl \{-\cos\vartheta (f_4-\cos\vartheta f_{11})+
\sin\vartheta (f_{22}+\sin\vartheta f_{30})-
\nn \\
&&
\sin\vartheta \cos\vartheta (f_{12}+f_{29})+ 
2\frac{E_1}{M_{\Delta}}\biggl [\cos\vartheta (f_{20}-\cos\vartheta f_{30})+
\nn \\
&&
\sin\vartheta (f_2-\sin\vartheta f_{11})-
\sin\vartheta \cos\vartheta (f_{12}+f_{29})\biggr ]\biggr \}, 
\nn \\
%%%%%%%%%
h_5&=&-\frac{1}{2\sqrt{6}}\biggl \{\cos\vartheta (f_8-f_3)+
\sin\vartheta (f_{21}-f_{26})-\cos^2\vartheta (f_9+f_5)-
 \nn \\
&&
\sin^2\vartheta (f_{28}+f_{24})+
\sin\vartheta \cos\vartheta (f_{10}+f_6+f_{27}+f_{23})+
 \nn \\
&&
2\frac{E_1}{M_{\Delta}}\biggl [\cos\vartheta (f_{19}-f_{25})+
\sin\vartheta (f_1-f_7)+ 
\cos^2\vartheta (f_{28}+f_{24})+
\nn \\
&&
\sin^2\vartheta (f_{9}+f_{5})+
\sin\vartheta \cos\vartheta (f_{10}+f_6+f_{27}+f_{23})\biggr ]\biggr \}, 
\nn \\
%%%%%%%%%%
h_6&=&-\frac{1}{2\sqrt{6}}\biggl \{\cos\vartheta (f_3-f_8)-
\sin\vartheta (f_{21}-f_{26})-\cos^2\vartheta (f_9+f_5)-
\nn \\
&&
\sin^2\vartheta (f_{28}+f_{24})+ 
\sin\vartheta \cos\vartheta (f_{10}+f_6+f_{27}+f_{23})+
\nn \\
&&
2\frac{E_1}{M_{\Delta}}\biggl [\cos\vartheta (f_{25}-f_{19})+
\sin\vartheta (f_7-f_1)+ 
\cos^2\vartheta (f_{28}+f_{24})+\nn \\
&&
\sin^2\vartheta (f_{9}+f_{5})+
\sin\vartheta \cos\vartheta (f_{10}+f_6+f_{27}+f_{23})\biggr ]\biggr \}, 
\nn \\
%%%%%%%%%%%%%%%%%%%
h_7&=&-\frac{1}{2\sqrt{6}}\biggl \{-\cos\vartheta (f_1+f_7)+
\sin\vartheta (f_{19}+f_{25})+\cos^2\vartheta (f_{10}-f_6)-\nn \\
&&
\sin^2\vartheta (f_{27}-f_{23})+ 
\sin\vartheta \cos\vartheta (f_{9}-f_5+f_{24}-f_{28
})+\nn \\
&&
2\frac{E_1}{M_{\Delta}}\biggl [-\cos\vartheta (f_{21}+f_{26})-
\sin\vartheta (f_3+f_8)+ 
\cos^2\vartheta (f_{27}-f_{23})-
\nn \\
&&
\sin^2\vartheta (f_{10}-f_{6})+
\sin\vartheta \cos\vartheta (f_{9}-f_5+f_{24}-f_{28})\biggr ]\biggr \}, 
\nn \\
%%%%%%%%%%%%%%%%%%%%%%%
h_8&=&-\frac{1}{2\sqrt{6}}\biggl \{-\cos\vartheta (f_1+f_7)+
\sin\vartheta (f_{19}+f_{25})-\cos^2\vartheta (f_{10}-f_6)+
\nn \\
&&
\sin^2\vartheta (f_{27}-f_{23})+ 
\sin\vartheta \cos\vartheta (f_{5}-f_9+f_{28}-f_{24})-
\nn \\
&&
2\frac{E_1}{M_{\Delta}}\biggl [\cos\vartheta (f_{21}+f_{26})+
\sin\vartheta (f_3+f_8)+ 
\cos^2\vartheta (f_{27}-f_{23})-
\nn \\
&&
\sin^2\vartheta (f_{10}-f_{6})-
\sin\vartheta \cos\vartheta (f_{5}-f_9+f_{28}-f_{24})\biggr ]\biggr \}, 
\nn \\
%%%%%%%%%%%%%%%%%%
h_9&=&-\frac{1}{2\sqrt{3}}\frac{\omega}{M}
\biggl \{-\cos\vartheta (f_2-\cos\vartheta f_{12})+
\sin\vartheta (f_{20}-\sin\vartheta f_{29})+
\nn \\
&&
\sin\vartheta \cos\vartheta (f_{11}-f_{30})+ 
 2\frac{E_1}{M_{\Delta}}\biggl [-\cos\vartheta (f_{22}-\cos\vartheta f_{29})-
 \nn \\
&&
\sin\vartheta (f_4+\sin\vartheta f_{12})+
\sin\vartheta \cos\vartheta (f_{11}-f_{30})\biggr ]\biggr \}, 
\nn \\
%%%%%%%%%%%%%%%%%%%%%%%
h_{10}&=&-\frac{1}{2\sqrt{3}}\frac{\omega}{M}
\biggl \{-\cos\vartheta (f_2+\cos\vartheta f_{12})+
\sin\vartheta (f_{20}+\sin\vartheta f_{29})+
\nn \\
&&
\sin\vartheta \cos\vartheta (f_{30}-f_{11})+ 
2\frac{E_1}{M_{\Delta}}\biggl [-\cos\vartheta (f_{22}+\cos\vartheta f_{29})-
\nn \\
&&
\sin\vartheta (f_4-\sin\vartheta f_{12})-
\sin\vartheta \cos\vartheta (f_{11}-f_{30})\biggr ]\biggr \}, 
\nn \\
%%%%%%%%%%%%%%%%%%%
h_{11}&=&-\frac{1}{2\sqrt{6}}\biggl \{\cos\vartheta (f_1-f_7)+
\sin\vartheta (f_{25}-f_{19})-\cos^2\vartheta (f_{10}+f_6)+
\nn \\
&&
\sin^2\vartheta (f_{27}+f_{23})- 
\sin\vartheta \cos\vartheta (f_{5}+f_9-f_{28}-f_{24})+
\nn \\
&&
2\frac{E_1}{M_{\Delta}}\biggl [\cos\vartheta (f_{21}-f_{26})+
\sin\vartheta (f_3-f_8)-
\cos^2\vartheta (f_{27}+f_{23})+
\nn \\
&&
\sin^2\vartheta (f_{10}+f_{6})-
\sin\vartheta \cos\vartheta (f_{5}+f_9-f_{28}-f_{24})\biggr ]\biggr \}, 
\nn \\
%%%%%%%%%%%%%%%%%%%%%%%%%%%%
h_{12}&=&-\frac{1}{2\sqrt{6}}\biggl \{\cos\vartheta (f_1-f_7)+
\sin\vartheta (f_{25}-f_{19})+\cos^2\vartheta (f_{10}+f_6)-
\nn \\
&&
\sin^2\vartheta (f_{27}+f_{23})
+\sin\vartheta \cos\vartheta (f_{5}+f_9-f_{28}-f_{24})+
\nn \\
&&
2\frac{E_1}{M_{\Delta}}\biggl [\cos\vartheta (f_{21}-f_{26})+
\sin\vartheta (f_3-f_8)+ \cos^2\vartheta (f_{27}+f_{23})-
\nn \\
&&
\sin^2\vartheta (f_{10}+f_{6})+
\sin\vartheta \cos\vartheta (f_{5}+f_9-f_{28}-f_{24})\biggr ]\biggr \}, 
\nn \\
%%%%%%%%%%%%%%%%%%%%%%%%%
h_{13}&=&-\frac{Q}{2\sqrt{3}}\biggl \{\cos\vartheta f_{15}-
\sin\vartheta f_{33}-\sin^2\vartheta f_{36}-\cos^2\vartheta f_{17}+
\nn \\
&&
\sin\vartheta \cos\vartheta (f_{18}+f_{35})+ 
2\frac{E_1}{M_{\Delta}}\biggl [-\cos\vartheta f_{31}+\cos^2\vartheta f_{36}-
\nn \\
&&
\sin\vartheta f_{13}+\sin^2\vartheta f_{17}+
\sin\vartheta \cos\vartheta (f_{18}+f_{35})\biggr ]\biggr \}, 
\nn \\
%%%%%%%%%%%%%%%%%%%
h_{14}&=&\frac{Q}{\sqrt{6}}\frac{\omega}{M}
\biggl [-\cos\vartheta f_{16}+\sin\vartheta f_{34}
+2\frac{E_1}{M_{\Delta}}\biggl (\cos\vartheta f_{32}+
\sin\vartheta f_{14}\biggr )\biggr ], 
\nn \\
%%%%%%%%%%%%%%%%%%%
h_{15}&=&\frac{Q}{2\sqrt{3}}
\biggl \{\cos\vartheta (f_{15}+\cos\vartheta f_{17})-
\sin\vartheta (f_{33}-\sin\vartheta f_{36})-
\nn \\
&&
\sin\vartheta \cos\vartheta (f_{18}+f_{35})- 
2\frac{E_1}{M_{\Delta}}\biggl [\cos\vartheta (f_{31}+\cos\vartheta f_{36})+
\nn \\
&&
\sin\vartheta (f_{13}+\sin\vartheta f_{17})+
\sin\vartheta \cos\vartheta (f_{18}+f_{35})\biggr ]\biggr \}, 
\nn \\
%%%%%%%%%%%%%%%%%%%
h_{16}&=&-\frac{Q}{2\sqrt{3}}
\biggl \{-\cos\vartheta (f_{13}-\cos\vartheta f_{18})+
\sin\vartheta (f_{31}-\sin\vartheta f_{35})+
\nn \\
&&
\sin\vartheta \cos\vartheta (f_{17}-f_{36})+ 
2\frac{E_1}{M_{\Delta}}\biggl [-\cos\vartheta (f_{33}-\cos\vartheta f_{35})-
\nn \\
&&
\sin\vartheta (f_{15}+\sin\vartheta f_{18})+
\sin\vartheta \cos\vartheta (f_{17}-f_{36})\biggr ]\biggr \}, 
\nn \\
%%%%%%%%%%%%%%%%%%%
h_{17}&=&-\frac{Q}{\sqrt{6}}\frac{\omega}{M}
\biggl [-\cos\vartheta f_{14}+\sin\vartheta f_{32}
-2\frac{E_1}{M_{\Delta}}\biggl (\cos\vartheta f_{34}+
\sin\vartheta f_{16}\biggr )\biggr ], 
\nn \\
%%%%%%%%%%%%%%%%%%%
h_{18}&=&\frac{Q}{2\sqrt{3}}
\biggl \{-\cos\vartheta (f_{13}+\cos\vartheta f_{18})+
\sin\vartheta (f_{31}+\sin\vartheta f_{35})+
\nn \\
&&
\sin\vartheta \cos\vartheta (f_{36}-f_{17})+ 
2\frac{E_1}{M_{\Delta}}\biggl [-\cos\vartheta (f_{33}+\cos\vartheta f_{35})-
\nn \\
&&
\sin\vartheta (f_{15}-\sin\vartheta f_{18})+
\sin\vartheta \cos\vartheta (f_{36}-f_{17})\biggr ]\biggr \}, 
\nn \\
%%%%%%%%%%%%%%%%%%%
h_{19}&=&-\frac{1}{2\sqrt{2}}
\biggl [\cos\vartheta (f_{1}+f_{7})-
\sin\vartheta (f_{19}+f_{25})+
\nn \\
&&
\sin\vartheta \cos\vartheta (f_{9}-f_{5}+f_{24}-f_{28})+ 
\cos^2\vartheta (f_{10}-f_{6})+
\nn \\
&&\sin^2\vartheta (f_{23}-f_{27})\biggr ], 
\nn \\
%%%%%%%%%%%%%%%%%%%
h_{20}&=&-\frac{1}{2\sqrt{2}}
\biggl [\cos\vartheta (f_{1}+f_{7})-
\sin\vartheta (f_{19}+f_{25})+
\nn \\
&&
\sin\vartheta \cos\vartheta (f_{5}-f_{9}+f_{28}-f_{24})- 
\cos^2\vartheta (f_{10}-f_{6})+
\nn \\
&&
\sin^2\vartheta (f_{27}-f_{23})\biggr ], \nn \\
%%%%%%%%%%%%%%%%%%%%%%%%%%
h_{21}&=&\frac{1}{2}\frac{\omega}{M}
\biggl [-\cos\vartheta f_{2}+\sin\vartheta f_{20}
-\cos^2\vartheta f_{12}+\sin^2\vartheta f_{29}+
\nn \\
&&
\sin\vartheta \cos\vartheta (f_{30}-f_{11})\biggr ], 
\nn \\
%%%%%%%%%%%%%%%%%%%
h_{22}&=&-\frac{1}{2}\frac{\omega}{M}
\biggl [\cos\vartheta f_{2}-\sin\vartheta f_{20}
-\cos^2\vartheta f_{12}+\sin^2\vartheta f_{29}+
\nn \\
&&
\sin\vartheta \cos\vartheta (f_{30}-f_{11})\biggr ], 
\nn \\
%%%%%%%%%%%%%%%%%%%
h_{23}&=&-\frac{1}{2\sqrt{2}}
\biggl [\cos\vartheta (f_{7}-f_{1})+
\sin\vartheta (f_{19}-f_{25})+
\sin\vartheta \cos\vartheta (f_{24}+f_{28}
\nn \\
&&
-f_{5}-f_{9})- 
\cos^2\vartheta (f_{10}+f_{6})+
\sin^2\vartheta (f_{27}+f_{23})\biggr ], \nn \\
%%%%%%%%%%%%%%%%%%%
h_{24}&=&-\frac{1}{2\sqrt{2}}
\biggl [\cos\vartheta (f_{7}-f_{1})+
\sin\vartheta (f_{19}-f_{25})-
\sin\vartheta \cos\vartheta (f_{24}+f_{28}
\nn \\
&&-f_{5}-f_{9})+ 
\cos^2\vartheta (f_{10}+f_{6})-
\sin^2\vartheta (f_{27}+f_{23})\biggr ], 
\nn \\
%%%%%%%%%%%%%%%%%%%%%%%
h_{25}&=&-\frac{1}{2\sqrt{2}}
\biggl [\cos\vartheta (f_{3}+f_{8})-
\sin\vartheta (f_{21}+f_{26})+
\sin\vartheta \cos\vartheta (f_{27}+f_{10}
\nn \\
&&
-f_{6}-f_{23})- 
\cos^2\vartheta (f_{9}-f_{5})-
\sin^2\vartheta (f_{28}-f_{24})\biggr ], 
\nn \\
%%%%%%%%%%%%%%%%%%%%%%%%%%%%%
h_{26}&=&-\frac{1}{2\sqrt{2}}
\biggl [-\cos\vartheta (f_{3}+f_{8})+
\sin\vartheta (f_{21}+f_{26})+
\sin\vartheta \cos\vartheta (f_{27}+f_{10}
\nn \\
&&
-f_{6}-f_{23})- 
\cos^2\vartheta (f_{9}-f_{5})-
\sin^2\vartheta (f_{28}-f_{24})\biggr ], 
\nn \\
%%%%%%%%%%%%%%%%%%%
h_{27}&=&-\frac{1}{2}\frac{\omega}{M}
\biggl [\cos\vartheta f_{4}-\sin\vartheta f_{22}
-\cos^2\vartheta f_{11}-\sin^2\vartheta f_{30}+
\nn \\
&&
\sin\vartheta \cos\vartheta (f_{12}+f_{29})\biggr ], 
\nn \\
%%%%%%%%%%%%%%%%%%%
h_{28}&=&-\frac{1}{2}\frac{\omega}{M}
\biggl [-\cos\vartheta f_{4}+\sin\vartheta f_{22}
-\cos^2\vartheta f_{11}-\sin^2\vartheta f_{30}+
\nn \\
&&
\sin\vartheta \cos\vartheta (f_{12}+f_{29})\biggr ], 
\nn \\
%%%%%%%%%%%%%%%%%%%
h_{29}&=&-\frac{1}{2\sqrt{2}}
\biggl [-\cos\vartheta (f_{3}-f_{8})+
\sin\vartheta (f_{21}-f_{26})-
\sin\vartheta \cos\vartheta (f_{27}+f_{10}
\nn \\
&&
+f_{6}+f_{23})+ 
\cos^2\vartheta (f_{9}+f_{5})+
\sin^2\vartheta (f_{28}+f_{24})\biggr ],
\nn \\
%%%%%%%%%%%%%%%%%%%
h_{30}&=&-\frac{1}{2\sqrt{2}}
\biggl [\cos\vartheta (f_{3}-f_{8})-
\sin\vartheta (f_{21}-f_{26})-
\nn \\
&&
\sin\vartheta \cos\vartheta (f_{27}+f_{10}+f_{6}+f_{23})+ 
\cos^2\vartheta (f_{9}+f_{5})+
\nn \\
&&
\sin^2\vartheta (f_{28}+f_{24})\biggr ],
\nn \\ 
%%%%%%%%%%%%%%%%%%%
h_{31}&=&\frac{Q}{2}\biggl [-\cos\vartheta f_{13}+\sin\vartheta f_{31}
+\cos^2\vartheta f_{18}-\sin^2\vartheta f_{35}+
\nn \\
&&
\sin\vartheta \cos\vartheta (f_{17}-f_{36})\biggr ], 
\nn \\
%%%%%%%%%%%%%%%%%%%
h_{32}&=&\frac{Q}{\sqrt{2}}\frac{\omega}{M}
\biggl (\sin\vartheta f_{32}-\cos\vartheta f_{14}\biggr ), 
\nn \\
%%%%%%%%%%%%%%%%%%%
h_{33}&=&\frac{Q}{2}\biggl [\cos\vartheta f_{13}-\sin\vartheta f_{31}
+\cos^2\vartheta f_{18}-\sin^2\vartheta f_{35}+\nn \\
&&
\sin\vartheta \cos\vartheta (f_{17}-f_{36})\biggr ], 
\nn \\
%%%%%%%%%%%%%%%%%%%
h_{34}&=&\frac{Q}{2}\biggl [\cos\vartheta f_{15}-\sin\vartheta f_{33}
-\cos^2\vartheta f_{17}-\sin^2\vartheta f_{36}+
\nn \\
&&
\sin\vartheta \cos\vartheta (f_{18}+f_{35})\biggr ], 
\nn \\
%%%%%%%%%%%%%%%%%%%
h_{35}&=&-\frac{Q}{\sqrt{2}}\frac{\omega}{M}
\biggl (\sin\vartheta f_{34}-\cos\vartheta f_{16}\biggr ), 
\nn \\
%%%%%%%%%%%%%%%%%%%
h_{36}&=&\frac{Q}{2}\biggl [-\cos\vartheta f_{15}+\sin\vartheta f_{33}
-\cos^2\vartheta f_{17}-\sin^2\vartheta f_{36}+
\nn \\
&&
\sin\vartheta \cos\vartheta (f_{18}+f_{35})\biggr ], \nn
\ea
where $Q=\sqrt{-k^2}/k_0,$ $E_1=(W^2+M_{\Delta}^2-m^2)/2W$, and $k_0$,
$\omega $, $E_1$ are the energies of the virtual photon, deuteron, $\Delta $--
isobar, respectively, in the $\gamma^*+d\to \Delta +N$ reaction CMS, $k^2$
is the square of the virtual photon four--momentum, and $\vartheta $ is the
angle between the virtual photon and $\Delta $--isobar momenta.

Let us present here for completeness, the inverse relations, i.e., the expressions for the
scalar amplitudes in terms of the helicity amplitudes:
\ba
%%%%%%%%%
f_1&=&-\frac{1}{\sqrt{2}}\biggl \{\cos\vartheta (h_{19}+h_{20}-h_{23}-
h_{24})+y\sin\vartheta \biggl [\sqrt{3}(h_{2}-h_{1}+h_{5}- \nn \\
&&
h_{6})+h_{25}-h_{26}-h_{29}+h_{30}\biggr ]\biggr \}, 
\nn \\
%%%%%%%%%
f_2&=&\frac{M}{\omega}\biggl [-\cos\vartheta (h_{21}+h_{22})+
y\sin\vartheta (\sqrt{3}h_{3}-\sqrt{3}h_{4}+h_{28}-h_{27})\biggr ], 
\nn \\
%%%%%%%%%
f_3&=&-\frac{1}{\sqrt{2}}\biggl \{\cos\vartheta (h_{25}-h_{26}-h_{29}+
h_{30})-y\sin\vartheta \biggl [\sqrt{3}(h_{7}+h_{8}-h_{11}- 
\nn \\
&&
h_{12})+h_{19}+h_{20}-h_{23}-h_{24}\biggr ]\biggr \}, 
\nn \\
%%%%%%%%%
f_4&=&\frac{M}{\omega}\biggl [\cos\vartheta (h_{28}-h_{27})+
y\sin\vartheta (\sqrt{3}h_{9}+\sqrt{3}h_{10}+h_{21}+h_{22})\biggr ], 
\nn \\
%%%%%%%%%
f_5&=&\frac{1}{\sqrt{2}}\cos\vartheta \biggl [
\sin\vartheta (h_{19}-h_{20}+h_{23}-h_{24})-\cos\vartheta (h_{25}+h_{26}
+h_{29}+ \nn \\
&&
h_{30})\biggr ]+\frac{y}{\sqrt{2}}\sin\vartheta \biggl \{\cos\vartheta
\biggl [\sqrt{3}(h_{7}+h_{11}-h_{8}-h_{12})+
h_{20}+h_{24}-h_{19}- \nn \\
&&
h_{23}\biggr ]-\sin\vartheta \biggl [\sqrt{3}(h_{1}+h_{5}+h_{2}+h_{6})+
h_{25}+h_{26}+h_{29}+h_{30}\biggr ]\biggr \}, 
\nn \\
%%%%%%%%%
f_6&=&\frac{1}{\sqrt{2}}\cos\vartheta \biggl [
\cos\vartheta (h_{19}-h_{20}+h_{23}-h_{24})+\sin\vartheta (h_{25}+h_{26}
+h_{29}+
\nn \\
&&
h_{30})\biggr ]- \frac{y}{\sqrt{2}}\sin\vartheta \biggl \{\cos\vartheta
\biggl [\sqrt{3}(h_{1}+h_{2}+h_{5}+h_{6})+
h_{25}+h_{26}+h_{29}+\nn \\
&&
h_{30}\biggr ]+ \sin\vartheta \biggl [\sqrt{3}(h_{7}+h_{11}-h_{8}-h_{12})+
h_{20}+h_{24}-h_{19}-h_{23}\biggr ]\biggr \}, 
\nn \\
%%%%%%%%%
f_7&=&-\frac{1}{\sqrt{2}}\biggl \{\cos\vartheta (h_{19}+h_{20}+h_{23}+
h_{24})-y\sin\vartheta \biggl [\sqrt{3}(h_{1}-h_{2}+h_{5}- 
\nn \\
&&
h_{6})-h_{25}+h_{26}-h_{29}+h_{30}\biggr ]\biggr \}, 
\nn \\
%%%%%%%%%
f_8&=&\frac{1}{\sqrt{2}}\biggl \{\cos\vartheta (h_{26}-h_{25}-h_{29}+
h_{30})+y\sin\vartheta \biggl [\sqrt{3}(h_{7}+h_{8}+h_{11}+
\nn \\
&&
h_{12})+ h_{19}+h_{20}+h_{23}+h_{24}\biggr ]\biggr \}, 
\nn \\
%%%%%%%%%
f_9&=&-\frac{1}{\sqrt{2}}\cos\vartheta \biggl [
\sin\vartheta (h_{19}-h_{20}-h_{23}+h_{24})-\cos\vartheta (h_{25}+h_{26}
-h_{29}- \nn \\
&&
h_{30})\biggr ]-\frac{y}{\sqrt{2}}\sin\vartheta \biggl \{\cos\vartheta
\biggl [\sqrt{3}(h_{7}-h_{8}-h_{11}+h_{12})+
h_{20}-h_{19}+ 
\nn \\
&&
h_{23}-h_{24}\biggr ]-\sin\vartheta \biggl [\sqrt{3}(h_{1}+h_{2}-h_{5}-h_{6})+
h_{25}+h_{26}-h_{29}-h_{30}\biggr ]\biggr \}, 
\nn \\
%%%%%%%%%
f_{10}&=&-\frac{1}{\sqrt{2}}\cos\vartheta \biggl [
\cos\vartheta (h_{19}-h_{20}-h_{23}+h_{24})+\sin\vartheta (h_{25}+h_{26}
-h_{29}-
\nn \\
&&
h_{30})\biggr ]+ \frac{y}{\sqrt{2}}\sin\vartheta \biggl \{\cos\vartheta
\biggl [\sqrt{3}(h_{1}+h_{2}-h_{5}-h_{6})+
h_{25}+h_{26}-h_{29}- 
\nn \\
&&
h_{30}\biggr ]+\sin\vartheta \biggl [\sqrt{3}(h_{7}-h_{11}-h_{8}+h_{12})+
h_{20}-h_{24}-h_{19}+h_{23}\biggr ]\biggr \}, 
\nn \\
%%%%%%%%%
f_{11}&=&\frac{M}{\omega}\biggl \{\cos\vartheta \biggl [
\cos\vartheta (h_{27}+h_{28})+\sin\vartheta (h_{22}-h_{21})\biggr ]+ 
y\sin\vartheta \biggl [\sin\vartheta (\sqrt{3}h_{3}+
\nn \\
&&
\sqrt{3}h_{4}+h_{27}+h_{28})-\cos\vartheta (\sqrt{3}h_{9}-\sqrt{3}h_{10}+
h_{22}-h_{21})\biggr ]\biggr \}, 
\nn \\
%%%%%%%%%
f_{12}&=&\frac{M}{\omega}\biggl \{\cos\vartheta \biggl [
\cos\vartheta (h_{22}-h_{21})-\sin\vartheta (h_{27}+h_{28})\biggr ]+ 
y\sin\vartheta \biggl [\cos\vartheta (\sqrt{3}h_{3}+
\nn \\
&&
\sqrt{3}h_{4}+h_{27}+h_{28})+\sin\vartheta (\sqrt{3}h_{9}-\sqrt{3}h_{10}+
h_{22}-h_{21})\biggr ]\biggr \}, 
\nn \\
%%%%%%%%%
f_{13}&=&\frac{1}{Q}\biggl [\cos\vartheta (h_{33}-h_{31})+
y\sin\vartheta (\sqrt{3}h_{13}-\sqrt{3}h_{15}-h_{36}+h_{34})\biggr ], 
\nn \\
%%%%%%%%%
f_{14}&=&\frac{\sqrt{2}}{Q}\frac{M}{\omega}
\biggl [-\cos\vartheta h_{32}+
y\sin\vartheta (\sqrt{3}h_{14}+h_{35})\biggr ], 
\nn \\
%%%%%%%%%
f_{15}&=&\frac{1}{Q}\biggl [\cos\vartheta (h_{34}-h_{36})+
y\sin\vartheta (\sqrt{3}h_{16}-\sqrt{3}h_{18}-h_{33}+h_{31})\biggr ], 
\nn \\
%%%%%%%%%
f_{16}&=&\frac{\sqrt{2}}{Q}\frac{M}{\omega}
\biggl [\cos\vartheta h_{35}+
y\sin\vartheta (\sqrt{3}h_{17}+h_{32})\biggr ], 
\nn \\
f_{17}&=&\frac{1}{Q}\biggl \{\cos\vartheta \biggl [
\sin\vartheta (h_{31}+h_{33})-\cos\vartheta (h_{34}+h_{36})\biggr ]- 
\nn \\
&&
y\sin\vartheta \biggl [\sin\vartheta (\sqrt{3}h_{13}+
\sqrt{3}h_{15}+h_{34}+h_{36})+
\nn \\
&&
\cos\vartheta (\sqrt{3}h_{16}+\sqrt{3}h_{18}+
h_{31}+h_{33})\biggr ]\biggr \}, 
\nn \\
%%%%%%%%%
f_{18}&=&\frac{1}{Q}\biggl \{\cos\vartheta \biggl [
\cos\vartheta (h_{31}+h_{33})+\sin\vartheta (h_{34}+h_{36})\biggr ]+ 
\nn \\
&&
y\sin\vartheta \biggl [-\cos\vartheta (\sqrt{3}h_{13}+
\sqrt{3}h_{15}+h_{34}+h_{36})+
\nn \\
&&
\sin\vartheta (\sqrt{3}h_{16}+\sqrt{3}h_{18}+
h_{31}+h_{33})\biggr ]\biggr \}, 
\nn \\
%%%%%%%%%
f_{19}&=&\frac{1}{\sqrt{2}}\biggl \{\sin\vartheta (h_{19}+h_{20}-h_{23}-
h_{24})-y\cos\vartheta \biggl [\sqrt{3}(h_{2}-h_{1}+h_{5}-h_{6})+ 
\nn \\
&&
h_{25}-h_{26}-h_{29}+h_{30}\biggr ]\biggr \}, 
\nn \\
%%%%%%%%%
f_{20}&=&\frac{M}{\omega}\biggl [\sin\vartheta (h_{21}+h_{22})+
y\cos\vartheta (\sqrt{3}h_{3}-\sqrt{3}h_{4}+h_{28}-h_{27})\biggr ], 
\nn \\
%%%%%%%%%
f_{21}&=&\frac{1}{\sqrt{2}}\biggl \{\sin\vartheta (h_{25}-h_{26}-h_{29}+
h_{30})+y\cos\vartheta \biggl [\sqrt{3}(h_{7}+h_{8}-h_{11}
\nn \\
&&
-h_{12})+ h_{19}+h_{20}-h_{23}-h_{24}\biggr ]\biggr \}, 
\nn \\
%%%%%%%%%
f_{22}&=&\frac{M}{\omega}\biggl [\sin\vartheta (h_{27}-h_{28})+
y\cos\vartheta (\sqrt{3}h_{9}+\sqrt{3}h_{10}+h_{21}+h_{22})\biggr ], 
\nn \\
%%%%%%%%%
f_{23}&=&-\frac{1}{\sqrt{2}}\sin\vartheta \biggl [
\sin\vartheta (h_{19}-h_{20}+h_{23}-h_{24})-\cos\vartheta (h_{25}+h_{26}
+h_{29}+
\nn \\
&&
h_{30})\biggr ]- \frac{y}{\sqrt{2}}\cos\vartheta \biggl \{\sin\vartheta
\biggl [\sqrt{3}(h_{1}+h_{2}+h_{5}+h_{6})+
h_{25}+h_{26}+h_{29}+
\nn \\
&&
h_{30}\biggr ]- \cos\vartheta \biggl [\sqrt{3}(h_{7}+h_{11}-h_{8}-h_{12})+
h_{20}+h_{24}-h_{19}-h_{23}\biggr ]\biggr \}, 
\nn \\
%%%%%%%%%
f_{24}&=&-\frac{1}{\sqrt{2}}\sin\vartheta \biggl [
\cos\vartheta (h_{19}-h_{20}+h_{23}-h_{24})+\sin\vartheta (h_{25}+h_{26}
+h_{29}+ 
\nn \\
&&
h_{30})\biggr ]-\frac{y}{\sqrt{2}}\cos\vartheta \biggl \{\cos\vartheta
\biggl [\sqrt{3}(h_{1}+h_{2}+h_{5}+h_{6})+
h_{25}+h_{26}+h_{29}+
\nn \\
&&
h_{30}\biggr ]+ \sin\vartheta \biggl [\sqrt{3}(h_{7}+h_{11}-h_{8}-h_{12})+
h_{20}+h_{24}-h_{19}-h_{23}\biggr ]\biggr \}, 
\nn \\
%%%%%%%%%
f_{25}&=&\frac{1}{\sqrt{2}}\biggl \{\sin\vartheta (h_{19}+h_{20}+h_{23}+
h_{24})+y\cos\vartheta \biggl [\sqrt{3}(h_{1}+h_{5}-h_{2}- 
\nn \\
&&
h_{6})+h_{26}+h_{30}-h_{25}-h_{29}\biggr ]\biggr \}, 
\nn \\
%%%%%%%%%
f_{26}&=&-\frac{1}{\sqrt{2}}\biggl \{\sin\vartheta (h_{26}+h_{30}-h_{25}-
h_{29})-y\cos\vartheta \biggl [\sqrt{3}(h_{2}+h_{6}+h_{11}+
\nn \\
&&
h_{7})+ h_{20}+h_{24}+h_{25}+h_{29}\biggr ]\biggr \}, 
\nn \\
%%%%%%%%%
f_{27}&=&\frac{1}{\sqrt{2}}\sin\vartheta \biggl [
\sin\vartheta (h_{19}-h_{20}-h_{23}+h_{24})-\cos\vartheta (h_{25}+h_{26}
-h_{29}-
\nn \\
&&
h_{30})\biggr ]+ \frac{y}{\sqrt{2}}\cos\vartheta \biggl \{\sin\vartheta
\biggl [\sqrt{3}(h_{1}+h_{2}-h_{5}-h_{6})+
h_{25}+h_{26}-h_{29}- \nn \\
&&
h_{30}\biggr ]-\cos\vartheta \biggl [\sqrt{3}(h_{7}-h_{8}-h_{11}+h_{12})+
h_{20}-h_{19}-h_{24}+h_{23}\biggr ]\biggr \}, 
\nn \\
%%%%%%%%%
f_{28}&=&\frac{1}{\sqrt{2}}\sin\vartheta \biggl [
\cos\vartheta (h_{19}-h_{20}-h_{23}+h_{24})+\sin\vartheta (h_{25}+h_{26}
-h_{29}- 
\nn \\
&&
h_{30})\biggr ]+\frac{y}{\sqrt{2}}\cos\vartheta \biggl \{\cos\vartheta
\biggl [\sqrt{3}(h_{1}+h_{2}-h_{5}-h_{6})+
h_{25}+h_{26}-h_{29}-
\nn \\
&&
h_{30}\biggr ]+ \sin\vartheta \biggl [\sqrt{3}(h_{7}-h_{8}-h_{11}+h_{12})+
h_{20}-h_{19}-h_{24}+h_{23}\biggr ]\biggr \}, 
\nn \\
%%%%%%%%%
f_{29}&=&\frac{M}{\omega}\biggl \{\sin\vartheta \biggl [
\sin\vartheta (h_{21}-h_{22})-\cos\vartheta (h_{27}+h_{28})\biggr ]+ 
\nn \\
&&
y\cos\vartheta \biggl [\sin\vartheta (\sqrt{3}h_{3}+
\sqrt{3}h_{4}+h_{27}+h_{28})-\nn \\
&&
\cos\vartheta (\sqrt{3}h_{9}-\sqrt{3}h_{10}+
h_{22}-h_{21})\biggr ]\biggr \}, \nn \\
%%%%%%%%%
f_{30}&=&\frac{M}{\omega}\biggl \{\sin\vartheta \biggl [
\cos\vartheta (h_{21}-h_{22})+\sin\vartheta (h_{27}+h_{28})\biggr ]+ 
\nn \\
&&
y\cos\vartheta \biggl [\cos\vartheta (\sqrt{3}h_{3}+\sqrt{3}h_{4}+
h_{27}+h_{28})+
\nn \\
&&\sin\vartheta (\sqrt{3}h_{9}-\sqrt{3}h_{10}+
h_{22}-h_{21})\biggr ]\biggr \}, 
\nn \\
%%%%%%%%%
f_{31}&=&\frac{1}{Q}\biggl [\sin\vartheta (h_{31}-h_{33})+
y\cos\vartheta (\sqrt{3}h_{13}-\sqrt{3}h_{15}-h_{36}+h_{34})\biggr ], 
\nn \\
%%%%%%%%%
f_{32}&=&\frac{\sqrt{2}}{Q}\frac{M}{\omega}
\biggl [\sin\vartheta h_{32}+
y\cos\vartheta (\sqrt{3}h_{14}+h_{35})\biggr ], 
\nn \\
%%%%%%%%%
f_{33}&=&\frac{1}{Q}\biggl [\sin\vartheta (h_{36}-h_{34})+
y\cos\vartheta (\sqrt{3}h_{16}-\sqrt{3}h_{18}-h_{33}+h_{31})\biggr ], 
\nn \\
%%%%%%%%%
f_{34}&=&\frac{\sqrt{2}}{Q}\frac{M}{\omega}
\biggl [-\sin\vartheta h_{35}+
y\cos\vartheta (\sqrt{3}h_{17}+h_{32})\biggr ], 
\nn \\
%%%%%%%%%
f_{35}&=&\frac{1}{Q}\biggl \{\sin\vartheta \biggl [
\cos\vartheta (h_{34}+h_{36})-\sin\vartheta (h_{31}+h_{33})\biggr ]- 
\nn \\
&&
y\cos\vartheta \biggl [\sin\vartheta (\sqrt{3}h_{13}+\sqrt{3}h_{15}+
h_{34}+h_{36})+
\nn \\
&&
\cos\vartheta (\sqrt{3}h_{16}+\sqrt{3}h_{18}+
h_{31}+h_{33})\biggr ]\biggr \}, 
\nn \\
%%%%%%%%%
f_{36}&=&\frac{1}{Q}\biggl \{-\sin\vartheta \biggl [
\sin\vartheta (h_{34}+h_{36})+\cos\vartheta (h_{31}+h_{33})\biggr ]- 
\nn \\
&&
y\cos\vartheta \biggl [\cos\vartheta (\sqrt{3}h_{13}+\sqrt{3}h_{15}+
h_{34}+h_{36})-
\nn \\
&&
\sin\vartheta (\sqrt{3}h_{16}+\sqrt{3}h_{18}+
h_{31}+h_{33})\biggr ]\biggr \}, 
\nn 
\ea
where $y=M_{\Delta}W/(W^2+M^2_{\Delta}-m^2)$.

\section{Appendix D}

In this Appendix, we present the formulas for the structure functions which
determine the hadronic tensor $H_{ij}$ for various polarization states of the
deuteron target for the $\gamma +d\to \Delta +N$ reaction. The 
structure functions are expressed in terms of the scalar amplitudes
$g_i \ (i=1, ..., 24)$ determining the $\gamma +d\rightarrow \Delta +N$
reaction.

$\bullet $ \underline{Unpolarized deuteron target}.

The hadronic tensor $H_{ij}(0)$ is determined by two real structure 
functions $a_1$ and $a_2$:
\ba
a_1&=&\frac{2}{3}\biggl \{r_1\biggl[|g_{1}|^2+|g_{3}|^2+
|g_{5}|^2+|g_{6}|^2+r|g_{2}|^2+r|g_{4}|^2\biggr]+ \nn \\
&&
r_2\biggl[|g_{13}|^2+|g_{15}|^2+
|g_{17}|^2+|g_{18}|^2+r|g_{14}|^2+r|g_{16}|^2\biggr]+ \nn \\
&&
2r_3Re(g_{1}g_{13}^*+g_{3}g_{15}^*+g_{5}g_{17}^*+g_{6}g_{18}^*+
rg_{2}g_{14}^*+rg_{4}g_{16}^*)+ \nn 
\\
&&
2r_4Re(g_{6}g_{17}^*-g_{5}g_{18}^*+g_{15}g_{1}^*-g_{13}g_{3}^*+
rg_{16}g_{2}^*-rg_{14}g_{4}^*)\biggr \}, \nn \\
%%%%%%%%%%%%%%%%%
a_2&=&\frac{2}{3}\biggl \{r_1\biggl[|g_{7}|^2+|g_{8}|^2+
|g_{9}|^2+|g_{10}|^2+r|g_{11}|^2+r|g_{12}|^2\biggr]+ \nn \\
&&
r_2\biggl[|g_{19}|^2+|g_{20}|^2+
|g_{21}|^2+|g_{22}|^2+r|g_{23}|^2+r|g_{24}|^2\biggr]+ \nn \\
&&
2r_3Re(g_{7}g_{19}^*+g_{10}g_{22}^*+g_{9}g_{21}^*+g_{8}g_{20}^*+
rg_{11}g_{23}^*+rg_{12}g_{24}^*)+ \nn \\
&&
2r_4Re(g_{10}g_{21}^*-g_{9}g_{22}^*+g_{20}g_{7}^*-g_{19}g_{8}^*+
rg_{12}g_{23}^*-rg_{11}g_{24}^*)\biggr \}, 
\nn
\ea
where we introduce the notations
\ba
r_1&=&2\frac{1-(1-\gamma^2)\sin^2\vartheta}{2+\gamma^2},~ 
r_2=2\frac{1-(1-\gamma^2)\cos^2\vartheta}{2+\gamma^2},~ 
r_3=\frac{\gamma^2-1}{2+\gamma^2}\sin2\vartheta ,~ 
\nn \\
r_4&=&\frac{\gamma}{2+\gamma^2}, 
r=\frac{(W^2+M^2)^2}{4M^2W^2},~ 
\gamma =\frac{W^2+M^2_{\Delta}-m^2}{2M_{\Delta}W},
\ea
where $M_{\Delta}$, $M$ and $m$ are the masses of the $\Delta $--isobar,
deuteron and nucleon, respectively; $W$ is the total energy of the $\Delta N$ 
pair in CMS of the $\gamma +d\to \Delta +N$ reaction, $\vartheta $ is the 
angle between $\Delta $--isobar and photon momenta.

$\bullet $ \underline{Vector polarized deuteron target}.

The hadronic tensor $H_{ij}(\xi )$ is determined by six structure functions
$b_i \ (i=1-6)$
\ba
b_1&=&-2\sqrt{r}Im\biggl [r_1(g_{2}g_{1}^*+g_{4}g_{3}^*)+r_2(g_{14}g_{13}^*+
g_{16}g_{15}^*)+r_3(g_{2}g_{13}^*-g_{1}g_{14}^*+ 
\nn \\
&&
g_{4}g_{15}^*-g_{3}g_{16}^*)-r_4(g_{1}g_{16}^*+
g_{4}g_{13}^*-g_{2}g_{15}^*-g_{3}g_{14}^*)\biggr ], 
\nn \\
%%%%%%%%%%%%%%
b_2&=&-2\sqrt{r}Im\biggl [r_1(g_{11}g_{9}^*+g_{12}g_{10}^*)+r_2(g_{23}g_{21}^*+
g_{24}g_{22}^*)+r_3(g_{12}g_{22}^*+
\nn \\
&&
g_{24}g_{10}^*- g_{11}g_{21}^*-g_{9}g_{23}^*)+r_4(g_{12}g_{21}^*+
g_{9}g_{24}^*-g_{10}g_{23}^*-g_{11}g_{22}^*)\biggr ], 
\nn \\
%%%%%%%%%%%%%%
b_3&=&-ImE_1,~ b_4=-ReE_1, \nn \\
E_1&=&r_1(g_{9}g_{5}^*+g_{10}g_{6}^*-g_{7}g_{1}^*-g_{8}g_{3}^*)+
r_2(g_{21}g_{17}^*+g_{22}g_{18}^*-g_{19}g_{13}^*-
\nn \\
&&
g_{20}g_{15}^*)+ 
r_3(g_{10}g_{18}^*+g_{22}g_{6}^*-g_{7}g_{13}^*-g_{19}g_{1}^*+
g_{21}g_{5}^*-g_{8}g_{15}^*+
\nn \\
&&
g_{9}g_{17}^*-g_{20}g_{3}^*)+ 
r_4(g_{10}g_{17}^*+g_{21}g_{6}^*-g_{9}g_{18}^*-g_{22}g_{5}^*+
g_{8}g_{13}^*+g_{19}g_{3}^*-
\nn \\
&&
g_{7}g_{15}^*-g_{20}g_{1}^*), \nn \\
b_5&=&-\sqrt{r}ImE_2,~ b_6=-\sqrt{r}ReE_2, 
\nn \\
E_2&=&r_1(g_{7}g_{2}^*+g_{8}g_{4}^*-g_{11}g_{5}^*-g_{12}g_{6}^*)+
r_2(g_{19}g_{14}^*+g_{20}g_{16}^*-g_{23}g_{17}^*-
\nn \\
&&
g_{24}g_{18}^*)+ r_3(g_{7}g_{14}^*+g_{19}g_{2}^*-g_{24}g_{6}^*-g_{12}g_{18}^*+
g_{8}g_{16}^*-g_{23}g_{5}^*+ 
\nn \\
&&
g_{20}g_{4}^*-g_{11}g_{17}^*)+r_4(g_{24}g_{5}^*+g_{11}g_{18}^*-g_{23}g_{6}^*-g_{12}g_{17}^*+g_{20}g_{2}^*+
\nn \\
&&
g_{7}g_{16}^*-g_{19}g_{4}^*-g_{8}g_{14}^*). 
\nn
\ea

$\bullet $ \underline{Tensor polarized deuteron target.}

The hadronic tensor $H_{ij}(S)$ is determined, in this case,  by ten structure 
functions $c_i \ (i=1-10)$ which have the following expressions in terms of 
the $\gamma +d\rightarrow \Delta +N$ reaction amplitudes
\ba
c_1&=&2r_1\biggl [|g_{2}|^2+|g_{4}|^2-\frac{1}{r}
(|g_{5}|^2+|g_{6}|^2)\biggr ]+2r_2\biggl [|g_{14}|^2+|g_{16}|^2-
\nn \\
&&
\frac{1}{r}(|g_{17}|^2+|g_{18}|^2)\biggr ]+ 
4r_3Re\biggl [g_{4}g_{16}^*+g_{2}g_{14}^*-\frac{1}{r}
(g_{5}g_{17}^*+g_{6}g_{18}^*)\biggr ]- \nn \\
&&
4r_4Re\biggl [g_{14}g_{4}^*-g_{16}g_{2}^*-\frac{1}{r}
(g_{5}g_{18}^*-g_{6}g_{17}^*)\biggr ], 
\nn \\
c_2&=&2r_1\biggl [|g_{11}|^2+|g_{12}|^2-\frac{1}{r}
(|g_{7}|^2+|g_{8}|^2)\biggr ]+2r_2\biggl [|g_{23}|^2+|g_{24}|^2-
\nn \\
&&
\frac{1}{r}(|g_{19}|^2+|g_{20}|^2)\biggr ]+ 
4r_3Re\biggl [g_{11}g_{23}^*+g_{12}g_{24}^*-\frac{1}{r}
(g_{8}g_{20}^*+g_{7}g_{19}^*)\biggr ]+ 
\nn \\
&&
4r_4Re\biggl [g_{12}g_{23}^*-g_{11}g_{24}^*-\frac{1}{r}
(g_{20}g_{7}^*-g_{19}g_{8}^*)\biggr ], 
\nn \\
c_3&=&2r_1\biggl [|g_{1}|^2+|g_{3}|^2-|g_{5}|^2-|g_{6}|^2\biggr ]+
2r_2\biggl [|g_{13}|^2+|g_{15}|^2-|g_{17}|^2-
\nn \\
&&
|g_{18}|^2\biggr ]+ 4r_3Re\biggl [g_{3}g_{15}^*+g_{1}g_{13}^*-
g_{5}g_{17}^*-g_{6}g_{18}^*\biggr ]- 
\nn \\
&&
4r_4Re\biggl [g_{6}g_{17}^*-g_{5}g_{18}^*+
g_{13}g_{3}^*-g_{15}g_{1}^*\biggr ], \nn \\c_4&=&2r_1\biggl [|g_{9}|^2+|g_{10}|^2-|g_{7}|^2-|g_{8}|^2\biggr ]+
2r_2\biggl [|g_{21}|^2+|g_{22}|^2-|g_{19}|^2-
\nn \\
&&
|g_{20}|^2\biggr ]+ 4r_3Re\biggl [g_{9}g_{21}^*+g_{10}g_{22}^*-
g_{8}g_{20}^*-g_{7}g_{19}^*\biggr ]+ 
\nn \\
&&
4r_4Re\biggl [g_{10}g_{21}^*-g_{9}g_{22}^*+
g_{19}g_{8}^*-g_{20}g_{7}^*)\biggr ], 
\nn \\
c_{5}&=&2r_1Re\biggl [g_{1}g_{2}^*+g_{3}g_{4}^*\biggr ]+
2r_2Re\biggl [g_{13}g_{14}^*+g_{15}g_{16}^*\biggr ]+
2r_3Re\biggl [g_{3}g_{16}^*+  
\nn \\
&&g_{4}g_{15}^*+g_{2}g_{13}^*+g_{1}g_{14}^*\biggr ]-
2r_4Re\biggl [g_{13}g_{4}^*+g_{14}g_{3}^*-g_{15}g_{2}^*-
g_{16}g_{1}^*\biggr ], 
\nn \\
%%%%%%%%%%%%%%%%%%%%%
c_{6}&=&2r_1Re\biggl [g_{9}g_{11}^*+g_{10}g_{12}^*\biggr ]+
2r_2Re\biggl [g_{21}g_{23}^*+g_{22}g_{24}^*\biggr ]+
2r_3Re\biggl [g_{9}g_{23}^*+  
\nn \\
&&
g_{10}g_{24}^*+g_{11}g_{21}^*+g_{12}g_{22}^*\biggr ]+
2r_4Re\biggl [g_{10}g_{23}^*+g_{12}g_{21}^*-g_{9}g_{24}^*-
g_{11}g_{22}^*\biggr ], 
\nn \\
c_7&=&ReE_3,~ c_8=ImE_3, 
\nn \\
E_{3}&=&r_1(g_{2}g_{7}^*+g_{4}g_{8}^*+
g_{5}g_{11}^*+g_{6}g_{12}^*)+
r_2(g_{14}g_{19}^*+g_{16}g_{20}^*+ 
\nn \\
&&g_{17}g_{23}^*+g_{18}g_{24}^*)+r_3(
g_{5}g_{23}^*+g_{17}g_{11}^*+g_{4}g_{20}^*+g_{16}g_{8}^*+ 
\nn \\
&&
g_{2}g_{19}^*+g_{14}g_{7}^*+g_{6}g_{24}^*+g_{18}g_{12}^*)+
r_4(g_{6}g_{23}^*+g_{17}g_{12}^*-g_{5}g_{24}^*- 
\nn \\
&&
g_{18}g_{11}^*+g_{2}g_{20}^*+g_{16}g_{7}^*-g_{4}g_{19}^*-
g_{14}g_{8}^*\biggr ], 
\nn \\
c_9&=&ReE_4,~ c_{10}=ImE_4, 
\nn \\
E_{4}&=&r_1(g_{1}g_{7}^*+g_{3}g_{8}^*+
g_{5}g_{9}^*+g_{6}g_{10}^*)+
r_2(g_{13}g_{19}^*+g_{15}g_{20}^*+ 
\nn \\
&&g_{17}g_{21}^*+g_{18}g_{22}^*)+r_3(
g_{5}g_{21}^*+g_{17}g_{9}^*+g_{3}g_{20}^*+g_{15}g_{8}^*+ 
\nn \\
&&
g_{1}g_{19}^*+g_{13}g_{7}^*+g_{6}g_{22}^*+g_{18}g_{10}^*)+
r_4(g_{6}g_{21}^*+g_{17}g_{10}^*-g_{5}g_{22}^*- 
\nn \\
&&
g_{18}g_{9}^*+g_{1}g_{20}^*+g_{15}g_{7}^*-g_{3}g_{19}^*-
g_{13}g_{8}^*). 
\nn
\ea
%%%%%%%%%%%%%%%%%%%%%
\section{Appendix E}
%%%%%%%%%%%%%%%%%%%%%%%
In this Appendix, we present the formulas for the structure functions which
determine the hadronic tensor ${\vec P}_{ij}$ describing the nucleon
polarization in the $\gamma +d\to \Delta +N$ reaction. The structure functions 
$d_i$, $(i=1-6)$, are written in terms of the reaction scalar amplitudes: 
\ba d_1&=&ImR_1,~ d_2=ReR_1,  
\nn\\
R_{1}&=&\frac{2}{3}r_1\biggl [g_{9}g_{3}^*+g_{10}g_{1}^*-
g_{7}g_{6}^*-g_{8}g_{5}^*+rg_{11}g_{4}^*+rg_{12}g_{2}^*\biggr ]+
\frac{2}{3}r_2\biggl [g_{21}g_{15}^*+ 
\nn \\
&&
g_{22}g_{13}^*-g_{19}g_{18}^*-g_{20}g_{17}^*+rg_{23}g_{16}^*+
rg_{24}g_{14}^*\biggr ]+\frac{2}{3}r_3\biggl [
g_{9}g_{15}^*+g_{10}g_{13}^*+  
\nn \\
&&
g_{21}g_{3}^*+g_{22}g_{1}^*-g_{7}g_{18}^*-g_{8}g_{17}^*-g_{19}g_{6}^*-g_{20}g_{5}^*+rg_{11}g_{16}^*+
\nn \\
&&
rg_{12}g_{14}^*+rg_{23}g_{4}^*+rg_{24}g_{2}^*\biggr ]+  
\frac{2}{3}r_4\biggl [g_{8}g_{18}^*+g_{10}g_{15}^*+g_{19}g_{5}^*+
g_{21}g_{1}^*-
\nn \\
&&
g_{7}g_{17}^*-g_{9}g_{13}^*-g_{20}g_{6}^*-g_{22}g_{3}^*+rg_{12}g_{16}^*+ 
rg_{23}g_{2}^*-rg_{11}g_{14}^*-rg_{24}g_{4}^*\biggr ], 
\nn \\
d_3&=&ImR_2,~ d_4=ReR_2, 
\nn \\
R_{2}&=&\frac{2}{3}r_1\biggl [g_{9}g_{1}^*+g_{8}g_{6}^*-
g_{7}g_{5}^*-g_{10}g_{3}^*+rg_{11}g_{2}^*-rg_{12}g_{4}^*\biggr ]+
\frac{2}{3}r_2\biggl [g_{20}g_{18}^*+ 
\nn \\
&&
g_{21}g_{13}^*-g_{19}g_{17}^*-g_{22}g_{15}^*+rg_{23}g_{14}^*-
rg_{24}g_{16}^*\biggr ]+\frac{2}{3}r_3\biggl [
g_{9}g_{13}^*+g_{8}g_{18}^*+
\nn \\
&&
g_{20}g_{6}^*+g_{21}g_{1}^*-g_{7}g_{17}^*-g_{10}g_{15}^*-g_{19}g_{5}^*-g_{22}g_{3}^*+
rg_{11}g_{14}^*-rg_{12}g_{16}^*+
\nn \\
&&
rg_{23}g_{2}^*-rg_{24}g_{4}^*\biggr ]+ \frac{2}{3}r_4\biggl [g_{7}g_{18}^*+g_{8}g_{17}^*+g_{9}g_{15}^*+
g_{10}g_{13}^*-g_{19}g_{6}^*-
\nn \\
&&
g_{20}g_{5}^*-g_{21}g_{3}^*-
g_{22}g_{1}^*+rg_{12}g_{14}^*-rg_{23}g_{4}^*+rg_{11}g_{16}^*-rg_{24}g_{2}^*\biggr ], 
\nn \\
d_{5}&=&-\frac{4}{3}Im\biggl \{r_1\biggl [g_{1}g_{3}^*+g_{5}g_{6}^*
+rg_{2}g_{4}^*\biggr ]+r_2\biggl [g_{13}g_{15}^*+
g_{19}g_{18}^*+rg_{14}g_{16}^*\biggr ]+ 
\nn \\
&&
r_3\biggl [g_{1}g_{15}^*+g_{5}g_{18}^*+g_{13}g_{3}^*+
g_{17}g_{6}^*+rg_{2}g_{16}^*+rg_{14}g_{4}^*\biggr ]+
r_4\biggl [g_{6}g_{18}^*+
\nn \\
&&
g_{5}g_{17}^*+ g_{15}g_{3}^*+g_{13}g_{1}^*+rg_{14}g_{2}^*+
rg_{2}g_{4}^*\biggr ]\biggr \}, 
\nn \\
d_{6}&=&-\frac{4}{3}Im\biggl \{r_1\biggl [g_{7}g_{8}^*+g_{9}g_{10}^*
+rg_{11}g_{12}^*\biggr ]+r_2\biggl [g_{19}g_{20}^*+
g_{21}g_{22}^*+rg_{23}g_{24}^*\biggr ]+ 
\nn \\
&&
r_3\biggl [g_{7}g_{20}^*+g_{9}g_{22}^*+g_{19}g_{8}^*+
g_{21}g_{10}^*+rg_{11}g_{24}^*+rg_{23}g_{12}^*\biggr ]+
r_4\biggl [g_{9}g_{21}^*+
\nn \\
&&
g_{10}g_{22}^*+ g_{17}g_{7}^*+g_{20}g_{8}^*+rg_{12}g_{24}^*+
rg_{11}g_{23}^*\biggr ]\biggr \}. 
\nn\ea

\end{document}